\documentclass[rnote]{aa}

\usepackage{natbib}
\usepackage{graphicx}
\usepackage{txfonts}
\usepackage{url}

\begin{document}
%=======================================================================
% The header of the document:
%=======================================================================
\title{Chemistry in Disks.}

\subtitle{IV. Benchmarking gas-grain chemical models with surface reactions.}

   \author{D. Semenov\inst{1}
           \and
           F. Hersant\inst{2,3}
           \and
           V. Wakelam\inst{2,3}
           \and
           A. Dutrey\inst{2,3}
           \and
           E. Chapillon\inst{4}
           \and
           St. Guilloteau\inst{2,3}
           \and
           Th. Henning\inst{1}
           \and
           R. Launhardt\inst{1}
           \and
           V.  Pi\'etu\inst{5}
	   \and
           K.Schreyer\inst{6}
	}

   \offprints{D. Semenov, \email{semenov@mpia.de}}

 \institute{Max Planck Institute for Astronomy,
  K\"onigstuhl 17, D-69117 Heidelberg, Germany
  	\email{semenov@mpia.de}
                \and
Universit\'e de Bordeaux, Observatoire Aquitain des Sciences de
l'Univers, 2 rue de l'Observatoire, BP 89, F-33271 Floirac Cedex, France
\and
CNRS, UMR 5804, Laboratoire d'Astrophysique de Bordeaux,
2 rue de l'Observatoire, BP 89, F-33271 Floirac Cedex, France
\and
Max-Planck-Institut f\"ur
  Radioastronomie, Auf dem H\"ugel 69, D-53121 Bonn, Germany
               \and
IRAM, 300 rue de la piscine, F-38406 Saint Martin d'H\`eres,
France
               \and
Astrophysikalisches Institut und Universit\"ats-Sternwarte,
Schillerg\"asschen 2-3, D-07745 Jena, Germany
}

\date{Received ??? 2010 / Accepted ??? 2010}

\abstract
% Context:
{We detail and benchmark two sophisticated chemical models developed by the
Heidelberg and Bordeaux astrochemistry groups.}
% Aims:
{The main goal of this study is to elaborate on a few well-described tests for
state-of-the-art astrochemical codes covering a range of physical conditions and chemical
processes, in  particular those aimed at constraining current and future
interferometric observations of protoplanetary disks.}
% Method:
{We consider three physical models:
a cold molecular cloud core, a hot core, and an outer region of a T Tauri disk.
Our chemical network (for both models) is based on the original gas-phase osu\_03\_2008 ratefile
and includes gas-grain interactions and a set of surface reactions for the
H-, O-, C-, S-, and N-bearing molecules.
The benchmarking is performed with the increasing complexity of the considered
processes: (1) the pure gas-phase chemistry, (2) the gas-phase
chemistry with accretion and desorption, and (3) the full gas-grain model with
surface reactions. Using atomic initial abundances with heavily depleted
metals and hydrogen in its molecular form, the chemical evolution is
modeled within $10^9$~years.
% Two scenarios for surface hydrogenation are examined, namely,
% when atomic hydrogen is able to tunnel through a reaction barrier and when it
% is not.
}
% Results:
{The time-dependent abundances calculated with the two chemical models are
essentially the same for all considered physical cases and for all species,
including the most complex polyatomic ions and organic molecules. This result 
however required a lot of efforts to make all necessary details
consistent through the model runs, e.g. definition of the gas particle density,
density of grain surface sites, the strength and shape of the UV radiation
field, etc.}
% Conclusions:
{The reference models and the benchmark setup, along with the
two chemical codes and resulting time-dependent abundances are made
publicly available in the Internet. This will
facilitate and ease the development of other astrochemical models,
and provide to non-specialists a detailed description of the
model ingredients and requirements to analyze the cosmic chemistry as
studied, e.g., by (sub-) millimeter observations of molecular lines.}


\titlerunning{Benchmarking of chemical models.}
\authorrunning{D. Semenov et al.}

\maketitle

\section{Introduction}
\label{intro}
Astrochemistry plays an important role in our understanding of the
star- and planet-formation processes across the Universe
\citep[e.g.,][]{Bergin_ea07,Kennicutt_98,SVB_05,vDB98}. Molecules serve as
unique probes of physical conditions, evolutionary stage, kinematics and
chemical complexity.
% Chemical species and their interaction with dust
% grains contribute substantially into a thermal balance of the medium
% \citep[e.g.,][]{Tielens_05}.
In astrophysical objects molecular concentrations
are usually hard to predict analytically as it involves a multitude of chemical
processes that almost never reach a chemical steady-state. Consequently,
to calculate molecular concentrations one has to specify initial conditions
and abundances, and solve a set of chemical kinetics equations.

Since the first seminal studies on chemical modeling of the interstellar medium (ISM)
by \citet{Bates_Spitzer1951}, \citet{HerbstKlemperer73}, and \citet{WatsonSalpeter72},
numerous chemical models of molecular clouds \citep[e.g.,][]{HHL92},
protoplanetary disks \citep[e.g.,][]{Aea02}, shells around late-type stars
\citep[e.g.,][]{AC_06}, AGN tori \citep[e.g.,][]{Meijerink_ea07}, and planetary atmospheres
\citep[e.g.,][]{Gladstone_ea96} have been developed.
These models are mainly based on three widely applied ratefiles:
the UDFA (Umist Database For Astrochemistry\footnote{\url{http://www.udfa.net}})
\citep{umist95,umist99,Woodall_ea07},
the OSU database
(Ohio State University\footnote{\url{http://www.physics.ohio-state.edu/~eric/research.html}})
\citep{OSU03},
and a network incorporated in the ``Photo-Dissociation
Region'' code from
Meudon\footnote{\url{http://aristote.obspm.fr/MIS/pdr/charge.html}}
\citep{Lepetit_ea06}. Recently another database of chemical reactions for the 
interstellar medium and planetary atmospheres, KIDA\footnote{\url{http://kida.obs.u-bordeaux1.fr}}, 
has been opened to the astrochemical community. The main aim of KIDA is to group all kinetic data 
to model the chemistry of these environments and to allow physico-chemists to upload their data 
to the database and astrochemists to present their models.
Several compilations of surface reactions have also been presented
\citep[e.g.,][]{AllenRobinson77,TielensHagen82,HHL92,HH93,GH_06}.
The detailed physics and chemistry processes incorporated in the modern
theoretical models allow to predict and, at last, to fit the observational
interferometric data such as, e.g., molecular line maps of protoplanetary
disks and molecular
clouds \citep[e.g.,][]{Semenov_ea05,Dutrey_ea07,Goicoechea_ea09,PH_09,Hea_10}.
However, with the advent of forthcoming high-sensitivity, high-resolution
instruments like ALMA, ELT, and JWST, the degree of complexity of these models
will have to be increased and their foundations to be critically re-evaluated
to match the quality of the data
\citep[e.g.,][]{AsensioRamos_ea07,Lellouch_08,Semenov_ea08}.

There are other difficulties with cosmochemical models.
Apart from often poorly known physical properties of the object,
its chemical modeling suffer from uncertainties of the reaction rates,
of which only $\la 10\%$ have been accurately determined
\citep[see, e.g.,][]{calc_rate_uncert,Wakelam_ea06,Vasyunin_ea08}.
In contrast to these internal uncertainties, another major source of ambiguity
is the lack of detailed description of chemical models that often incorporate
different astrochemical ratefiles, initial abundances, dust grain properties,
etc., thus making results hard to interpret and compare.
% This is partly due to i) the lack of detailed description of adopted
% model parameters, ii) approaches
% utilized to calculate reaction rates, and iii) partly due to the use of
%modified ratefiles
% (with e.g. updated rate constants or removed/added reactions).
It is thus essential to perform consistent benchmarking of various advanced
chemical models under a variety of realistic physical conditions, in particular
those encountered in protoplanetary disks. To the best of our
knowledge, the only important benchmarking study attempted so far
focused on several PDR physico-chemical codes \citep{PDR_bench}.

The ultimate goal of present work is i) to provide a detailed description of our
time-dependent sophisticated chemical codes and ii) to establish a set of
reference models covering a wide range of physical conditions, from a cold
molecular cloud core to a hot corino and an outer region of a protoplanetary
disk. In comparison with the PDR benchmark, our study is based on a more
extended set of gas-grain and surface reactions. All the relevant software,
figures, reaction network, and calculated abundances are available in the
Internet\footnote{\url{http://www.mpia.de/homes/semenov/Chemistry_benchmark/home.html}}.

The organization of the paper is the following. In Section~\ref{chem_models},
the two codes and
the chemical network are presented in detail, along with our approach to
calculate the reaction rates. In Section~\ref{tests}, we describe
various benchmarking models, the physical conditions, and the initial chemical
abundances. The benchmarking runs and their results are presented and
compared in Section~\ref{res}. Final conclusions
are drawn in Section~\ref{sum}.

\section{Chemical Models}
\label{chem_models}

\subsection{Heidelberg and Bordeaux chemical codes}
\label{chem_codes}
In this study, we compare two chemical codes, ``ALCHEMIC'' and
``NAUTILUS'', developed independently over
the last several years by the Heidelberg and Bordeaux astrochemistry groups.
Both codes have been intensively utilized in various studies of molecular
cloud and protoplanetary disk chemistry, e.g. the influence of
the reaction rate uncertainties on the results of chemical modeling of cores
\citep{Vasyunin04,Wakelam_ea05,Wakelam_ea06} and disks \citep{Vasyunin_ea08},
modeling of the disk chemical evolution with turbulent diffusion
\citep{Semenov_ea06,Hersant_ea09}, interpretation of interferometric data
\citep{Semenov_ea05,Dutrey_ea07,Schreyer_ea08,Henning_ea10}, and predictions for
ALMA \citep{Semenov_ea08}. Both codes are optimized to model
time-dependent evolution of a large set of gas-phase and surface species linked
through thousands of gas-phase, gas-grain and surface processes. We found that
both these codes have comparable performance, accuracy, and fast convergence
speed thanks to the use of advanced ODE and sparse matrix linear system solvers.

The Heidelberg ``ALCHEMIC'' code is written in Fortran~77 and based on the
publicly available DVODPK
(Differential Variable-coefficient Ordinary Differential equation solver
with the Preconditioned Krylov method GMRES for the solution of linear
systems) ODE package\footnote{\url{http://www.netlib.org/ode/vodpk.f}}. The
full Jacobi matrix is generated automatically from the supplied chemical
network to be used as a left-hand preconditioner. For astrochemical models
dominated by reactions involving hydrogen, the Jacobi matrix is sparse (having $\la
1\%$ of non-zero elements), so to solve the corresponding linearized system of
equations a high-performance sparse unsymmetric MA28 solver from the Harwell
Mathematical Software Library\footnote{\url{http://www.hsl.rl.ac.uk/}} is used.
As ratefiles, both the recent OSU\,06 and
osu.2008 and the RATE\,06 networks can be utilized.
A typical run with relative and absolute accuracies of $10^{-5}$ and $10^{-15}$
for the full gas-grain network with surface chemistry ($\sim 650$ species,
$\sim 7\,000$ reactions) and 5~Myr of evolution takes only a few seconds of CPU
time on the Xeon~2.8GHz PC (with gfortran 4.3 compiler).

The Bordeaux ``NAUTILUS'' code is written in Fortran~90 and uses the LSODES (Livermore Solver for
Ordinary Differential Equations with general Sparse Jacobian matrix) solver, part of
ODEPACK\footnote{\url{http://www.netlib.org/odepack/}} \citep{Hindmarsh83}. ``NAUTILUS'' is adapted from the
original gas-grain model of \cite{HH93} and its subsequent evolutions made over the years at the Ohio State University. The
main differences with the OSU model rely in the numerical scheme and optimization (``NAUTILUS'' is about 20
times faster), the possibility to compute 1D structures with diffusive transport and the adaptation to disk
physics and chemistry. The full Jacobian is computed from the chemical network without preconditioning. For
historical reasons, only the OSU rate files are being used, although minor adjustments could permit to extend
the model to other networks. Similar performances with ``ALCHEMIC'' are achieved, the same typical run of the
full gas-grain network taking a few seconds on a standard desktop computer.\\

Below we detail how we model various gas-phase and surface reactions.

\subsection{Modeling of Chemical Processes}
\label{chem_rates}

Chemical codes solve numerically the equations of chemical kinetics describing
the formation and destruction of molecules:
\begin{equation}
\frac{dn_i}{dt} = \sum_{l,m}k_{lm}n_ln_m - n_i\sum_{i\neq l}k_ln_l +
k_i^{\rm des}n_i^{s} - k_i^{\rm acc}n_i
 \end{equation}
 \begin{equation}
\frac{dn_i^s}{dt} = \sum_{l,m}k^{s}_{lm}n_l^{s}n_m^{s} - n_i^{s}\sum_{i\neq
l}k_l^{s}n_l^{s} - k_i^{\rm des}n_i^{s}
+ k_i^{\rm acc}n_i
\end{equation}
where $n_i$ and $n_i^{s}$ are the gas-phase and surface concentrations of the
$i$-th species (cm$^{-3}$),
$k_{lm}$ and $k_{l}$ are the gas-phase
reaction rates (in units of s$^{-1}$ for the first-order kinetics and
cm$^3$\,s$^{-1}$ for the second-order
kinetics), $k_i^{\rm acc}$ and $k_i^{\rm des}$ denote the accretion and
desorption rates
(s$^{-1}$),
respectively, and $k_{lm}^s$ and $k_{l}^s$ are surface reaction rates (cm$^3$\,s$^{-1}$).
% We consider only first-order and second-order kinetics, without three-body
% reactions. The chemistry includes both gas phase and gas-grains and surface
% reactions.
% , and is calculated for a long time span of $10^9$~years.

\subsubsection{Gas-phase reactions}
\label{gas_phase}
For the benchmarking purposes, we utilize a recent osu\_03\_2008
gas-phase
ratefile\footnote{\url{http://www.physics.ohio-state.edu/~eric/research.html}}.
This ratefile incorporates the data for the 456 atomic, ionic, and molecular
species involved in the 4389 gas-phase reactions. The corresponding reaction
rates are calculated
as follows, using the standard Arrhenius representation:
\begin{equation}
\label{Arrhenius}
k(T) = \alpha \left( \frac{T}{300} \right)^\beta
\exp{\left(-\frac{\gamma}{T}\right)},
\end{equation}
where $\alpha$ is the value of the reaction rate at the room temperature of
300~K, the parameter $\beta$ characterizes the temperature dependence of the
rate, and $\gamma$ is the activation barrier (in Kelvin). We utilize this
expression for all gas-phase two-body processes, e.g. ion-molecular and
neutral-neutral reactions.

For the cosmic ray and FUV ionization and
dissociation the following prescriptions have been used.

\paragraph{Cosmic ray (CR) ionization \& dissociation}
\begin{equation}
\label{kcr}
k_{\rm CR} = \alpha \zeta_{\rm CR},
\end{equation}
where $\zeta_{\rm CR}=1.3\,10^{-17}$~s$^{-1}$ is the adopted CR
ionization rate. In all benchmarking models any additional sources of
ionization like the X-ray stellar radiation and the decay of short-living
radionucleides (e.g., $^{26}$Al and $^{40}$K) are not considered.
This small set of reactions includes
ionization of relevant atoms and dissociation of molecular hydrogen.
The same expression is used to compute photodissociation and ionization
by the CR-induced FUV photons. Also, the CR- and UV-driven dissociation of
surface species is calculated by the Eq.~\ref{kcr} ($616$ processes).

\paragraph{FUV photodissociation \& ionization}
To calculate photoreaction rates through the environment,
we adopt pre-computed fits of \citet{van_Dishoeck88} for a 1D plane-parallel
slab, using the Draine FUV IS radiation field. Unlike in the PDR benchmarking
study, the self-and mutual-shielding of CO and H$_2$ against UV
photodissociation are not taken into account for simplicity.
The corresponding rate is calculated as:
\begin{equation}
k_{\rm FUV} = \alpha \exp{\left(-\gamma A_{\rm V}\right)}\chi,
\end{equation}
where the A$_{\rm V}$ is the visual extinction (mag.) and the $\chi$
is the unattenuated FUV flux expressed in units of the FUV interstellar radiation field $\chi_0$ of
\citet{Draine_78}. We do not take the Ly$_\alpha$ radiation into account
as it requires sophisticated modeling of the radiation transport -- a
full-scale benchmarking study by itself
\citep[e.g.,][]{Pascucci_ea04,PDR_bench}.

\subsection{Gas-grain interactions}
\label{gas_grain}
% Gas-phase species interact with grains, they can stick on grain surfaces and
%be
% evaporated by thermal and non-thermal processes.
%
For simplicity, we assume that the dust grains are uniform
spherical particles, with a radius of $a_g=0.1\,\mu$m, made of amorphous olivine, with density of
$\rho_d=3$~g\,cm$^{-3}$ and a dust-to-gas mass ratio $m_{d/g}=0.01$. The surface
density of sites is $N_s=1.5\,10^{15}$~sites\,cm$^{-2}$, and the total number of
sites per such a grain is $S=1.885\,10^6$. The dust and gas temperatures are
assumed to be the same.

Gas-grain interactions start with the accretion of neutral molecules onto dust
surfaces with a sticking efficiency of 100$\%$ (195 processes). Molecules are
assumed to only physisorb on the grain surface (by van der Waals force) rather
than by forming a chemical bond with the lattice (chemisorption). The rate of
accretion of a gas-phase species $i$ (cm$^{-3}$ s$^{-1}$) is given by:
\begin{equation}
R_{\rm acc}(i)=k_{\rm acc}(i)n(i),
\end{equation}
where $n(i)$ is the density of gas-phase species $i$ (cm$^{-3}$), and
$k_{\rm acc}(i)=\sigma_d \langle v(i)\rangle n_d$ is the accretion rate. Here
$\sigma_d=\pi a_g^2$ is the geometrical cross section of the grain with
the radius $\rm a_g$, $n_d$ is the density of grains (cm$^{-3}$)
and $\rm \langle v(i)\rangle$ is the thermal velocity of species $i$ (cm\,s$^{-1}$). The latter
quantity is expressed as $\sqrt{8k_B T/(\pi\mu(i)m_p)}$, with $T$ being the gas
temperature (K), $m_p=1.66054\,10^{-24}$ g is the proton's mass, $\mu(i)$ is
the reduced mass of the molecule $i$ (in atomic mass units), and
$k_B=1.38054\,10^{-16}$~erg\,K$^{-1}$ is the Boltzmann's constant. All constants are summarized in Table 2.

In addition, electrons can stick to neutral grains, producing negatively charged
grains. Atomic ions radiatively recombine on these negatively
charged grains, leading to grain neutralization (13 reactions in total).
The corresponding
two-body reaction rate is calculated as:
\begin{equation}
k(T) = \alpha \left( \frac{T}{300} \right)^\beta \frac{n_{\rm H}}{n_d},
\end{equation}
where $n_{\rm H}$ is the total hydrogen nucleus density (cm$^{-3}$). This quantity is
calculated by the following expression:
\begin{equation}
n_{\rm H} = \frac{\rho}{\mu m_p},
\end{equation}
where $\rho$ is the gas mass density (g\,cm$^{-3}$), and $\mu = 1.43$ is the
mean mass per hydrogen nucleus.
Consequently, the density of grains $n_d$ is
expressed as:
\begin{equation}
n_d = \frac{\rho m_{d/g}}{4/3\pi a_g^3\rho_d\mu}.
\end{equation}
We consider neither interactions of molecular ions with grains nor
the photoelectric effect leading to positively charged grains.

In our simplified model used for benchmarking purposes, the surface molecules
can leave the grain by only two mechanisms. First, they can desorb back into
the gas phase when a grain is hit by a relativistic iron nucleus and heated
for a short while up to several tens of Kelvin (160 reactions, cosmic-ray induced desorption). Second, in
sufficiently warm regions thermal desorption becomes efficient.
The thermal desorption for the $i$-th surface molecule is calculated by the
Polanyi-Wigner equation:
\begin{equation}
\label{kdes}
k_{\rm des}(T_d) = \nu(i) \exp{\left(-\frac{E_{\rm des}}{T_d}\right)}.
\end{equation}
Here $\rm \nu(i)=\sqrt{\frac{2N_s k_B E_{\rm des}}{\pi^2 m m_p}}$
is the characteristic vibrational frequency of the $i$-th species, $E_{\rm des}$ is its desorption energy (Kelvin), $m$ the mass of the species and $T_d$ is the grain temperature. Desorption energies $E_{\rm
des}$ are taken from \citet{GH_06}. We do not distinguish between various
thermal evaporation scenarios for different molecules, e.g. via ``volcanic'' or
multilayer desorption \citep{Collings_ea04}.

The cosmic ray desorption rate is computed as suggested by Eq.~15 in \citet{HH93}. It is
based on the assumption that a cosmic ray particle (usually an iron nucleus)
deposits on average 0.4~MeV into a dust grain of the adopted radius, impulsively
heating it to a peak temperature $T_{\rm crp}=70$~K \citep[see
also][]{Leger_ea85}. The resulting rate is:
\begin{equation}
\label{kdescp}
k_{\rm crd} = f k_{\rm des}(70~K)
\end{equation}
$k_{\rm crd}$ is the thermal evaporation rate at 70~K times the 
fraction of time the grain temperature stays close to 70~K. 
The fraction $f$ is a ratio of a grain cooling timescale via desorption of molecules
to the timescale of subsequent heating events (e.g., for a $0.1\mu$m silicate particle and 
the standard CR ionization rate $1.3\,10^{-17}$~s$^{-1}$ these timescales are $\sim 10^{-5}$~s and
$\sim 3\,10^{13}$~s, respectively). 
Other 
non-thermal desorption mechanisms that can play an important role
in chemistry of protoplanetary disks like
photodesorption are not considered.

\subsection{Surface Reactions}
\label{surface}
Surface reactions (532 in the network) are treated in the standard rate
approximation, assuming
only Langmuir-Hinshelwood formation mechanism, as described, e.g., in \citet{HHL92}.
Surface species
are only allowed to move from one surface site to another by thermal hopping.
When two surface species find each other, they can recombine. We assume that the
products do not leave the surface as the excess of energy produced by such a
reaction will be immediately absorbed by the grain lattice, i.e. we did not include the desorption process proposed by \citet{Garrod_ea07}.

Rate coefficients for surface reactions between species $i$ and $j$ are
expressed as follows:
\begin{equation}
k_{i,j}=P (R_{\rm diff}(i)+R_{\rm diff}(j))/n_d.
\end{equation}
Here $P$ is the probability for the reaction to
occur. This parameter is $1$ for an exothermic
reaction without activation energy  and $1/2$ if the two reactants are of the same type. For exothermic reactions with activation energy $E_a$ (or endothermic reactions), this probability is $\alpha \exp{(-\frac{E_a}{T})}$, with $\alpha$ the branching ratio of the reaction ($\alpha = 1/3$ if there are three reaction channels). In some cases, tunneling effects can increase this probability\footnote{This process is not included in ALCHEMIC and thus has not been tested in this benchmark.}. When this is the case, $P$ is computed through the following formula \citep{HHL92}:
\begin{equation}
P = \alpha \exp[-2 (b/\hbar)(2 k_B \mu(i) mp E_a)^{1/2}]
\end{equation}
with $b$ the barrier thickness (1 \AA) and $\hbar$ the Planck constant times $2\pi$ ($1.05459\times 10^{-27}$~erg~s$^{-1}$).

%$\alpha \exp{(-\frac{E_a}{T})} $ is the probability for the reaction to
%occur. This parameter is $1$ for an exothermic
%reaction without activation energy, $1/2$ for a homogeneous barrierless
%reaction, and smaller for an endothermic reaction.
The thermal diffusion rate of species $i$ is:
\begin{equation}
R_{\rm diff}(i)=\nu(i)\exp(-\frac{T_{\rm diff}(i)}{T_d})/S .
\end{equation}
Here $k_B T_{\rm diff}$ is the activation energy of diffusion for the $i$-th
molecule.

The diffusion and desorption energies of a limited set of molecular species
have been derived \citep[e.g.,][]{Katzea99,Bisschop_ea06,Oeberg_ea07}. Moreover,
these energies depend on the type of the surface lattice and its
structural properties (porosity, crystallinity)
\citep[e.g.,][]{Acharyya_ea07,Thrower_ea09}. While diffusion energy of a
species must be lower than its desorption energy, the exact ratio is not well
constrained for a majority of molecules in our network \citep[except for
H$_2$, see][]{Katzea99}. Following \citet{RuffleHerbst00}, we adopted the
$T_{\rm diff}/T_{\rm des}$ ratio of 0.77 for the all relevant species in the
model.

It has been proposed that light surface species like H, H$_2$ and isotopes can
also quantum tunnel through a potential well of the surface site and thus be
able to quickly scan the surface, greatly increasing the efficiency of surface
recombination even at very low temperatures \citep[e.g.,][]{DW_86,HHL92}.
However, according to the theoretical interpretation of Katz et
al. (1999) tunneling of atomic hydrogen on amorphous surfaces does not occur.
Therefore, it is not considered in our benchmarking study.

Finally, at very low influx of reacting species having a high recombination
rate, concentrations on a grain can become very low, leading to a stochastic
regime. In this case surface chemistry cannot be reliably described by the
standard rate equation method, which tends to overestimate the rates. Other
approaches like Monte Carlo techniques \citep[e.g.,][]{Vasyunin_ea09}, master
equations \citep[e.g.,][]{Greenea01}, and modified rate equations
\citep[e.g.,][]{Caselliea98,Garrod_ea09} should then be utilized.
As it has been shown by \citet{Vasyunin_ea09}, this only happens
for rather dilute, warm gas and very small grains, and when quantum
tunneling of hydrogen is allowed. Thus, the use of the standard rate equation
approach to model surface chemistry is fully justified in our benchmarking
study.

\subsection{Initial abundances \& other parameters}
\label{ia}
We use an up-to-date set of elemental abundances from \citet{Wakelam_ea08}. The 12 elemental species include H, He, N,
O, C, S, Si, Na, Mg, Fe, P and Cl.
Except for hydrogen, which is assumed to be entirely locked in molecular form, all elements are initially atomic. They are also ionized except for He, N and O (see Table~\ref{tab:init_abund}). All heavy elements are
heavily depleted from the gas phase similar to the ``low metals'' abundances of
\citet{Lee_ea98}. All grains are initially neutral. Since large-scale chemical
models with an extended set of surface reactions rarely reach a chemical steady
state, all benchmarking tests run over a long evolutionary time span of
$10^9$~years (with 60 logarithmic time steps, starting from 1~year). Both
chemical codes use the same absolute and relative accuracy parameters for the
solution, $10^{-20}$ and $10^{-6}$, respectively.

\section{Benchmarking models}
\label{tests}
The physical conditions of the five benchmarking cases are summarized in
Table~\ref{tab:phys_mods}, \citep[see also][]{SnowMcCall06,Hassel_ea08}.
These are chosen to represent realistic astrophysical object yet to be
relatively simple.
We decide to focus first on physical models representative of less complex
astrophysical media: a cold core (``TMC1'')
and a hot corino (``HOT CORE''). The ``TMC1'' model has a temperature of 10~K, a
hydrogen nucleus density of $2\times 10^4$~cm$^{-3}$, and a visual extinction of 10 mag.
The ``HOT CORE'' model has a temperature of 100~K, a hydrogen nucleus density
of $2\times 10^7$~cm$^{-3}$, and a visual extinction of 10 mag. Both
models have FUV IS RF $\chi = 1 \chi_0$ of \citet{Draine_78} and $\zeta_{CR} =
1.3\,10^{-17}$~s$^{-1}$.

Protoplanetary disks are more complex objects from
chemical point of view. Basically, an outer disk ($r\ga 50-100$~AU)
observable with modern radio-interferometers can be divided in three distinct
parts from top to bottom. The hot and tenuous disk surface (usually called
``disk atmosphere'') is located at above 4-6 gas scale heights. In this region
disk chemistry is similar to that of H{\sc II} and PDR region, with a
limited set of primal ionization-recombination processes. {There ionization
is mainly governed by the stellar X-ray radiation and stellar and interstellar FUV
radiation.} 
Closer to the
midplane, a partly X-ray and FUV-shielded region called the ``warm molecular
layer'' is located
(at about 1-3 scale heights). This mildly ionized, dense and warm layer is a
harbor of complex chemistry, where gas-grain interactions and endothermic
reactions are particularly active. There, a plethora of more complex species
is formed and reside in the gas. Below is the highly dense, cold and dark
midplane, where most molecules are frozen out onto grains, enabling a
variety of slow surface processes to be active.

Instead of simulating disk chemistry in the entire disk, we decide to pick up a
few representative disk regions with highly distinct physical conditions,
and within the reach of spatial resolution of modern interferometers.
We consider three representative layers taken at a radius of
$98$~AU (which requires a sub-arcsecond resolution even for nearby objects).
Those are designated ``DISK1'' (the disk midplane), ``DISK2'' (the warm
molecular layer), and ``DISK3'' (the disk atmosphere).
We take physical conditions similar to those encountered in the DM Tau disk,
for which a lot of high-resolution molecular data are available.
In our simulations, we adopted the 1+1D steady-state irradiated disk model with
vertical temperature gradient that represents
the low-mass Class~II protoplanetary disk surrounding the young T Tauri star DM
Tau \citep{DAea99}.
The disk has a radius of 800~AU, an accretion rate
$\dot{M}=10^{-8}\,M_{\odot}$\,yr$^{-1}$, a viscosity
parameter $\alpha = 0.01$, and a mass $M\simeq0.07\,M_{\odot}$
\citep{Dutrey_ea97,Pietu_ea07}.
The central T Tau star has an effective temperature $T_*=4\,000$~K, mass
$M_*=0.5M_{\odot}$, and radius
$R_*=2R_{\odot}$.

We assumed that the disk is illuminated by the FUV radiation from the central
star with an intensity $\chi=410\,\chi_0$ at the distance of $100$~AU and by the interstellar
UV radiation with intensity $\chi_0$ in plane-parallel geometry
\citep[][]{Draine_78,van_Dishoeck88,Bea03,Dutrey_ea07}. The visual extinction
for stellar light at a given disk cell is calculated as:
\begin{equation}
A_\mathrm{V}={N_\mathrm{H}\over1.59\cdot10^{21}}~\frac{\mathrm{mag}}{\mathrm{cm}^{-2}},
\end{equation}
where $N_\mathrm{H}$ is the vertical column density of hydrogen nuclei between
the point and the disk atmosphere. Note that, according to our definition, the 
unattenuated stellar FUV intensity for a fixed disk radius is the largest in the midplane 
and gets lower for upper disk heights as the distance to the star increases.

Consequently, the ``DISK1'' model is located at $6.768$~AU above the midplane,
has a temperature of 11.4~K, a hydrogen nucleus density of $5.413\times 10^8$~cm$^{-3}$, 
a visual extinction toward the star of
$40.35$ mag. and an extinction of $37.07$ mag. in the vertical direction, and
the FUV RF intensity of $\chi_*=428.3\chi_0$.
 The ``DISK2'' model is located at
$29.97$~AU above the midplane, has a temperature of 45.9~K, a hydrogen nucleus density
of $2.588\times 10^7$~cm$^{-3}$, a visual extinction toward the star of
$23.23$ mag. and the vertical extinction of $1.939$ mag., and
the FUV RF intensity of $\chi_*=393.2\chi_0$. Finally, the ``DISK3'' cell is
located at $45.44$~AU above the midplane, has a temperature of 55.2~K, a
hydrogen nucleus density of $3.669\times 10^6$~cm$^{-3}$, a visual extinction toward the
star of $1.608$ mag. and the vertical extinction of $0.217$ mag., and
the FUV RF intensity of $\chi=353.5\chi_0$. The remaining parameters of all
the models are described above. All those parameters are summarized in Table 1.

\section{Benchmarking Results}
\label{res}
\begin{figure*}
\includegraphics[width=0.21\textwidth,clip=,angle=90]{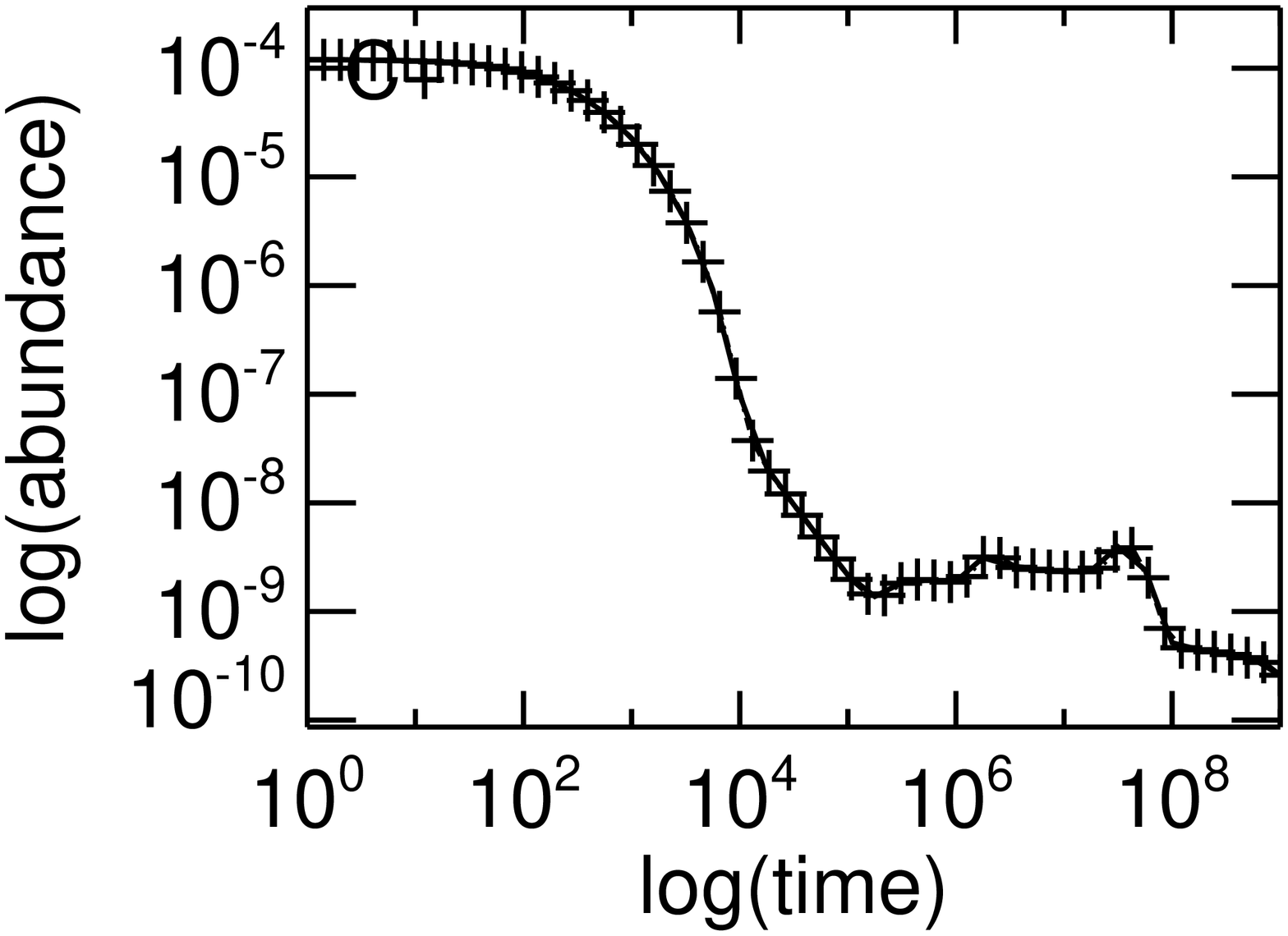}
\includegraphics[width=0.21\textwidth,clip=,angle=90]{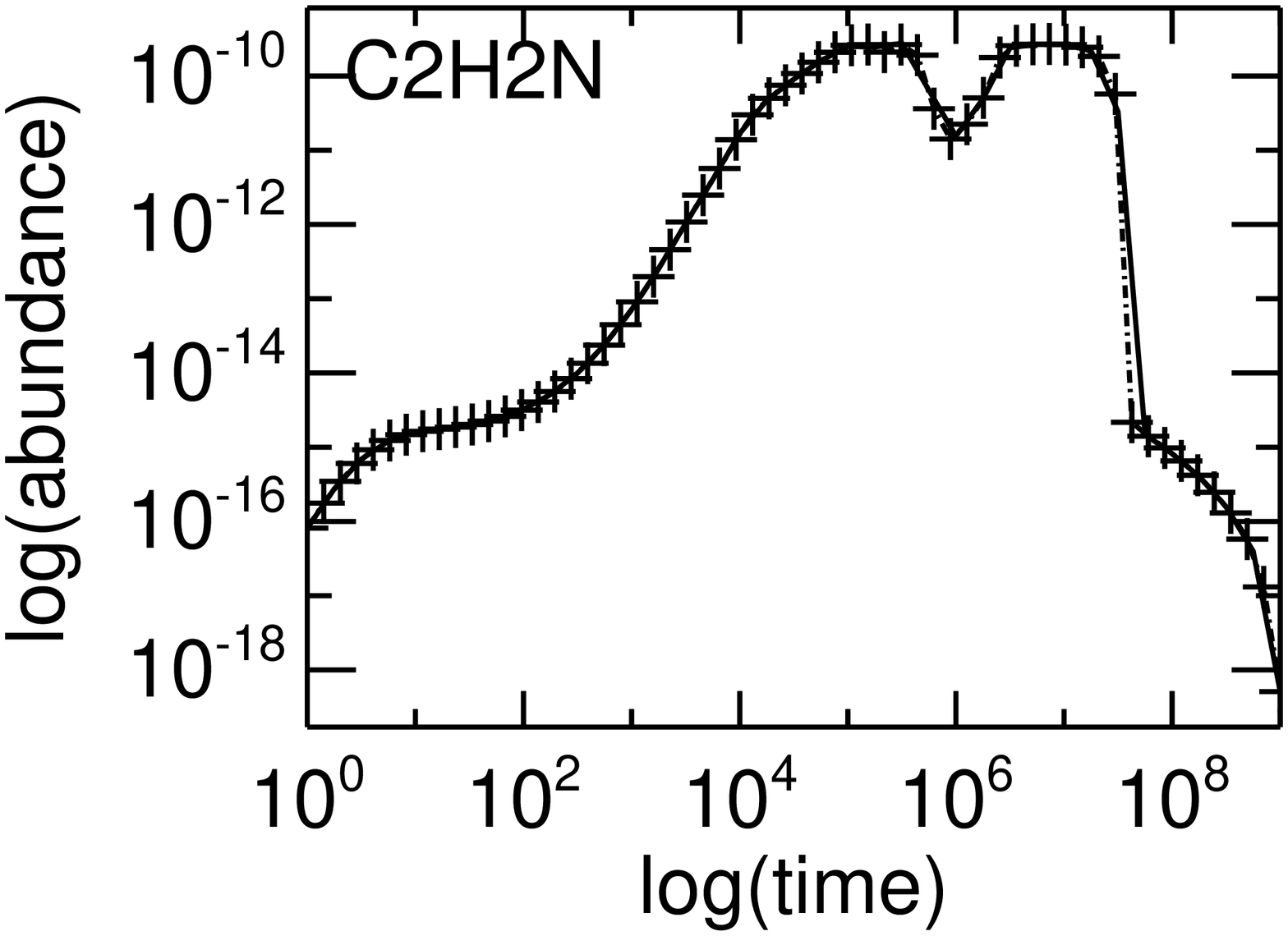}
\includegraphics[width=0.21\textwidth,clip=,angle=90]{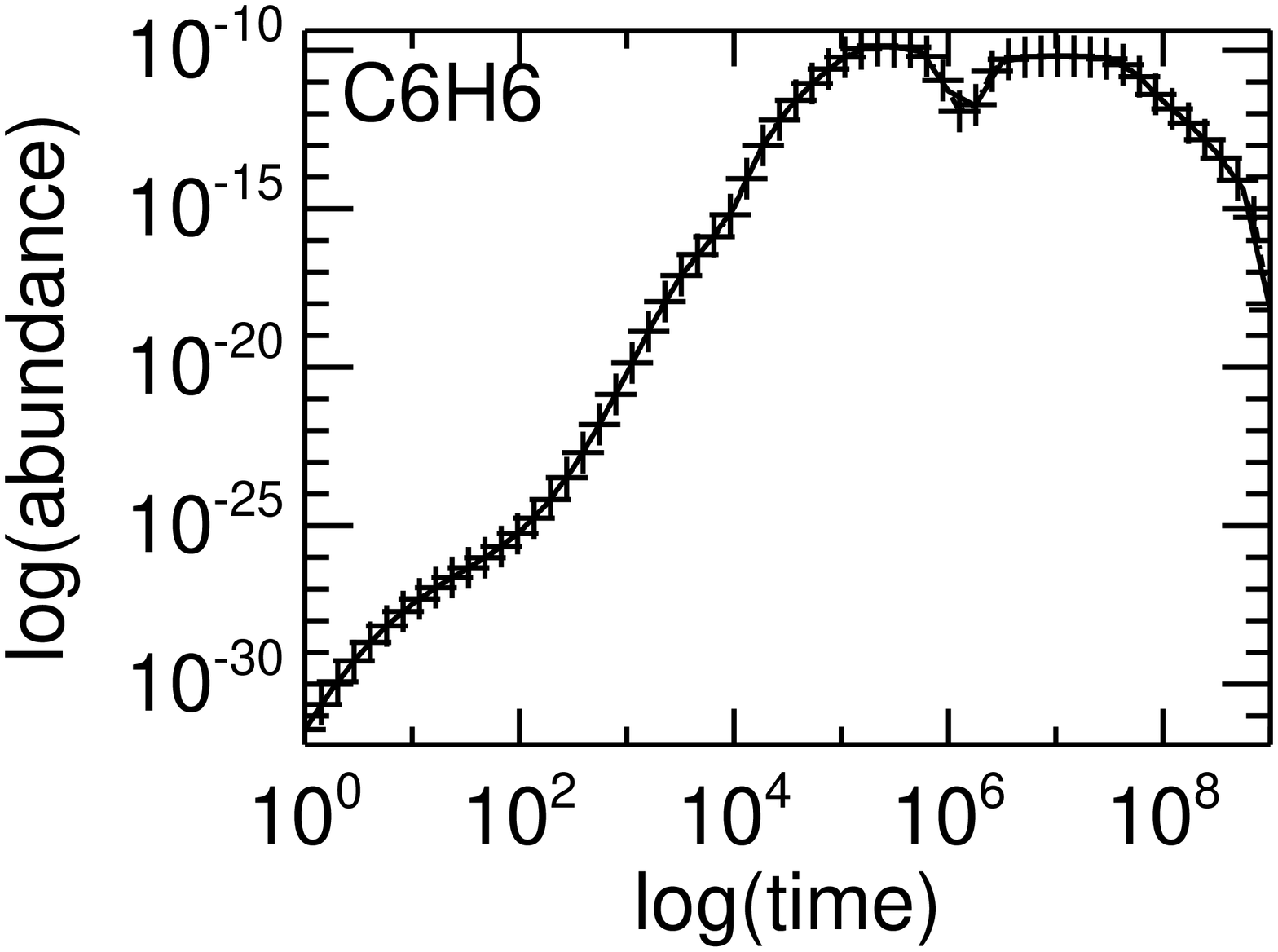}\\
\includegraphics[width=0.21\textwidth,clip=,angle=90]{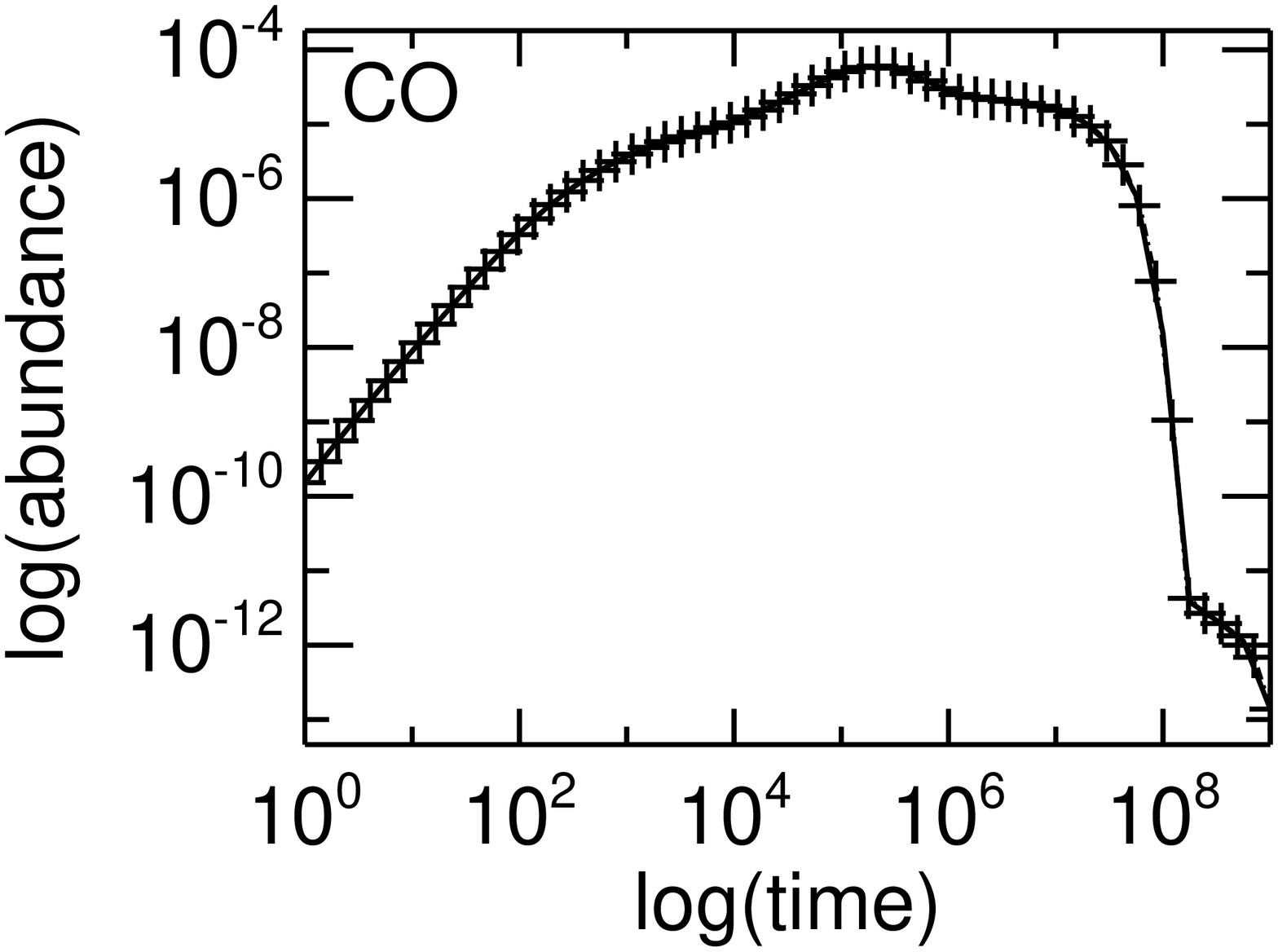}
\includegraphics[width=0.21\textwidth,clip=,angle=90]{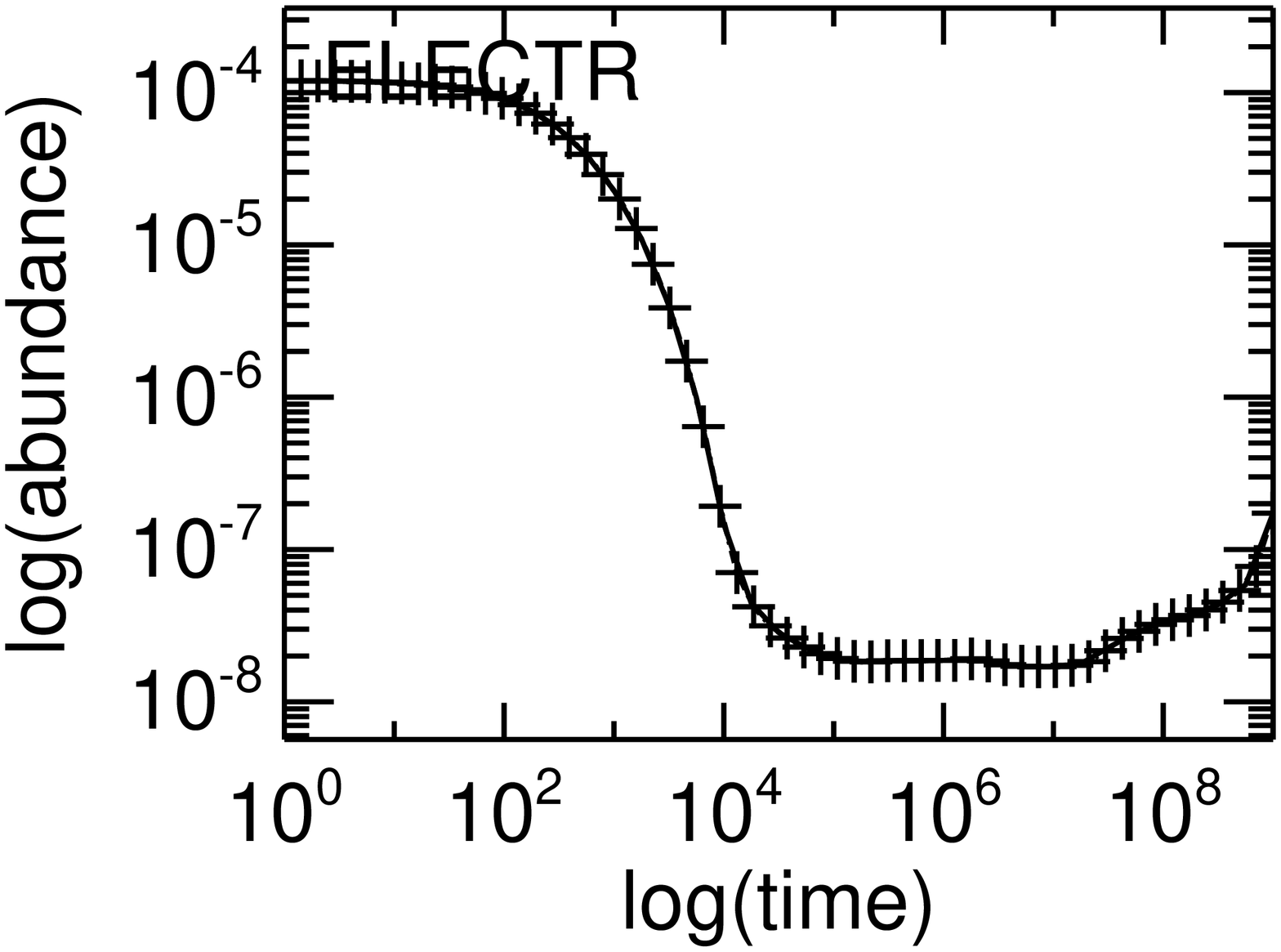}
\includegraphics[width=0.21\textwidth,clip=,angle=90]{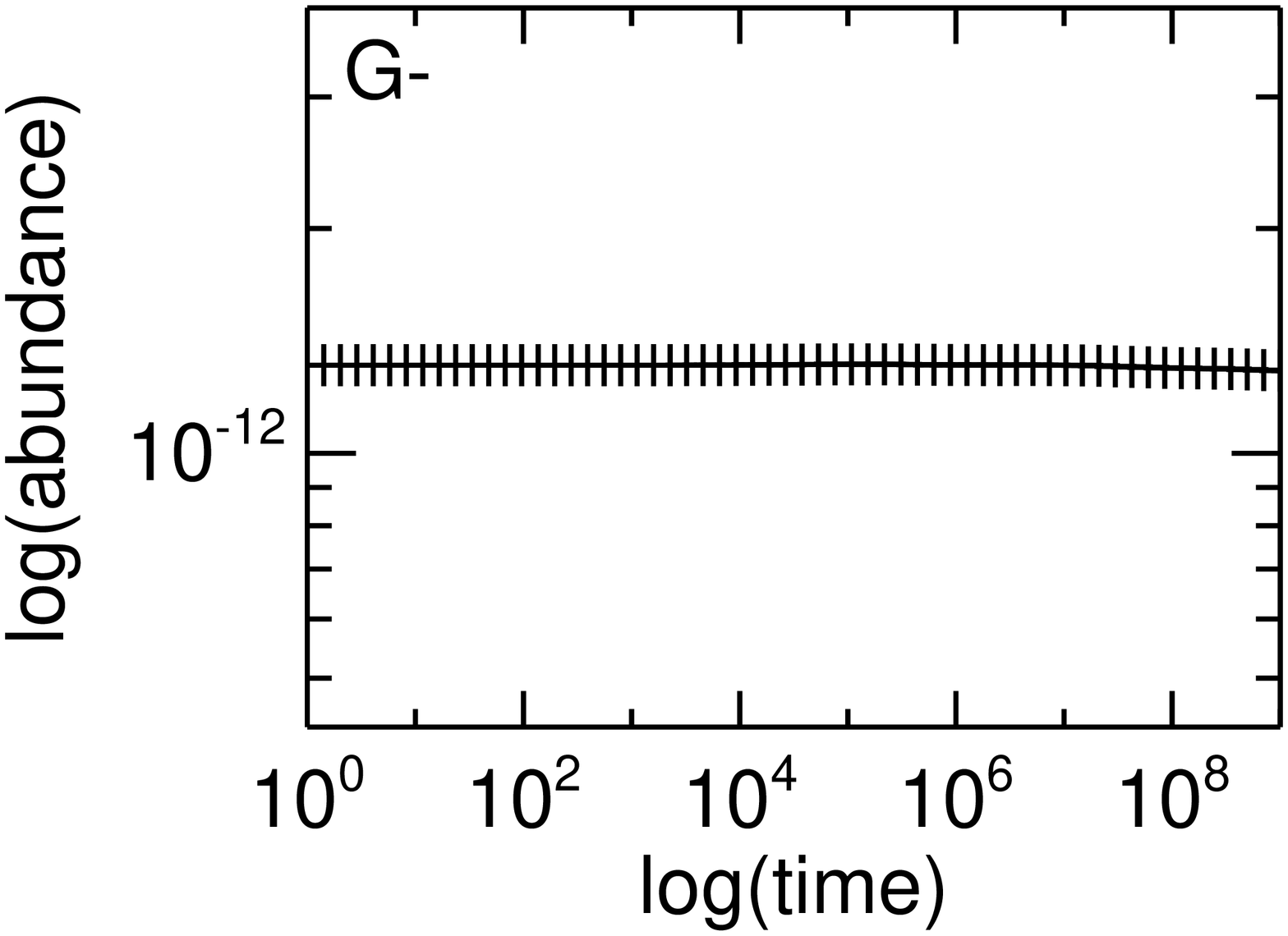}\\
\includegraphics[width=0.21\textwidth,clip=,angle=90]{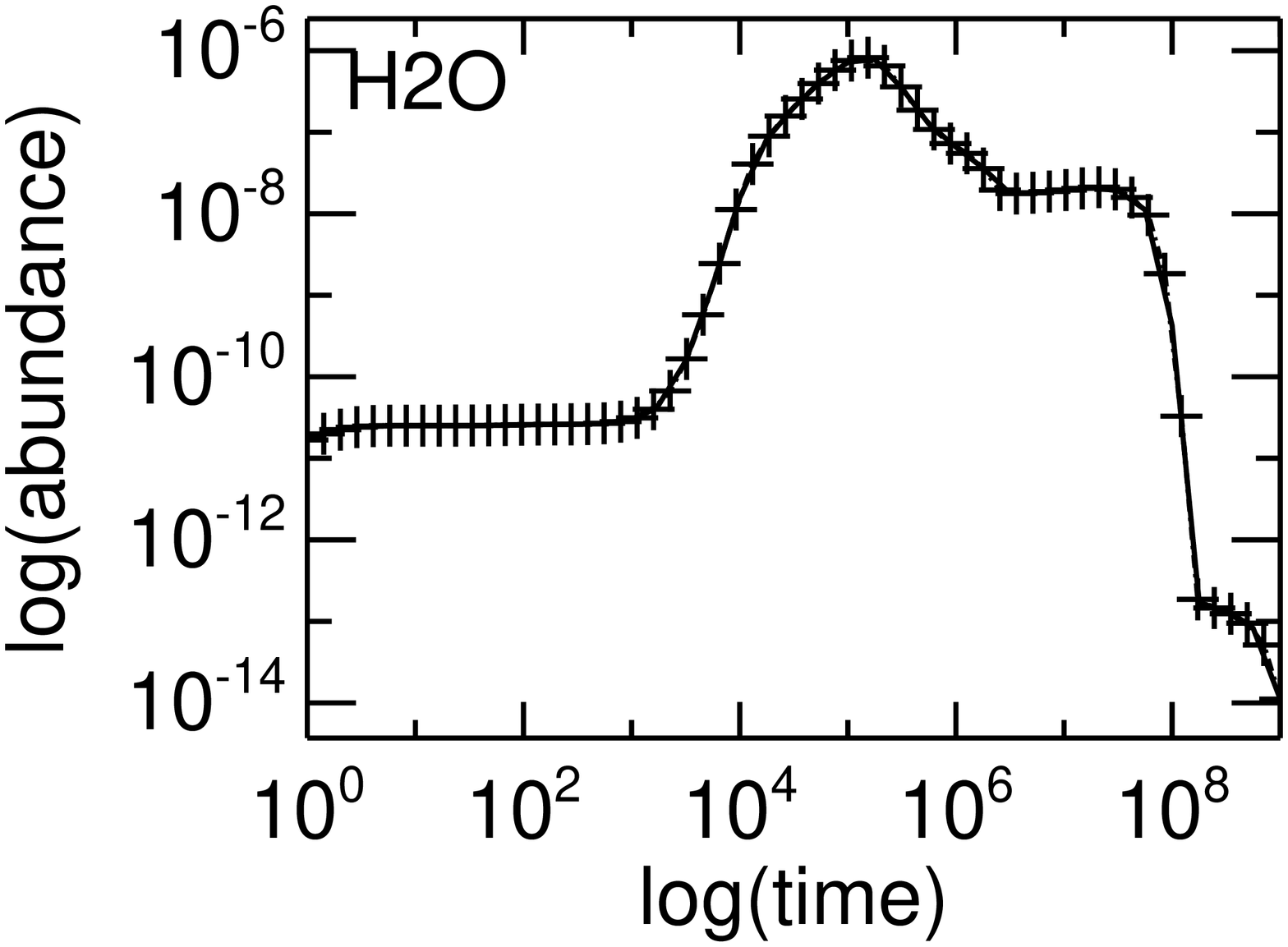}
\includegraphics[width=0.21\textwidth,clip=,angle=90]{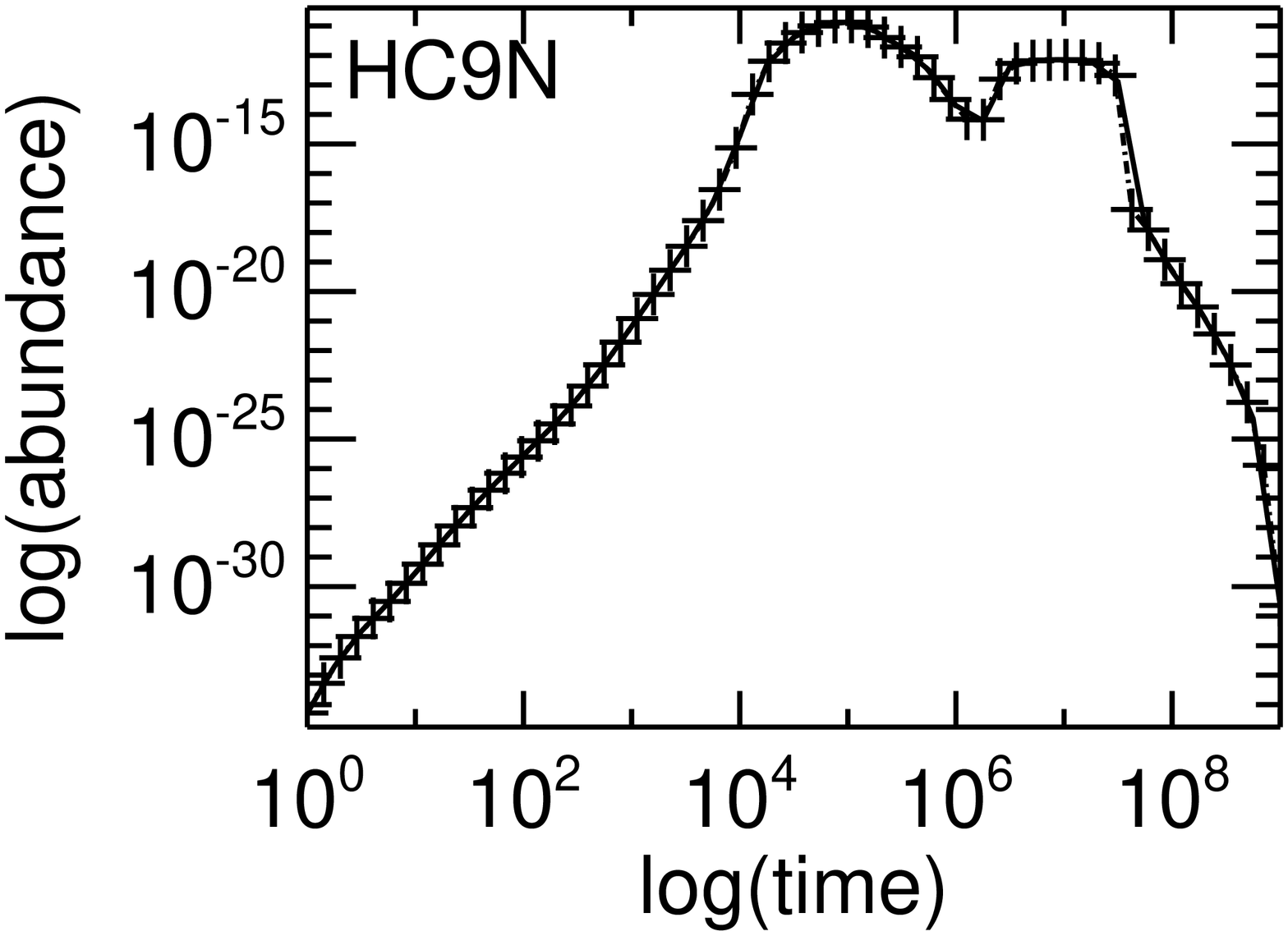}
\includegraphics[width=0.21\textwidth,clip=,angle=90]{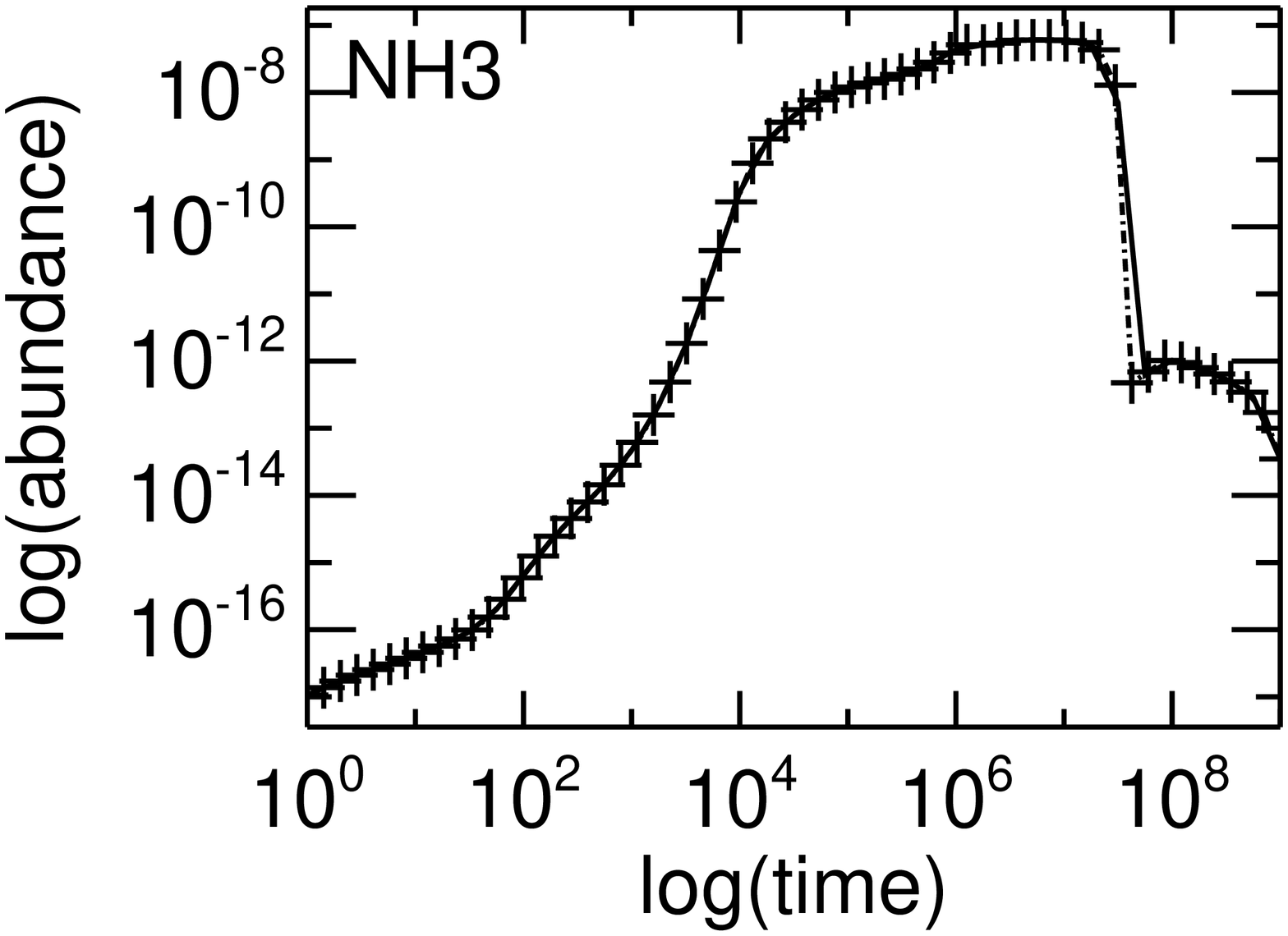}\\
\includegraphics[width=0.21\textwidth,clip=,angle=90]{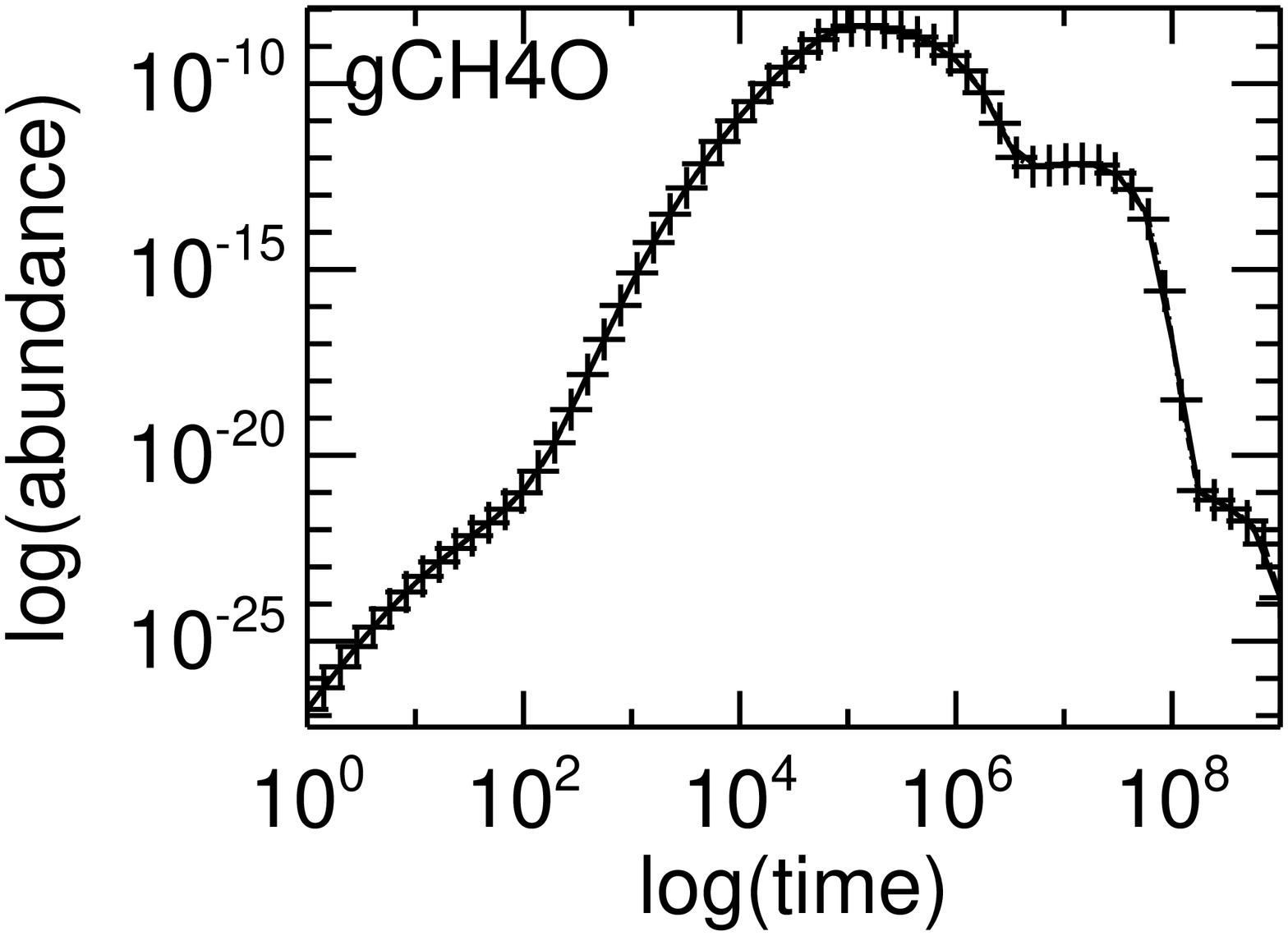}
\includegraphics[width=0.21\textwidth,clip=,angle=90]{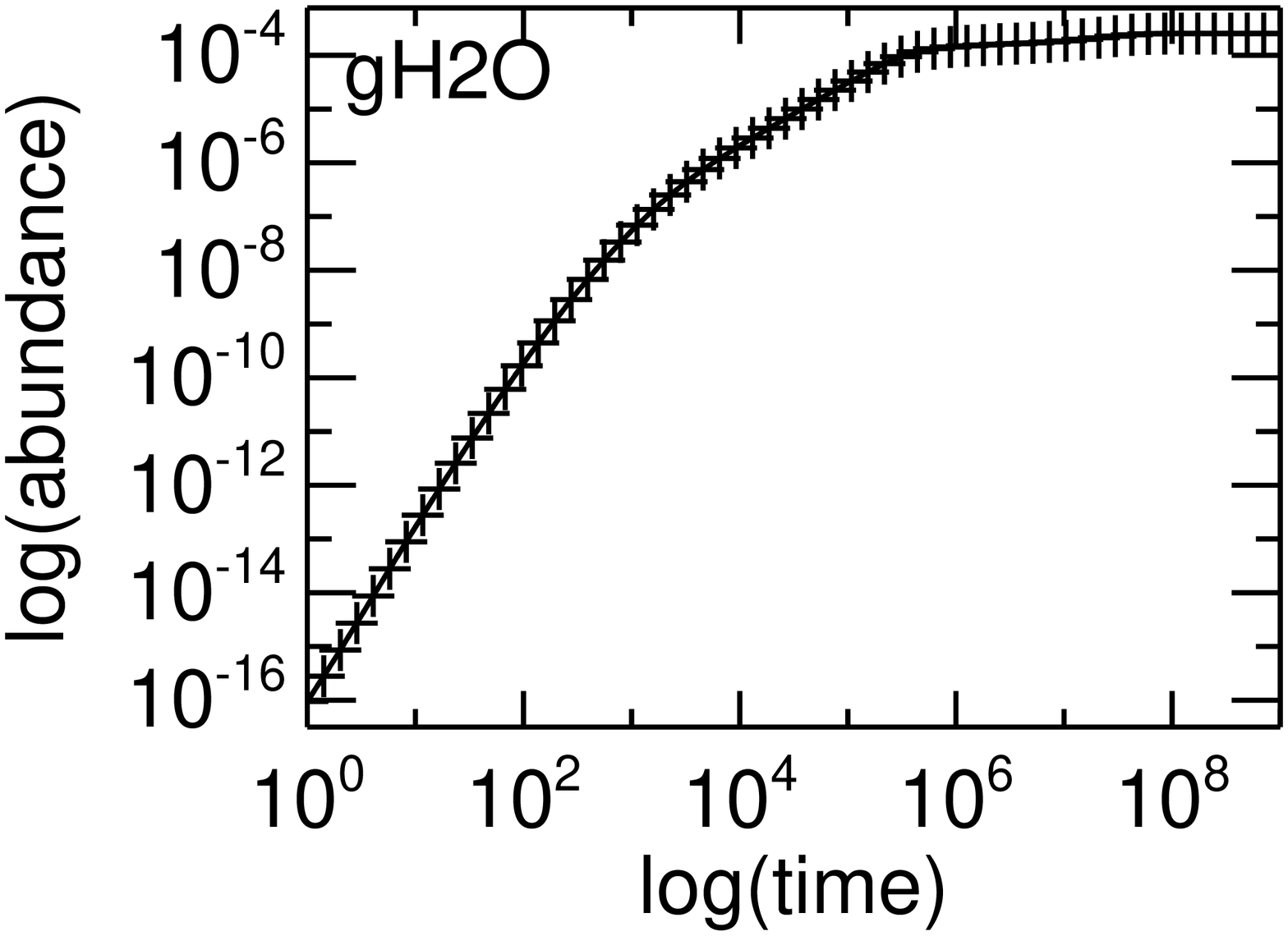}
\includegraphics[width=0.21\textwidth,clip=,angle=90]{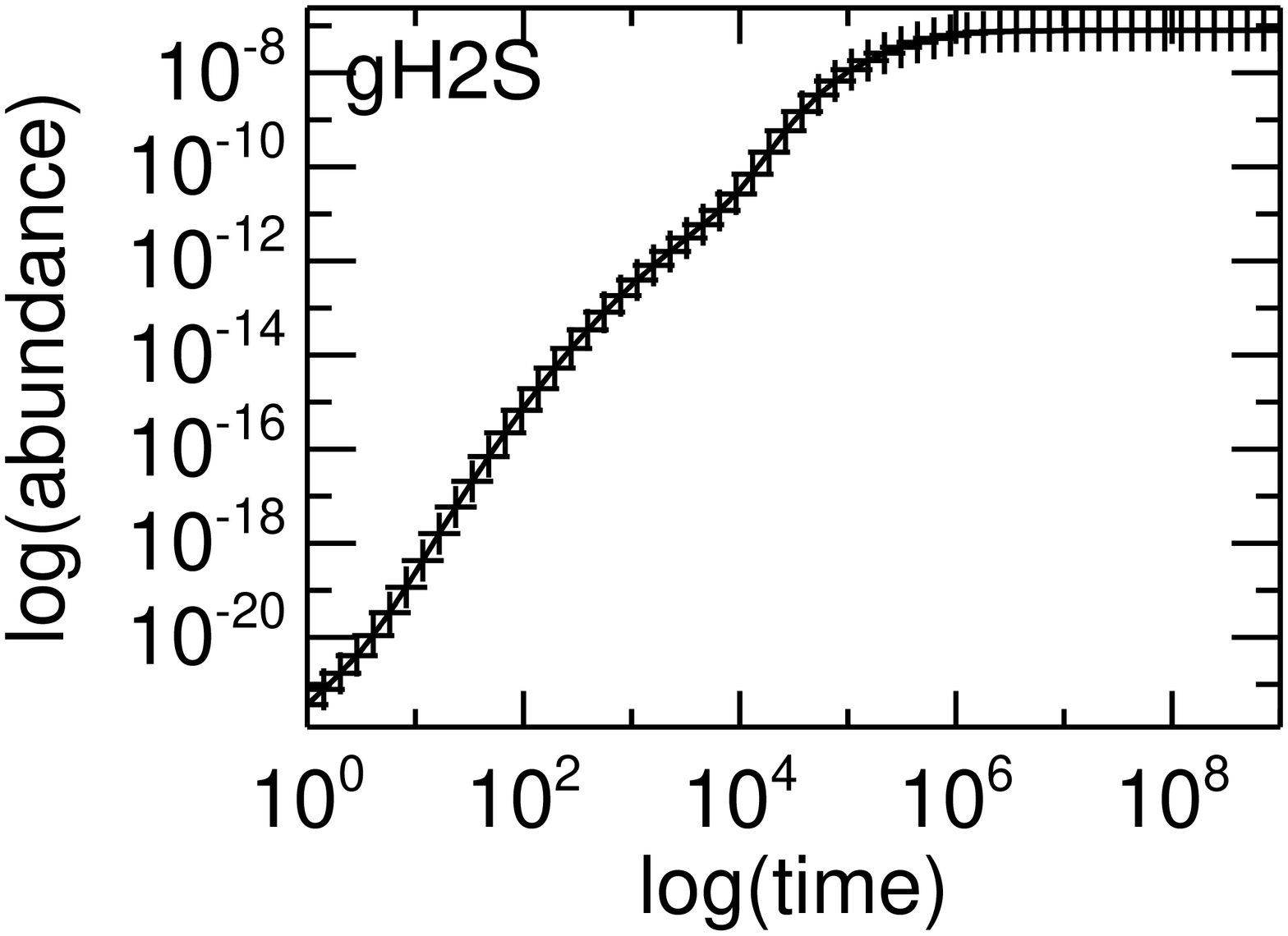}\\
\includegraphics[width=0.21\textwidth,clip=,angle=90]{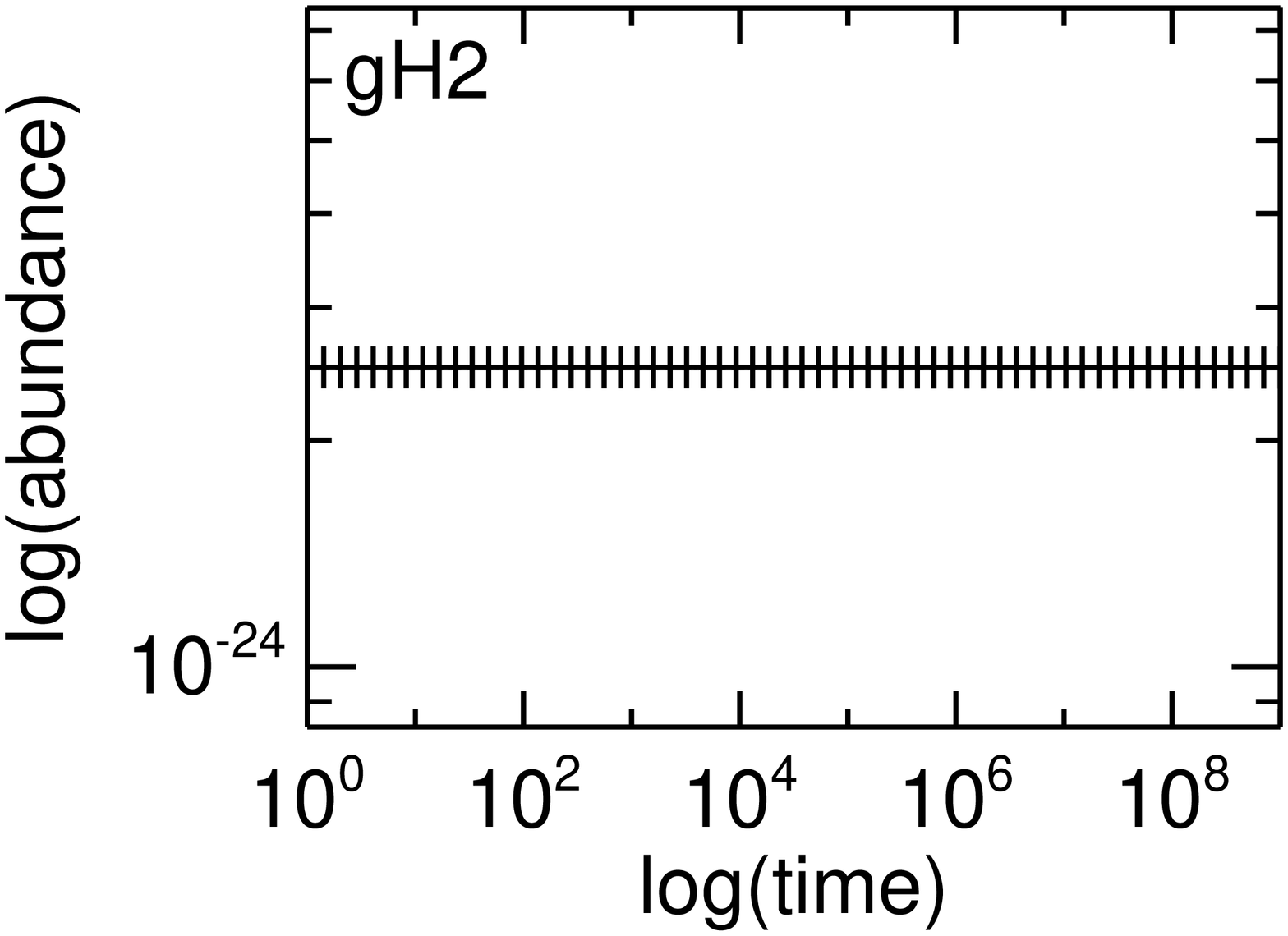}
\includegraphics[width=0.21\textwidth,clip=,angle=90]{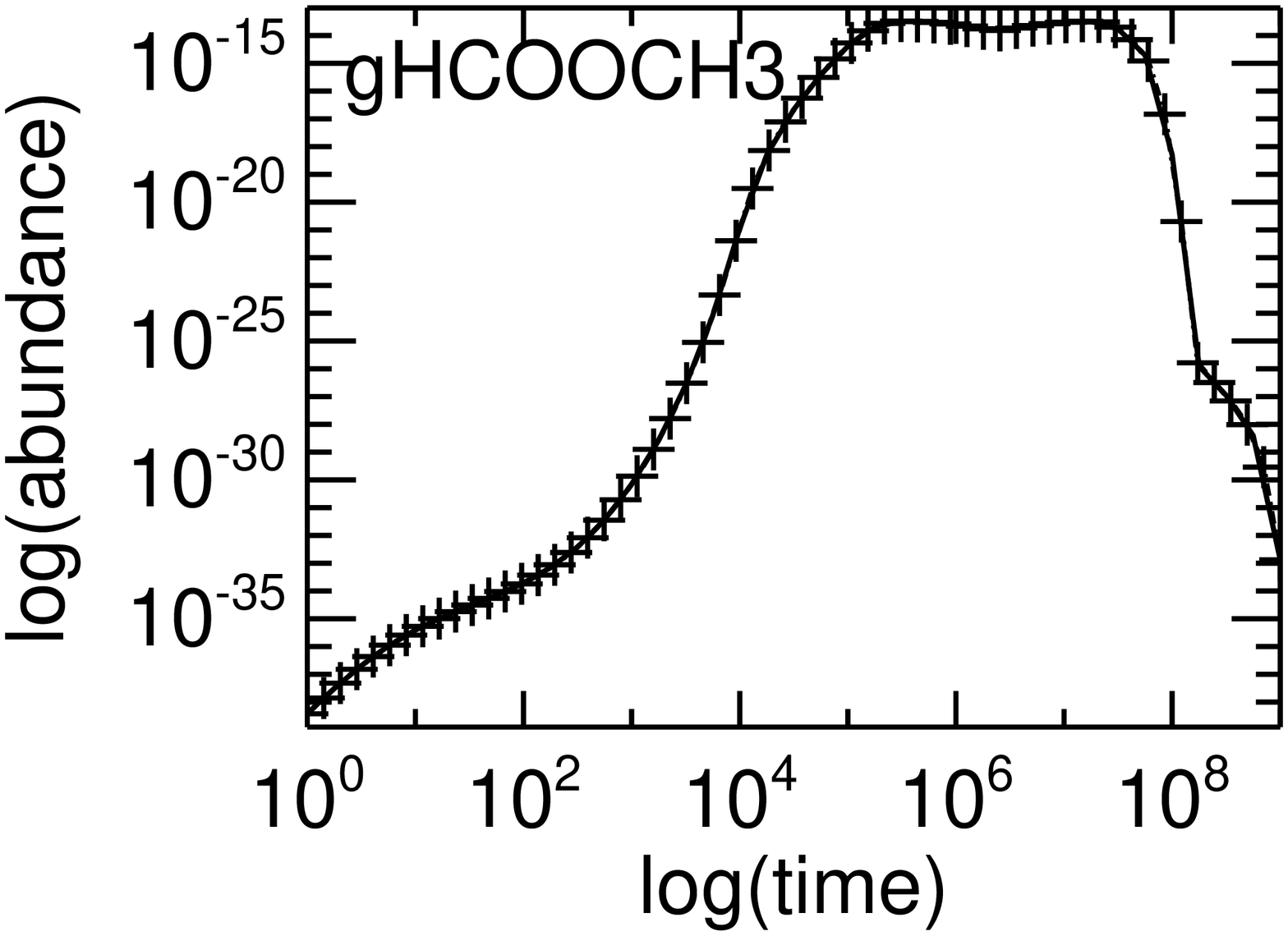}
\includegraphics[width=0.21\textwidth,clip=,angle=90]{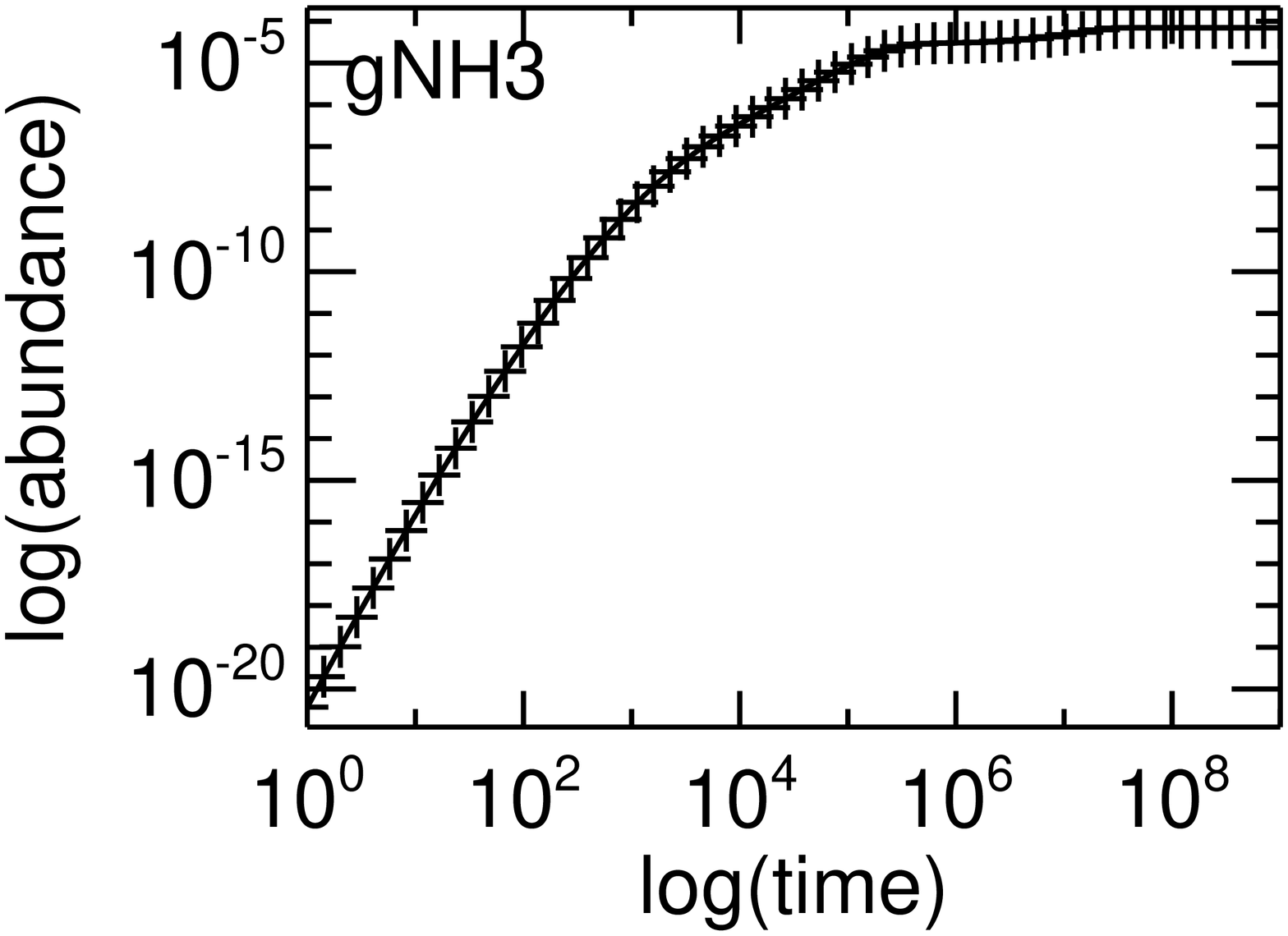}
\caption{Time-dependent abundances
as computed with the Heidelberg (crosses) and Bordeaux (solid line) chemical models
for the ``TMC1'' case. The X-axes show time in years. The Y-axes are abundances
relative to the total amount of hydrogen nuclei. The species names are given in each panel.
The prefix ``g'' denotes surface molecules, ``CH4O'' is methanol,
and ``G-'' stands for negatively charged grains.}
\label{fig:TMC1}
\end{figure*}

\begin{figure*}
\includegraphics[width=0.21\textwidth,clip=,angle=90]{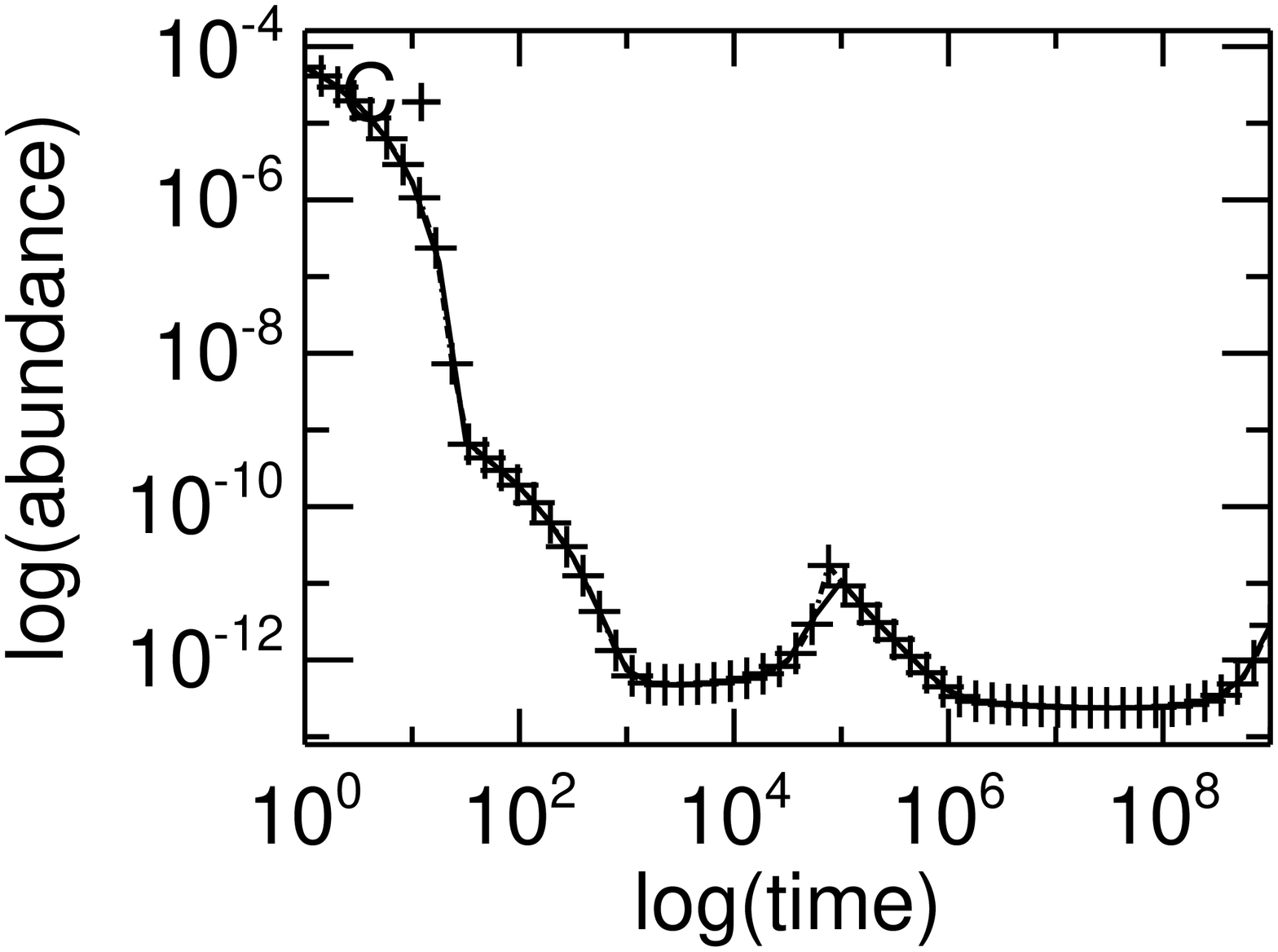}
\includegraphics[width=0.21\textwidth,clip=,angle=90]{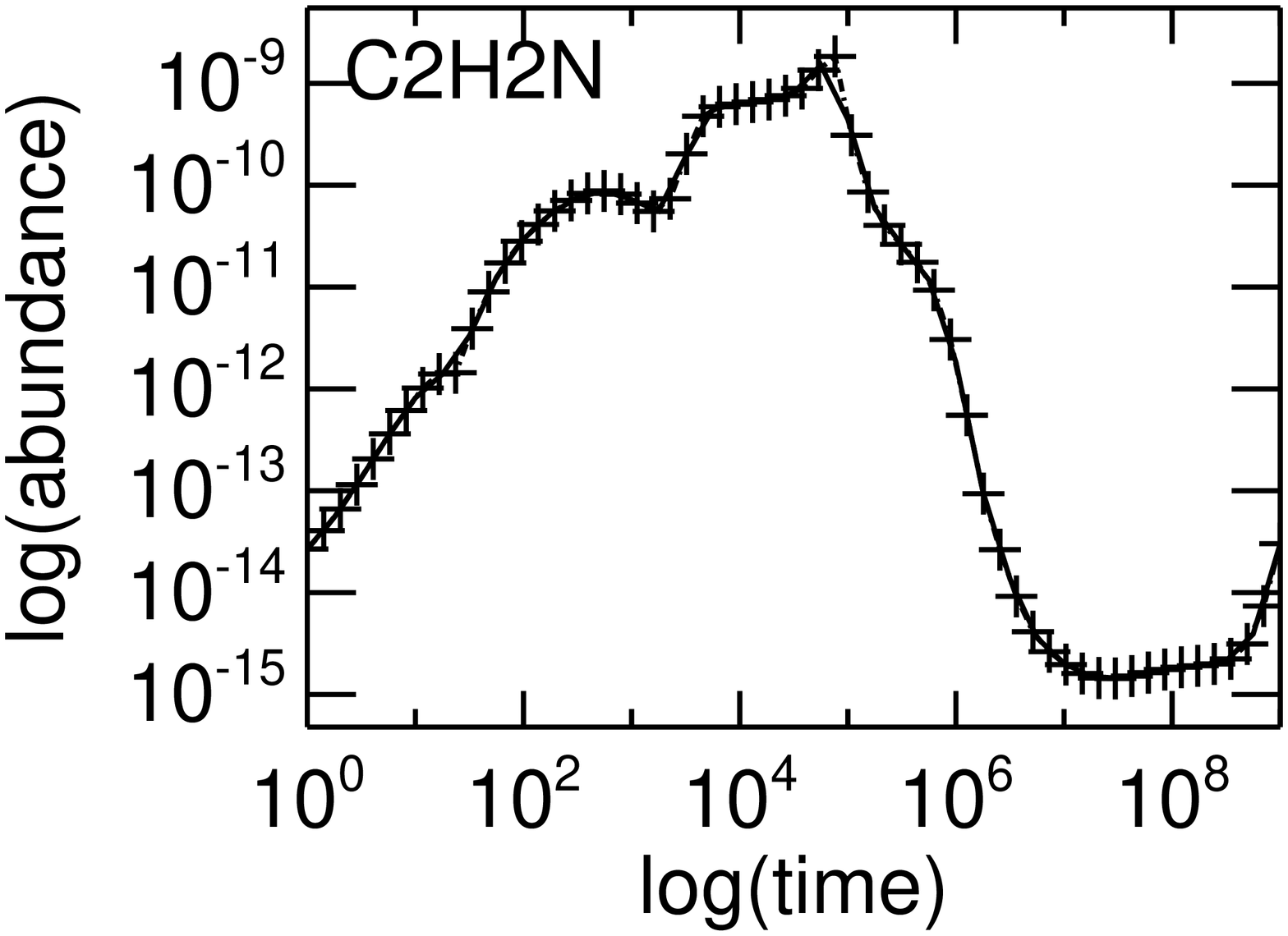}
\includegraphics[width=0.21\textwidth,clip=,angle=90]{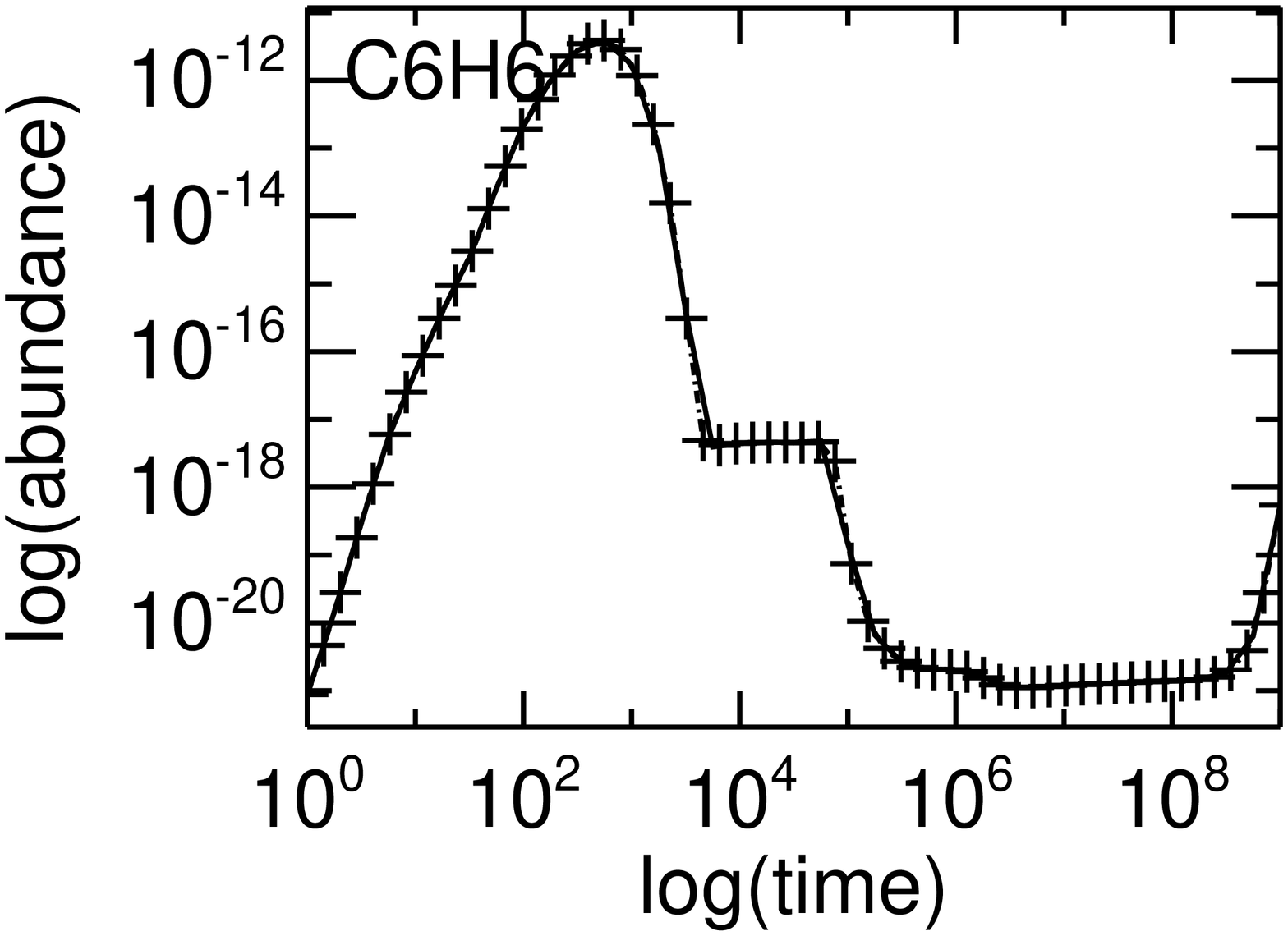}\\
\includegraphics[width=0.21\textwidth,clip=,angle=90]{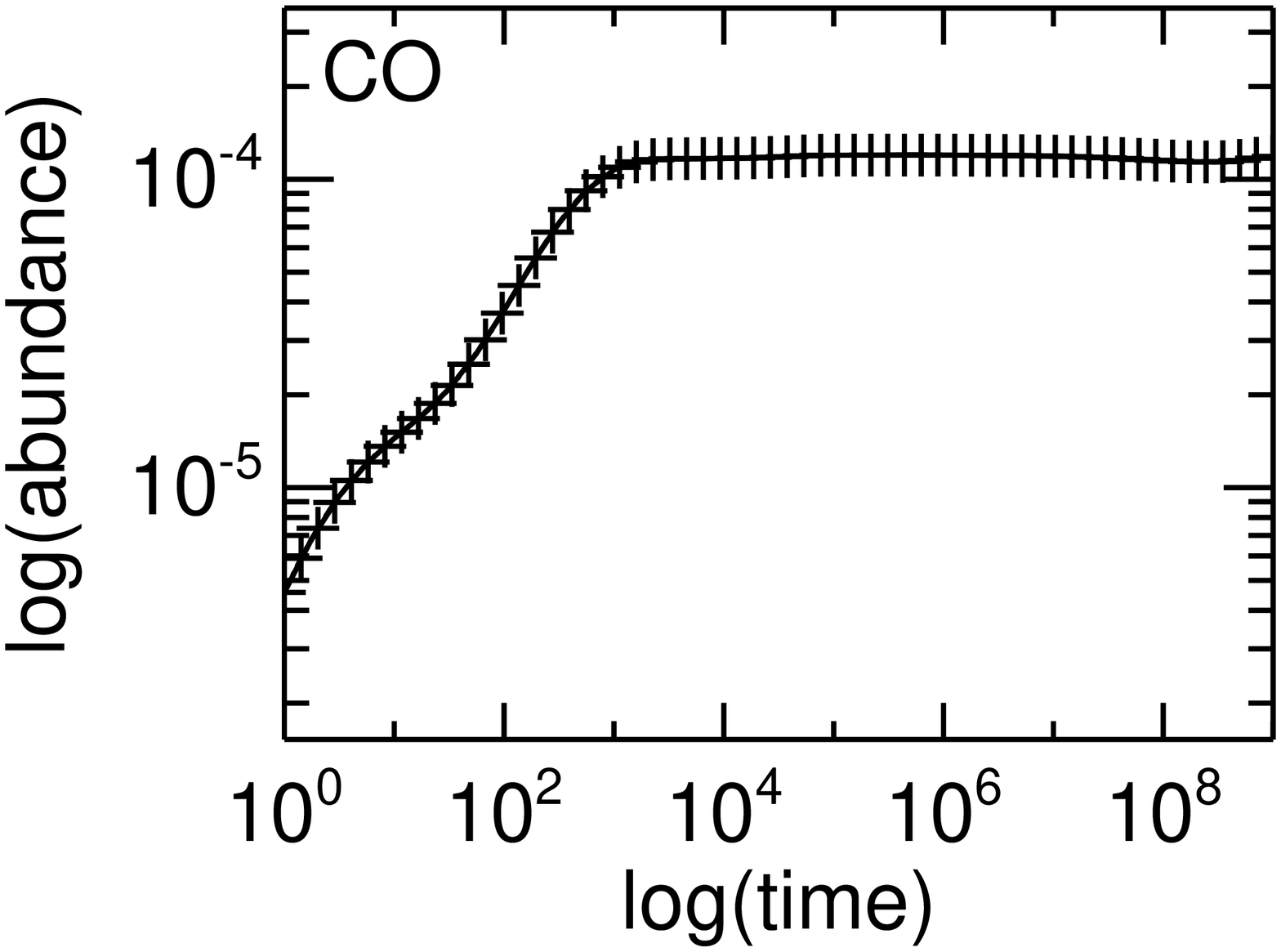}
\includegraphics[width=0.21\textwidth,clip=,angle=90]{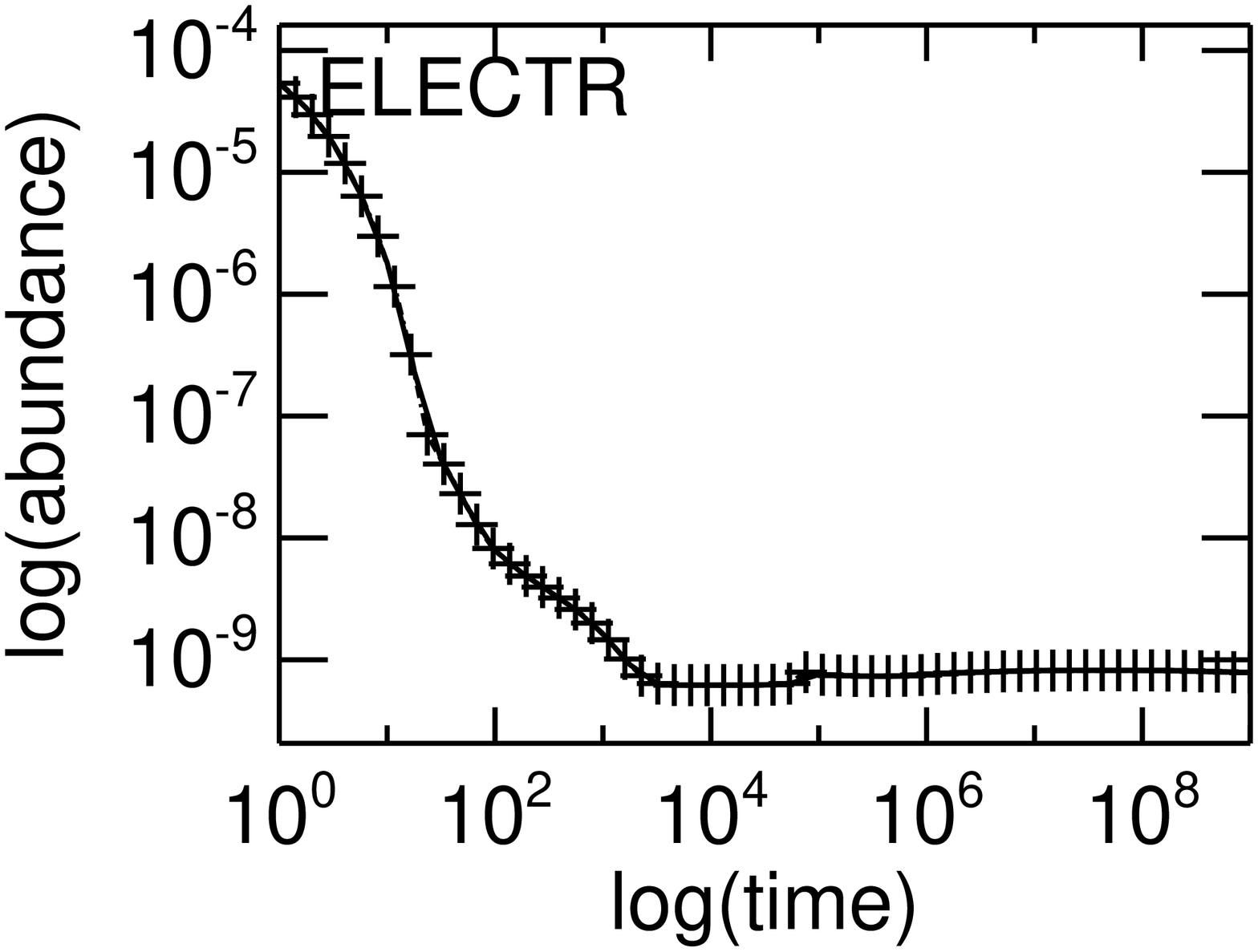}
\includegraphics[width=0.21\textwidth,clip=,angle=90]{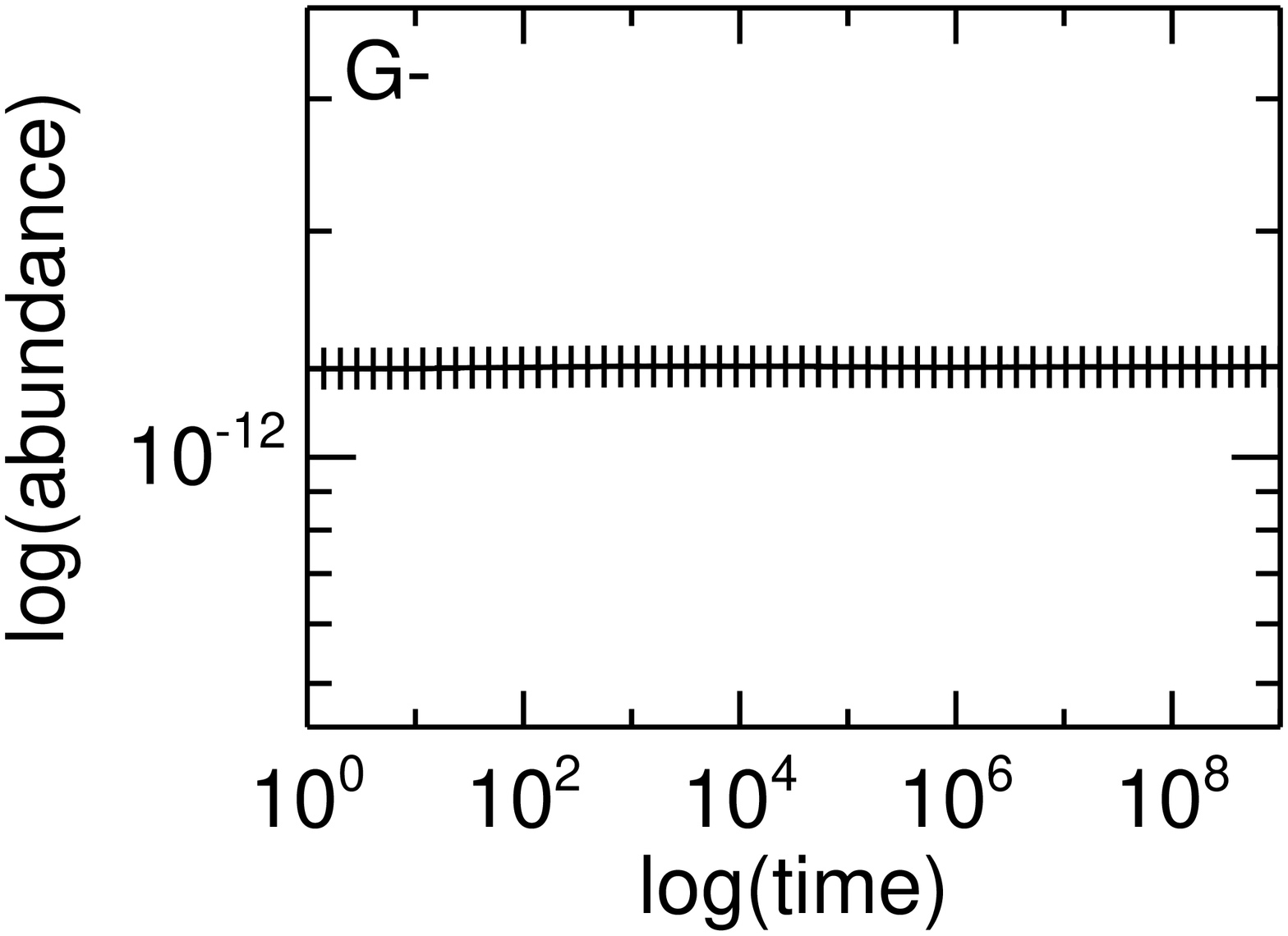}\\
\includegraphics[width=0.21\textwidth,clip=,angle=90]{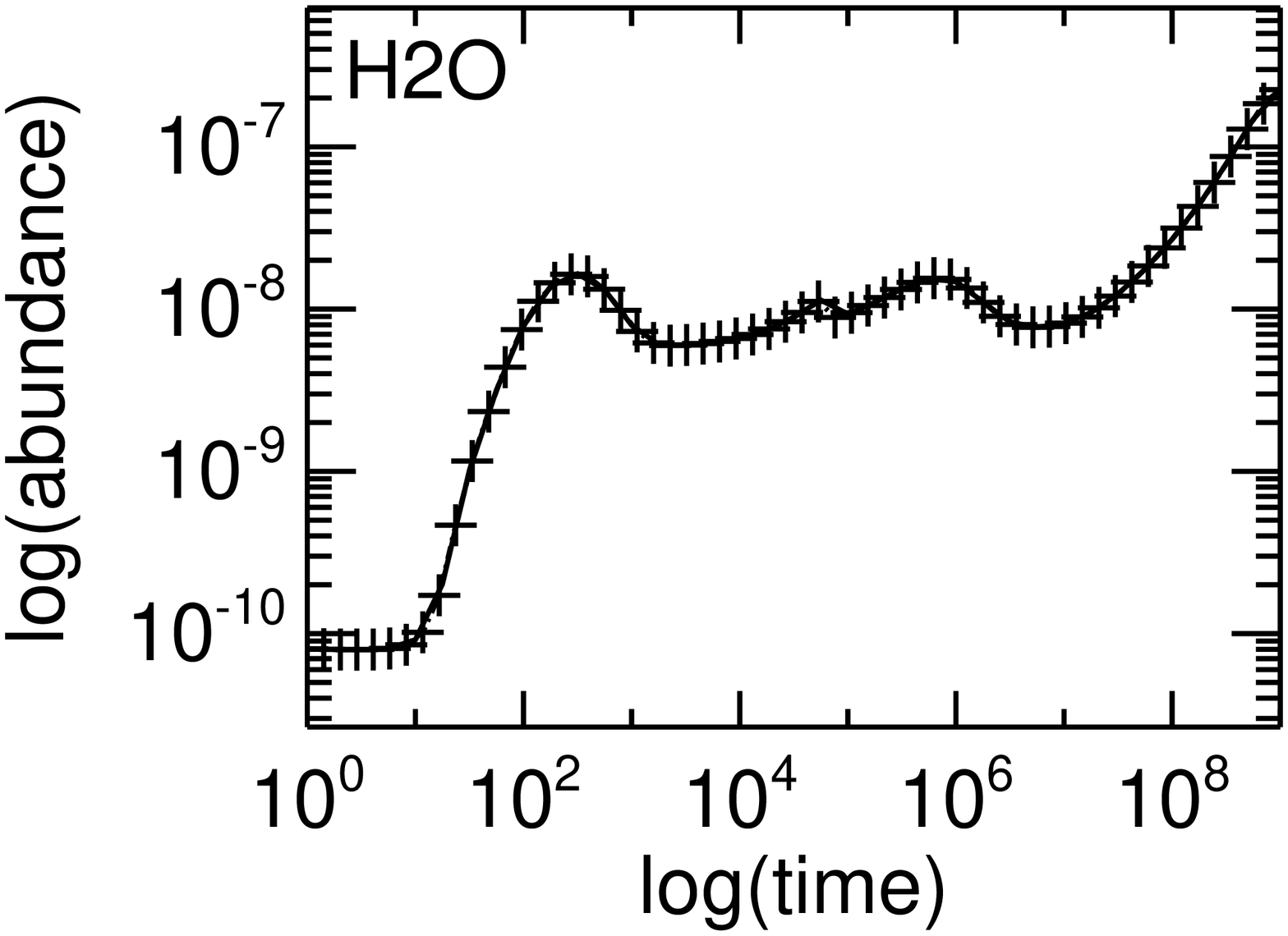}
\includegraphics[width=0.21\textwidth,clip=,angle=90]{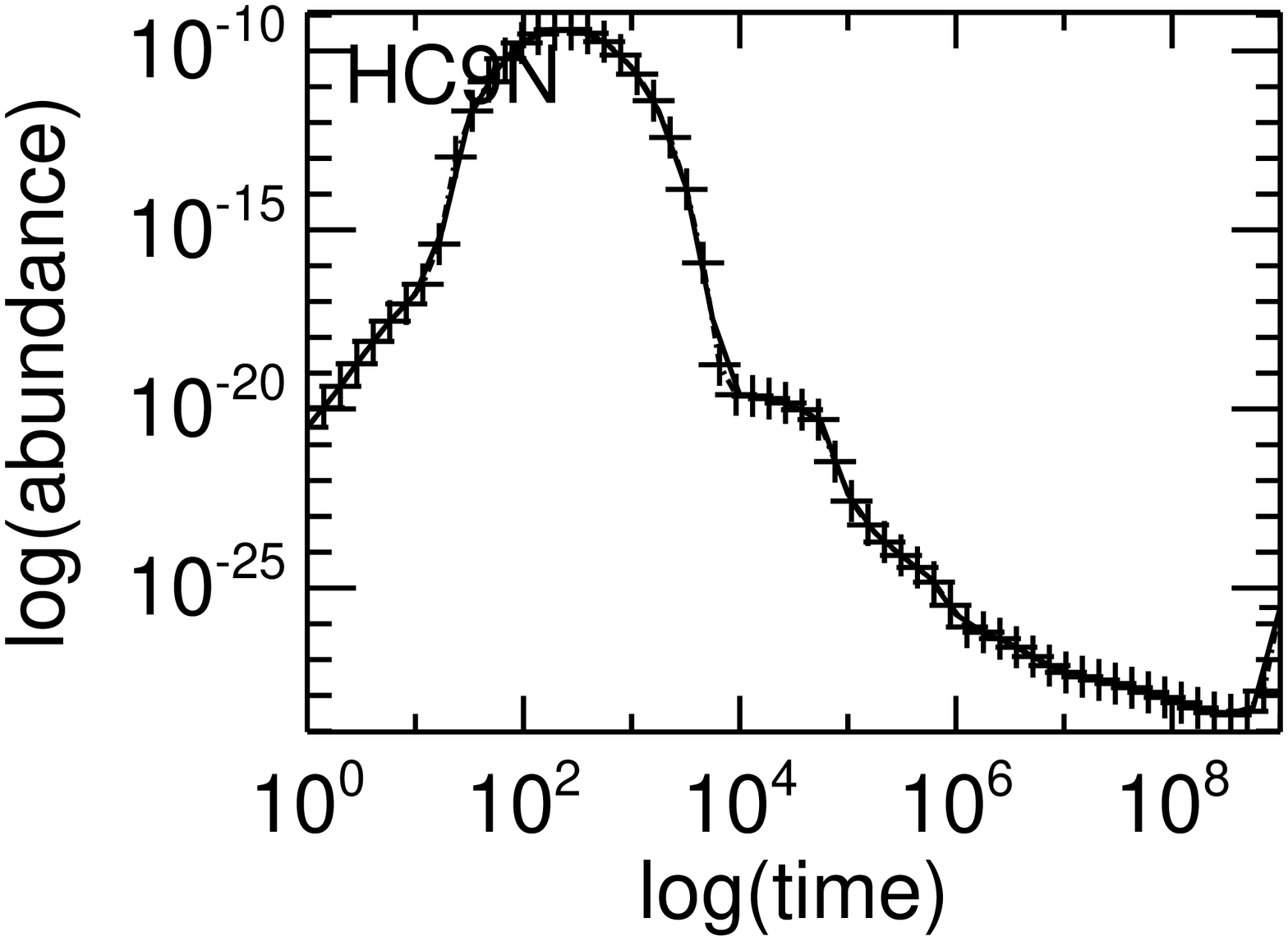}
\includegraphics[width=0.21\textwidth,clip=,angle=90]{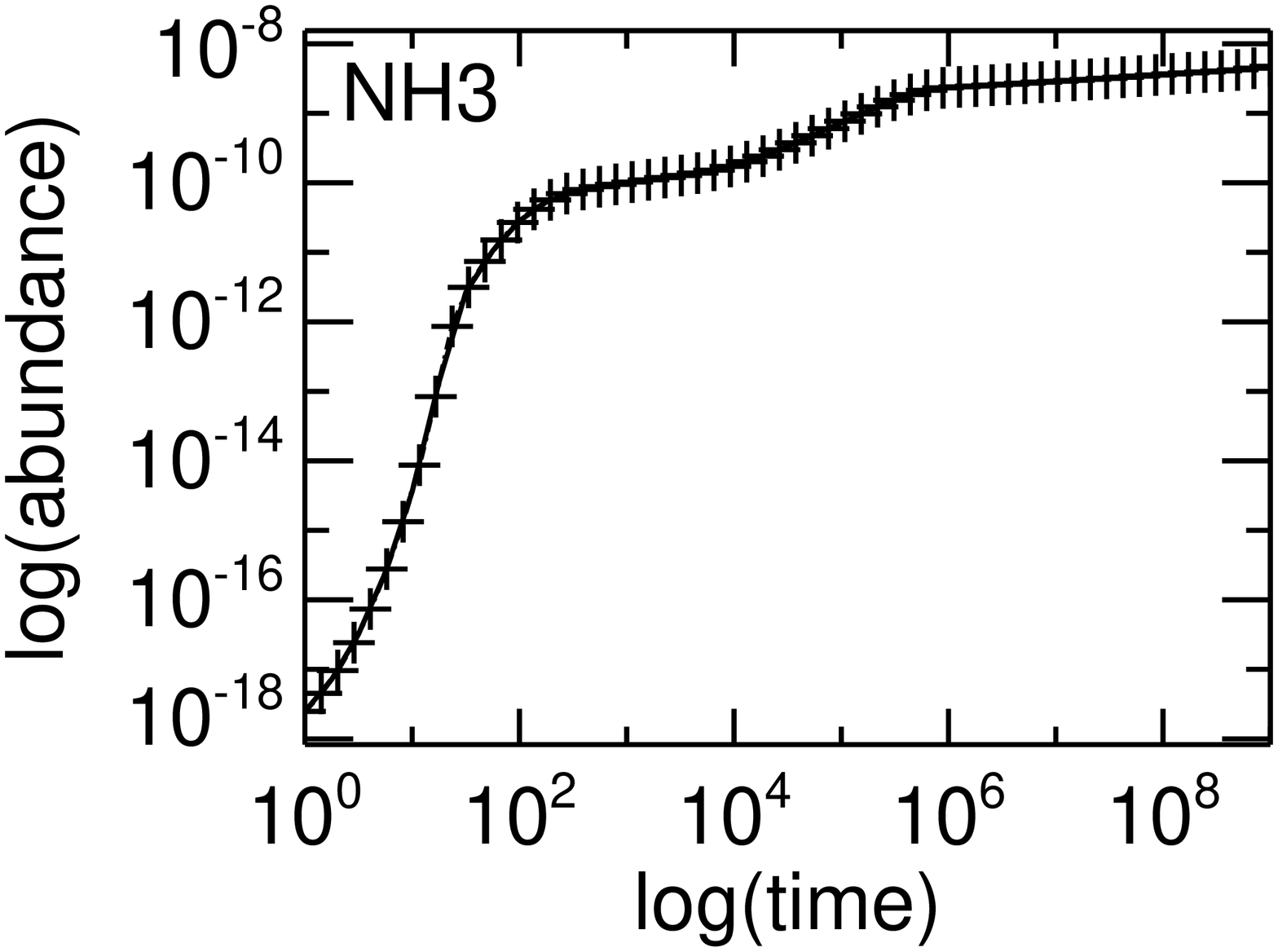}\\
\includegraphics[width=0.21\textwidth,clip=,angle=90]{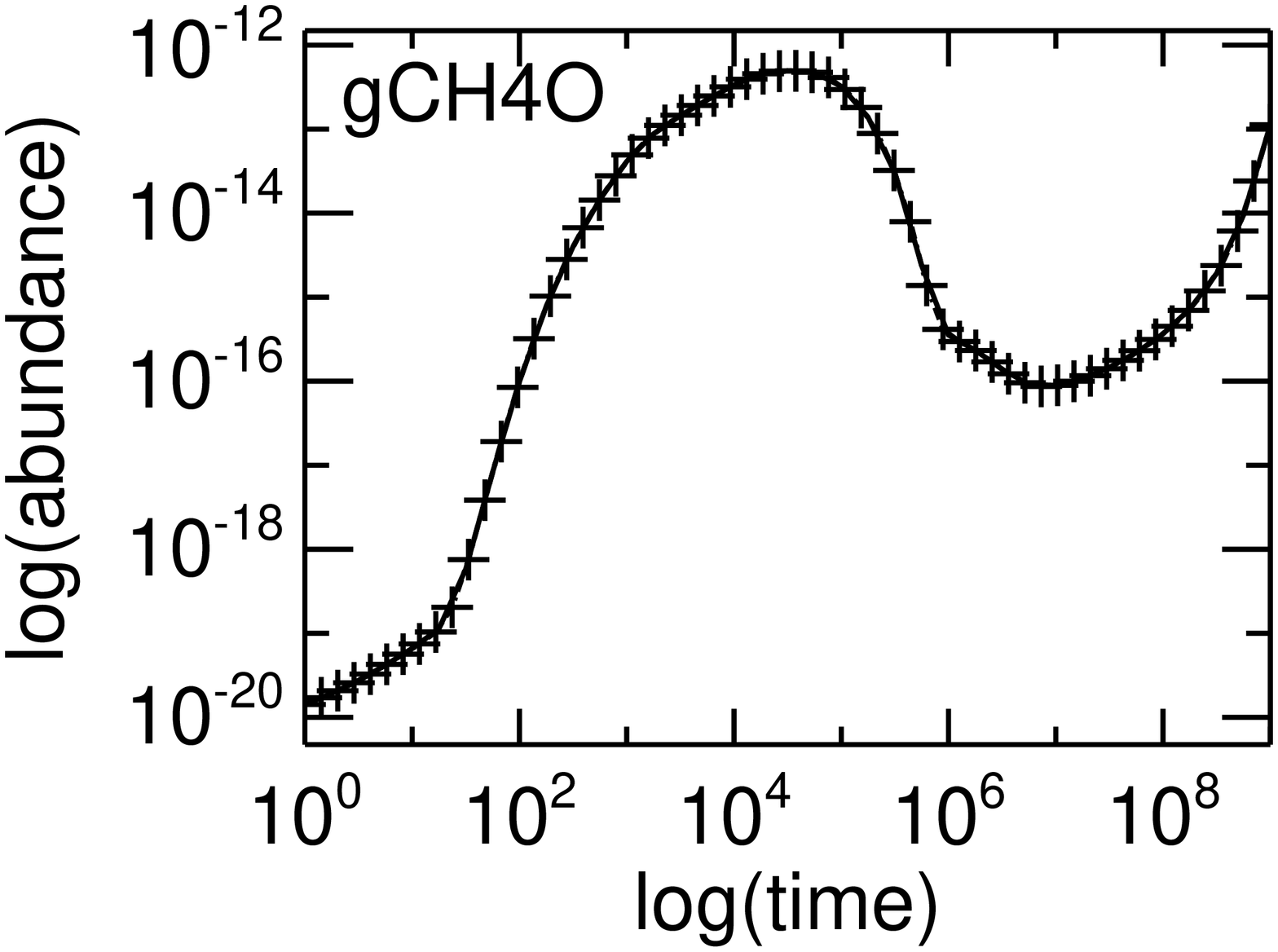}
\includegraphics[width=0.21\textwidth,clip=,angle=90]{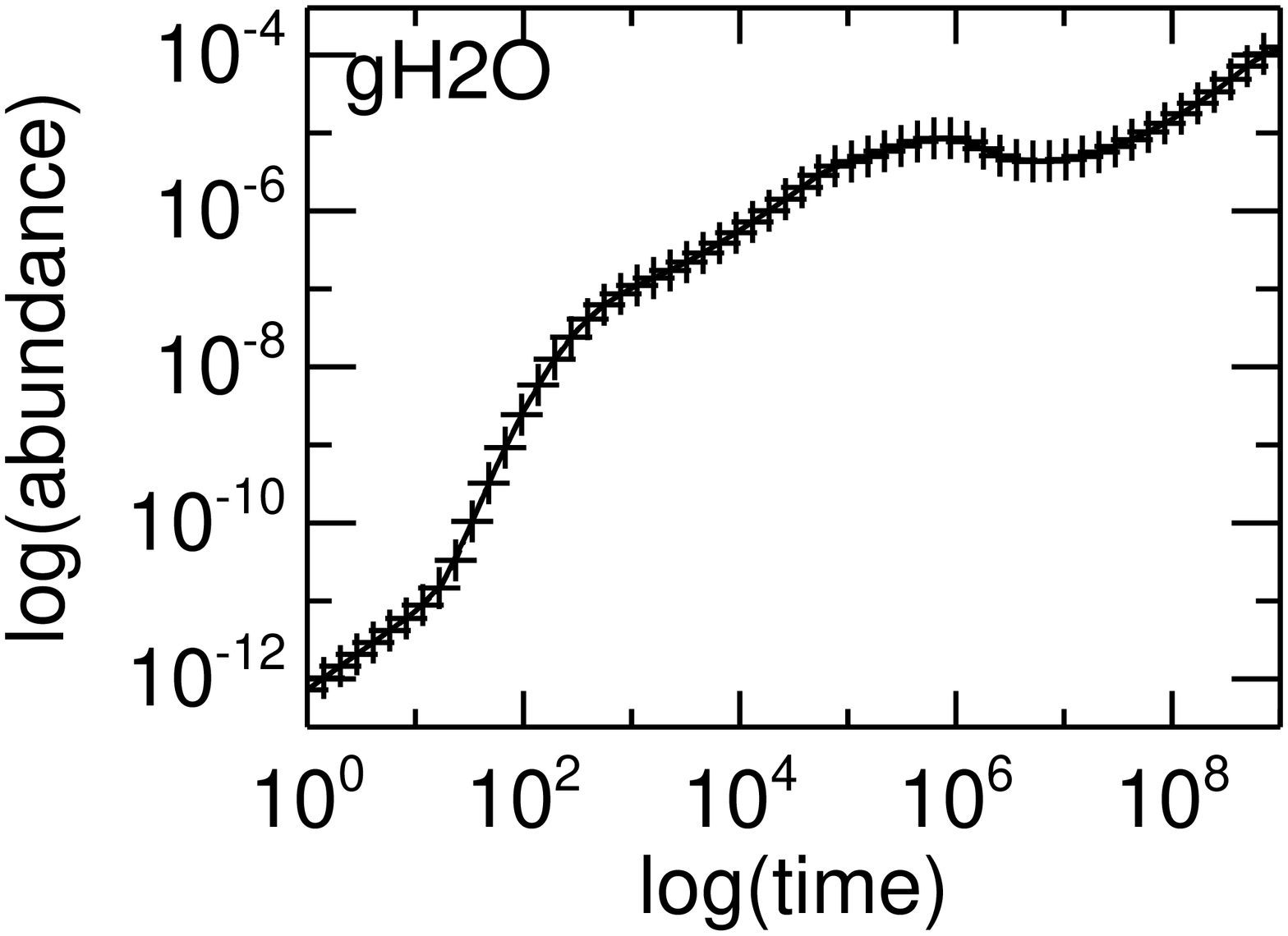}
\includegraphics[width=0.21\textwidth,clip=,angle=90]{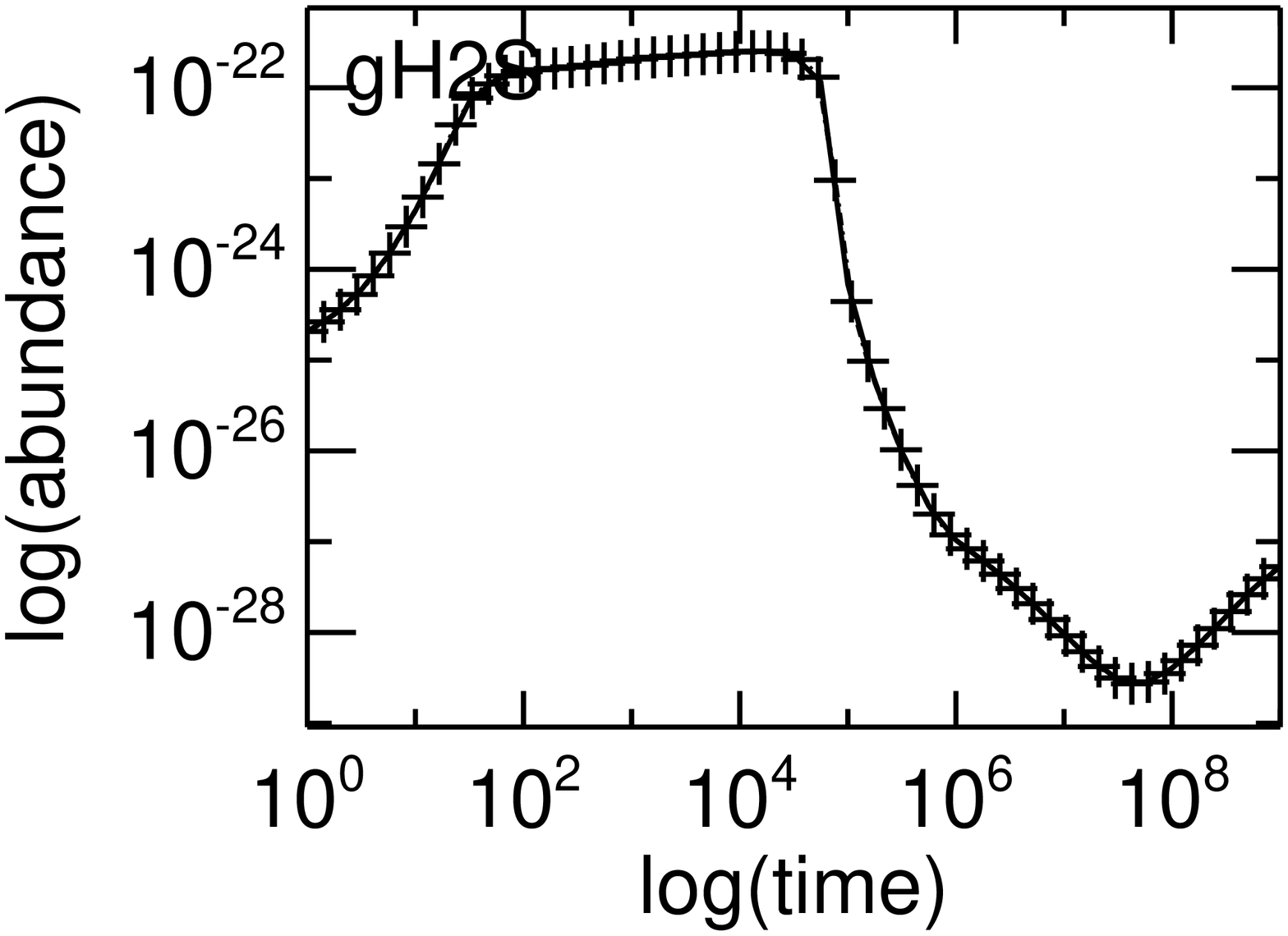}\\
\includegraphics[width=0.21\textwidth,clip=,angle=90]{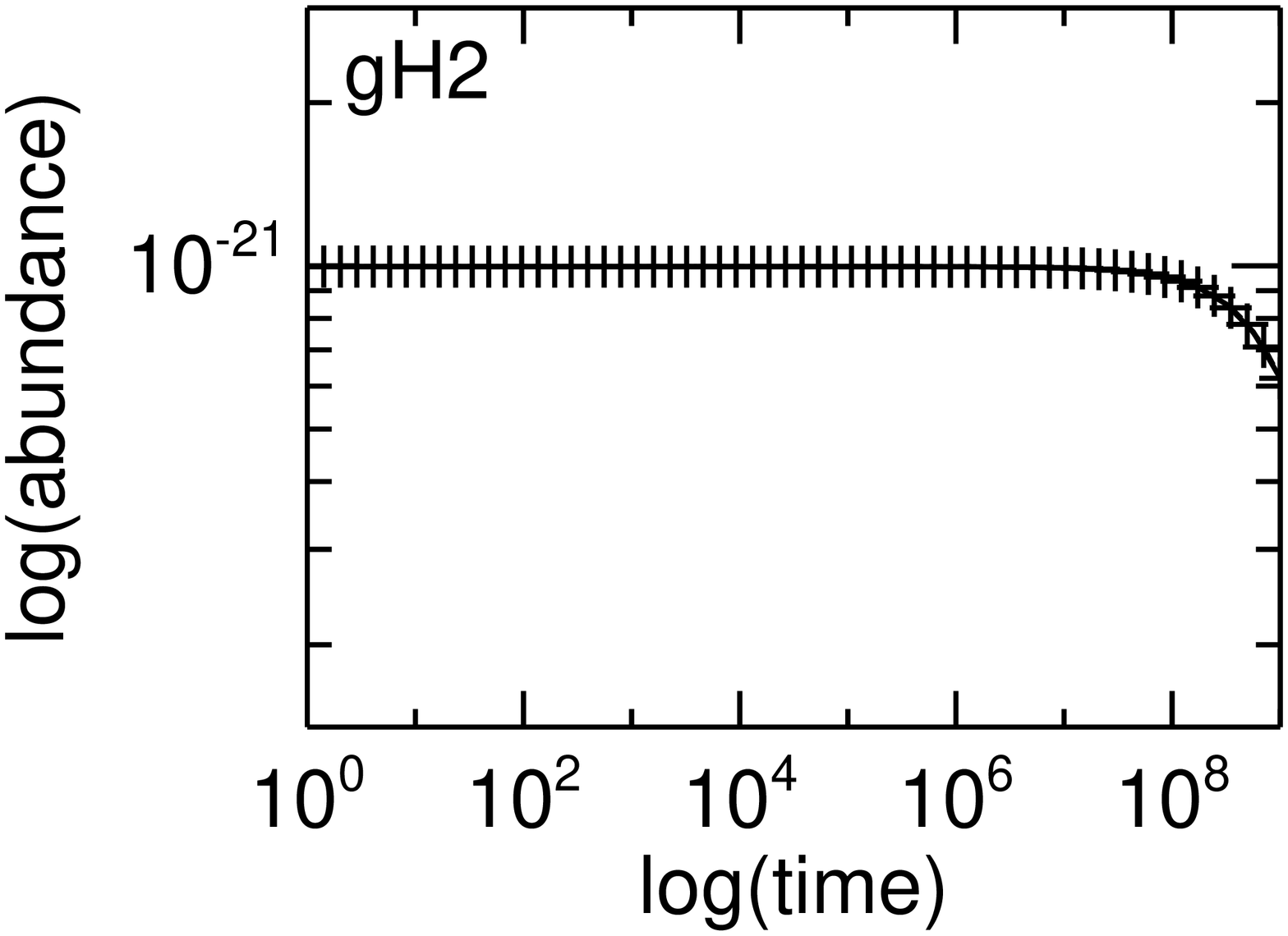}
\includegraphics[width=0.21\textwidth,clip=,angle=90]{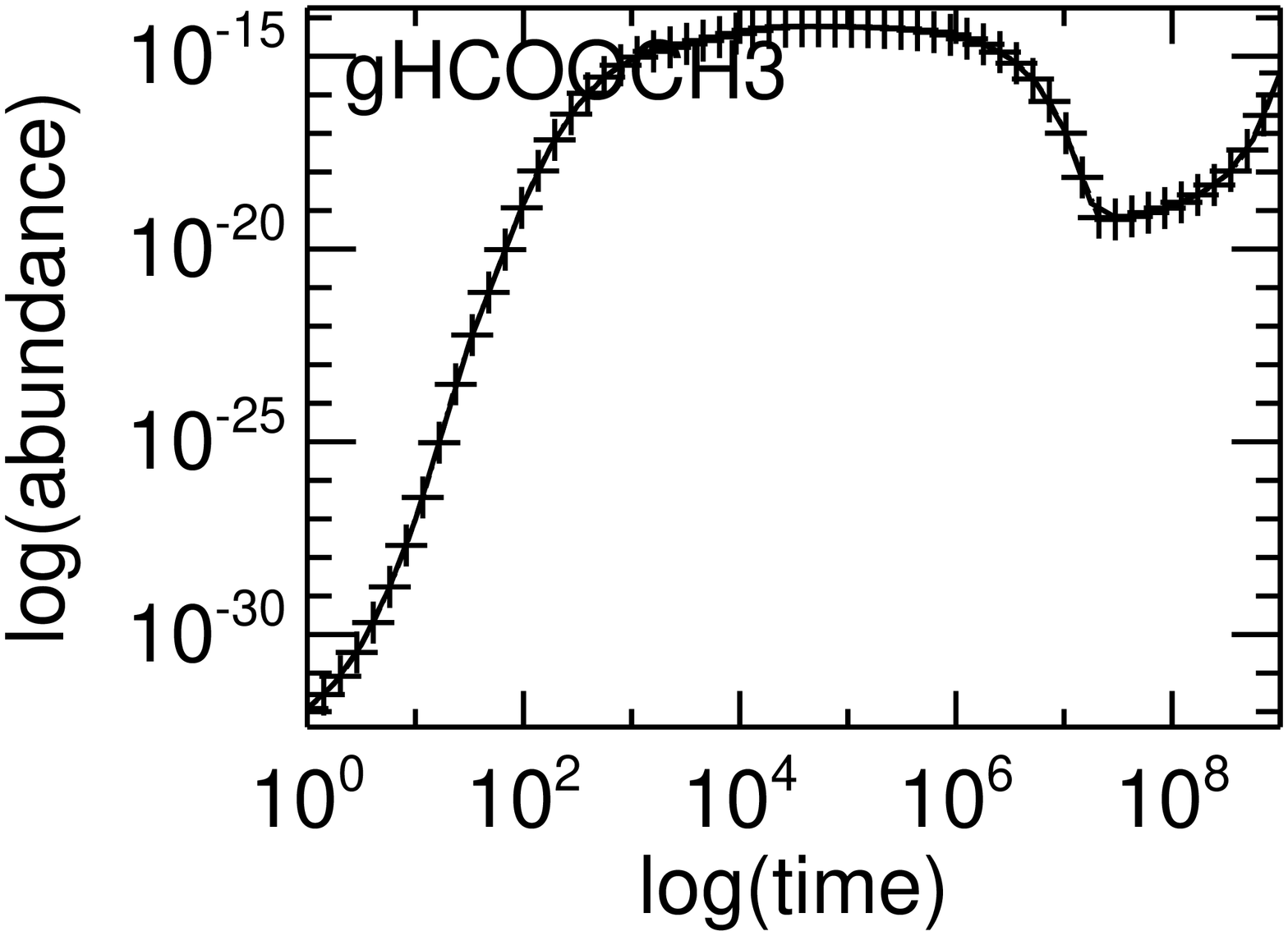}
\includegraphics[width=0.21\textwidth,clip=,angle=90]{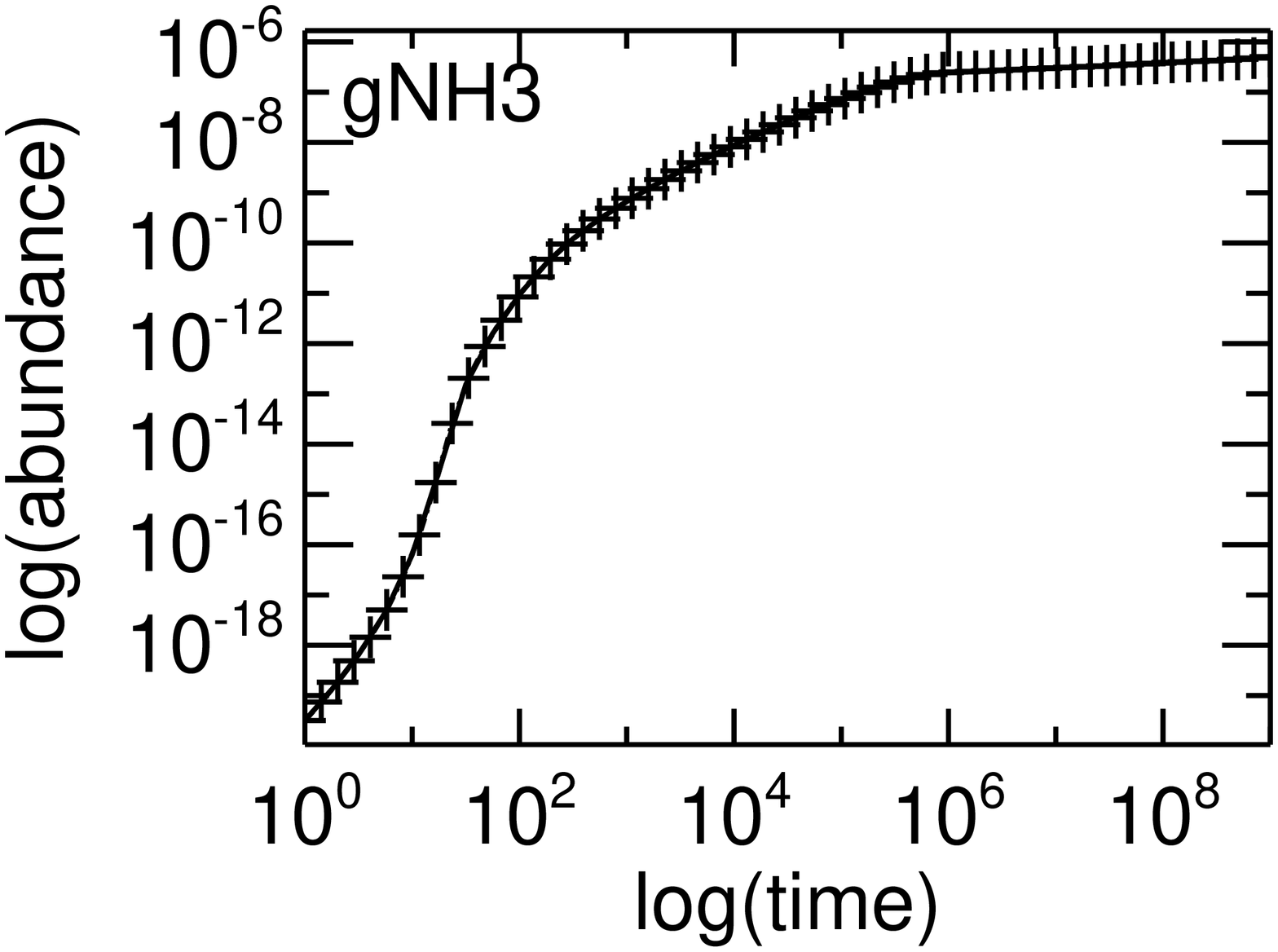}
\caption{The same as in Fig.~\ref{fig:TMC1} but
for the ``Hot Corino'' case.}
\label{fig:HOT_CORE}
\end{figure*}

\begin{figure*}
\includegraphics[width=0.21\textwidth,clip=,angle=90]{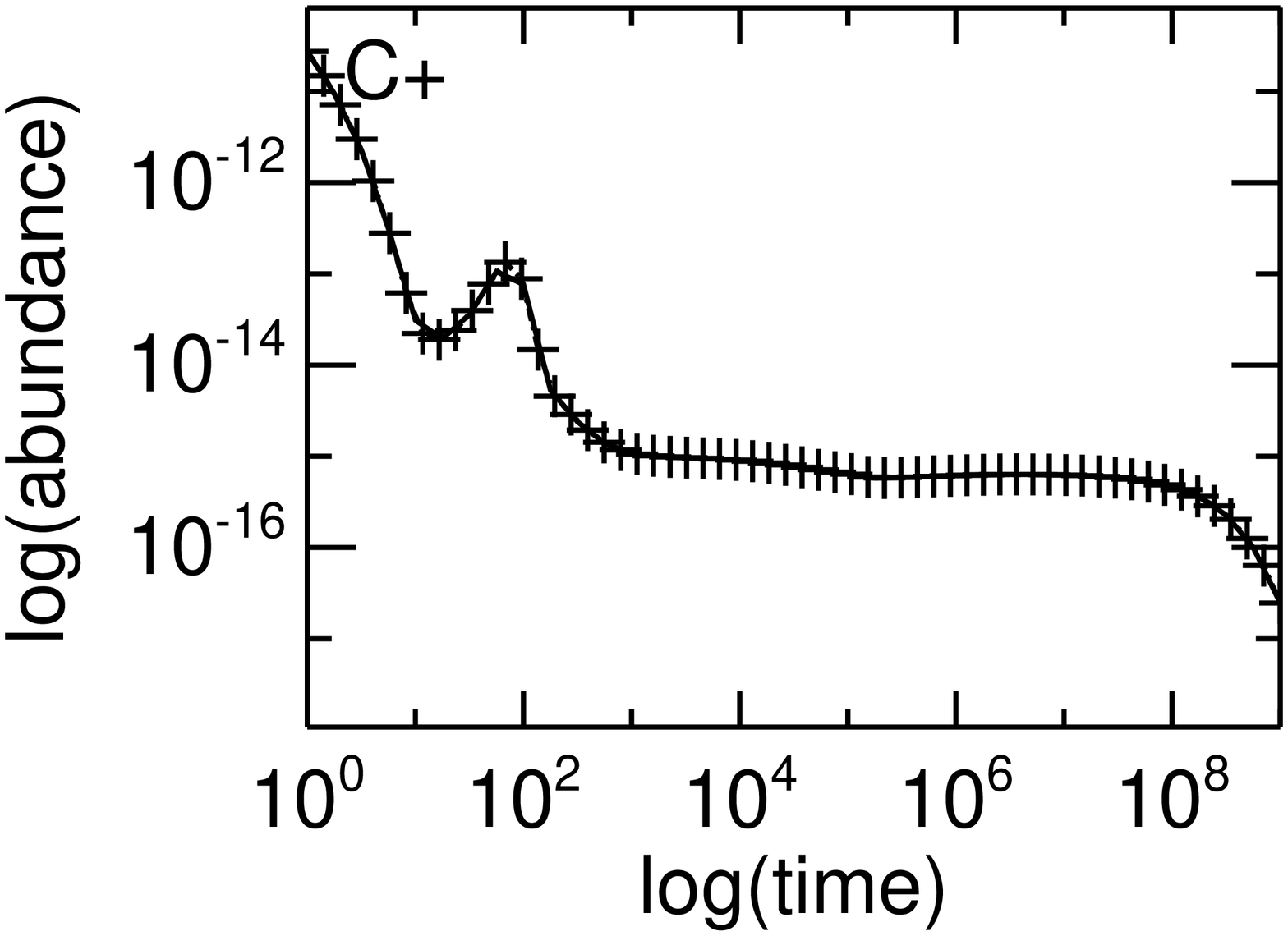}
\includegraphics[width=0.21\textwidth,clip=,angle=90]{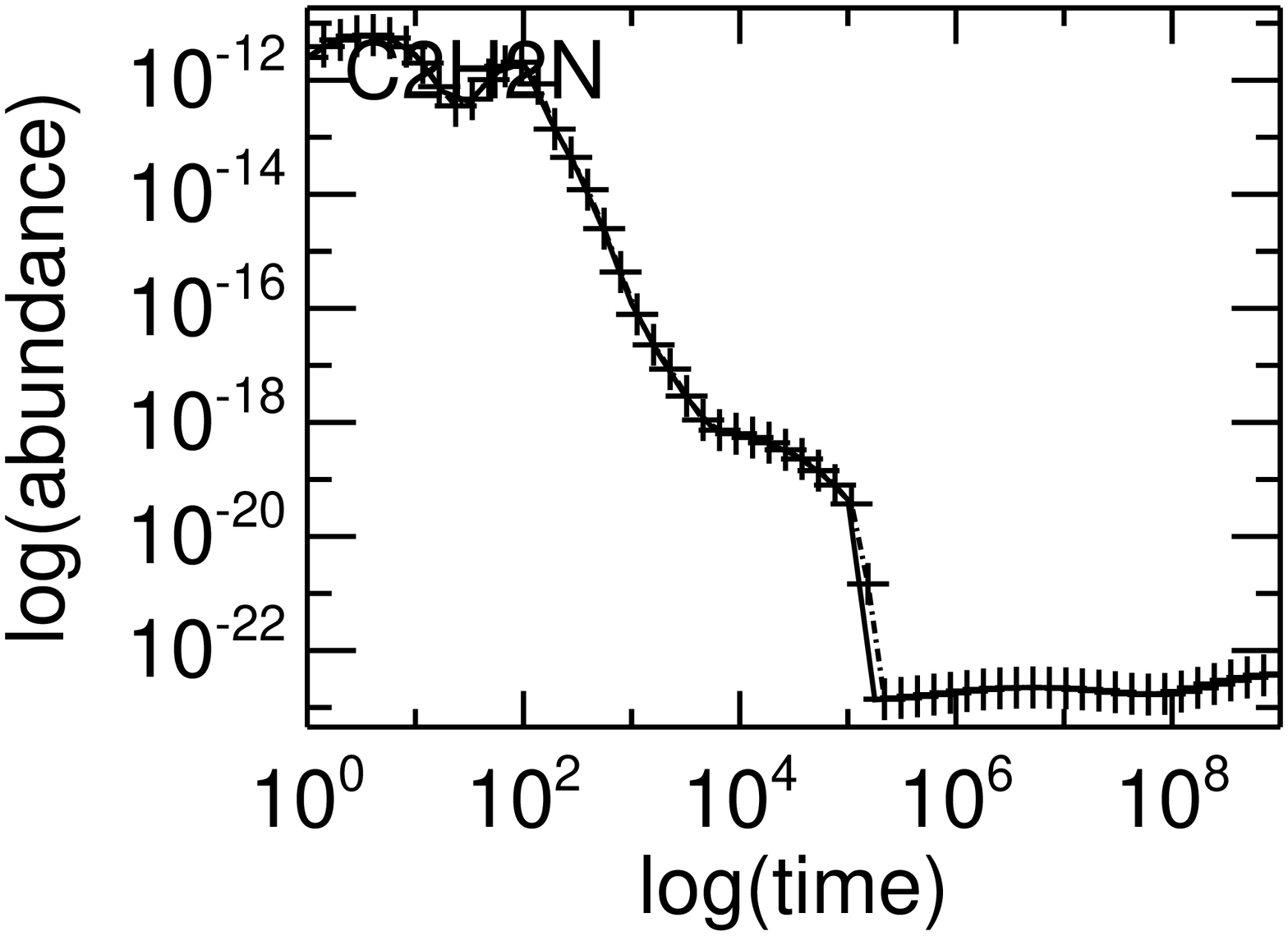}
\includegraphics[width=0.21\textwidth,clip=,angle=90]{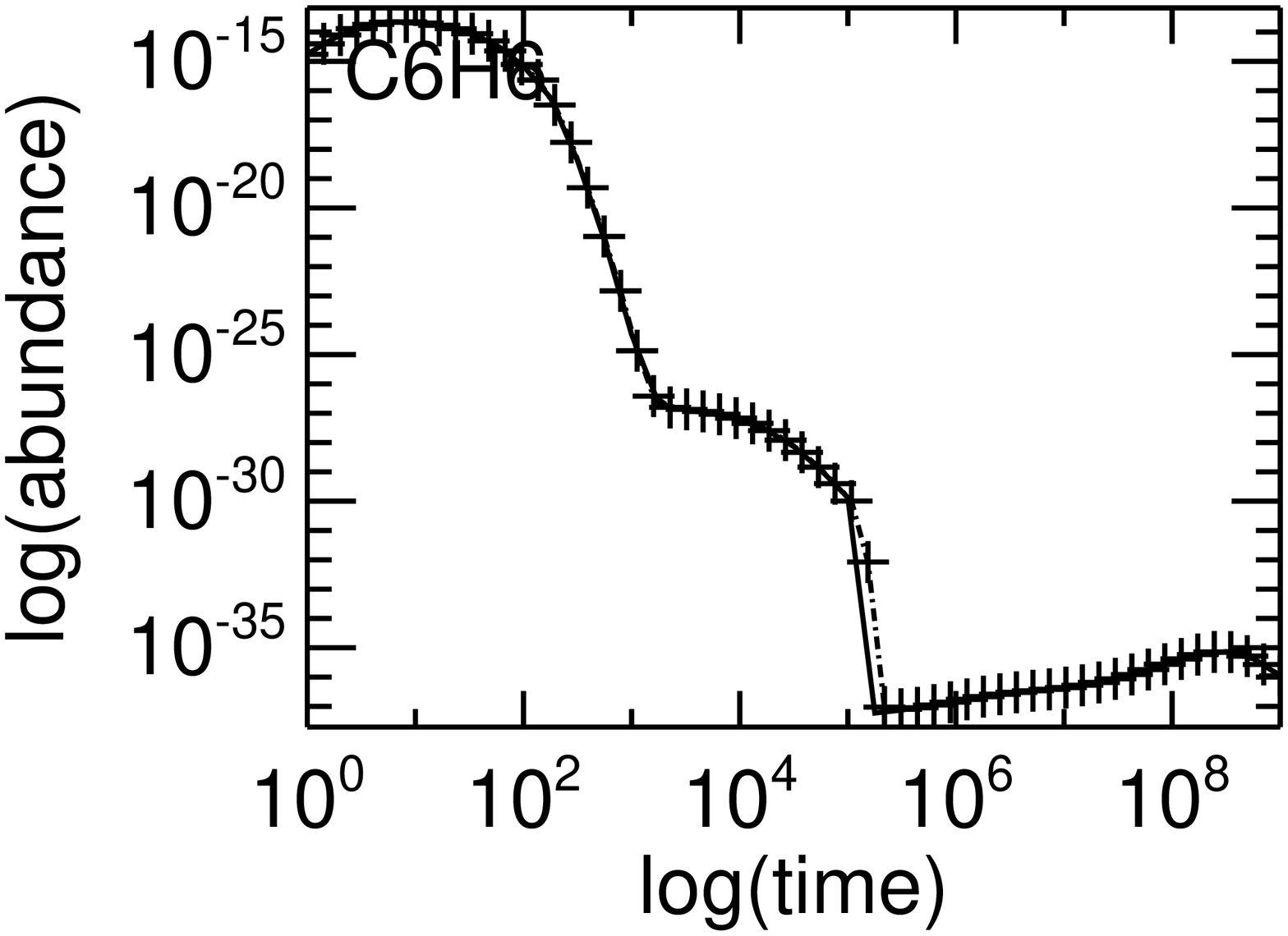}\\
\includegraphics[width=0.21\textwidth,clip=,angle=90]{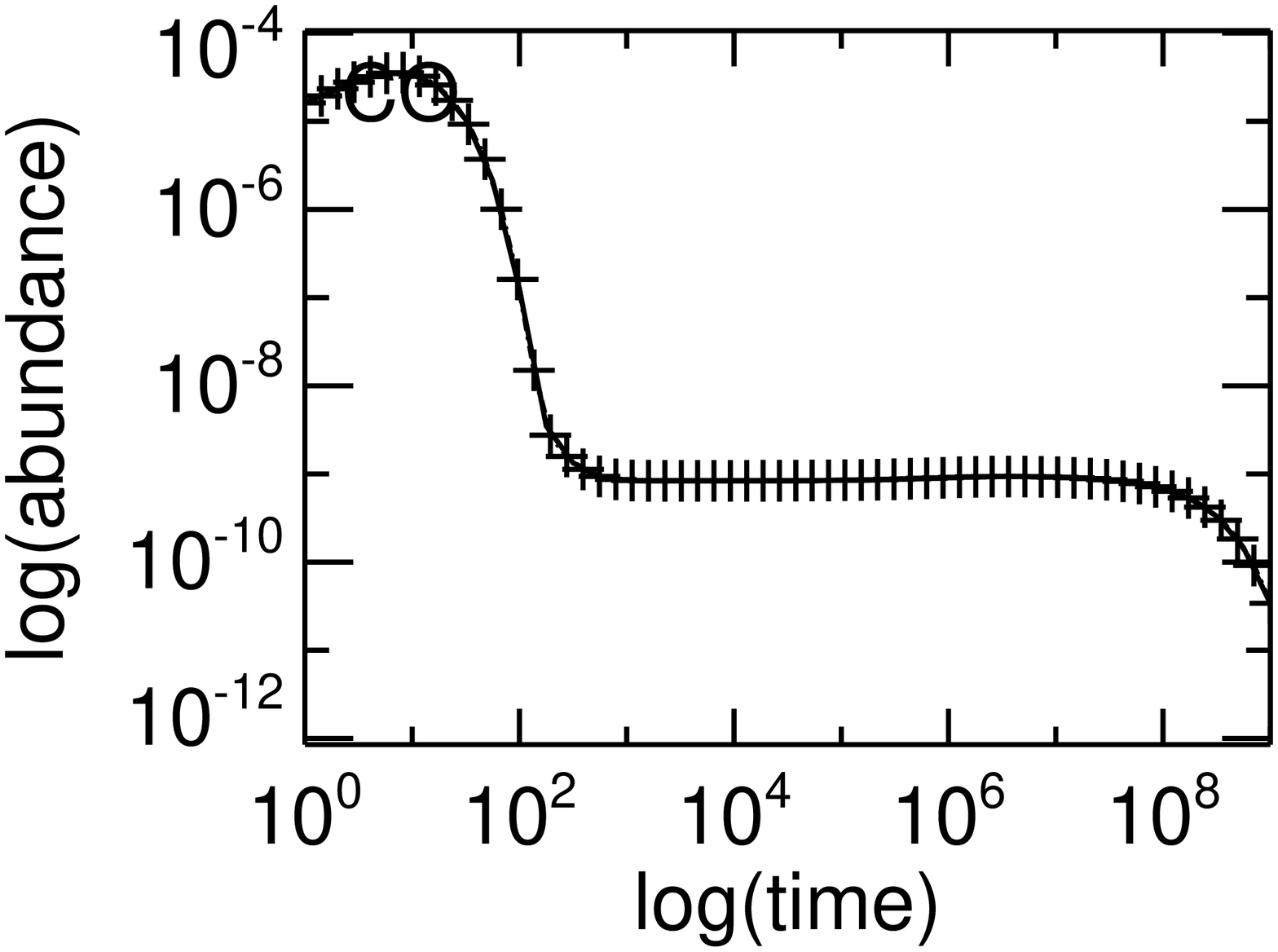}
\includegraphics[width=0.21\textwidth,clip=,angle=90]{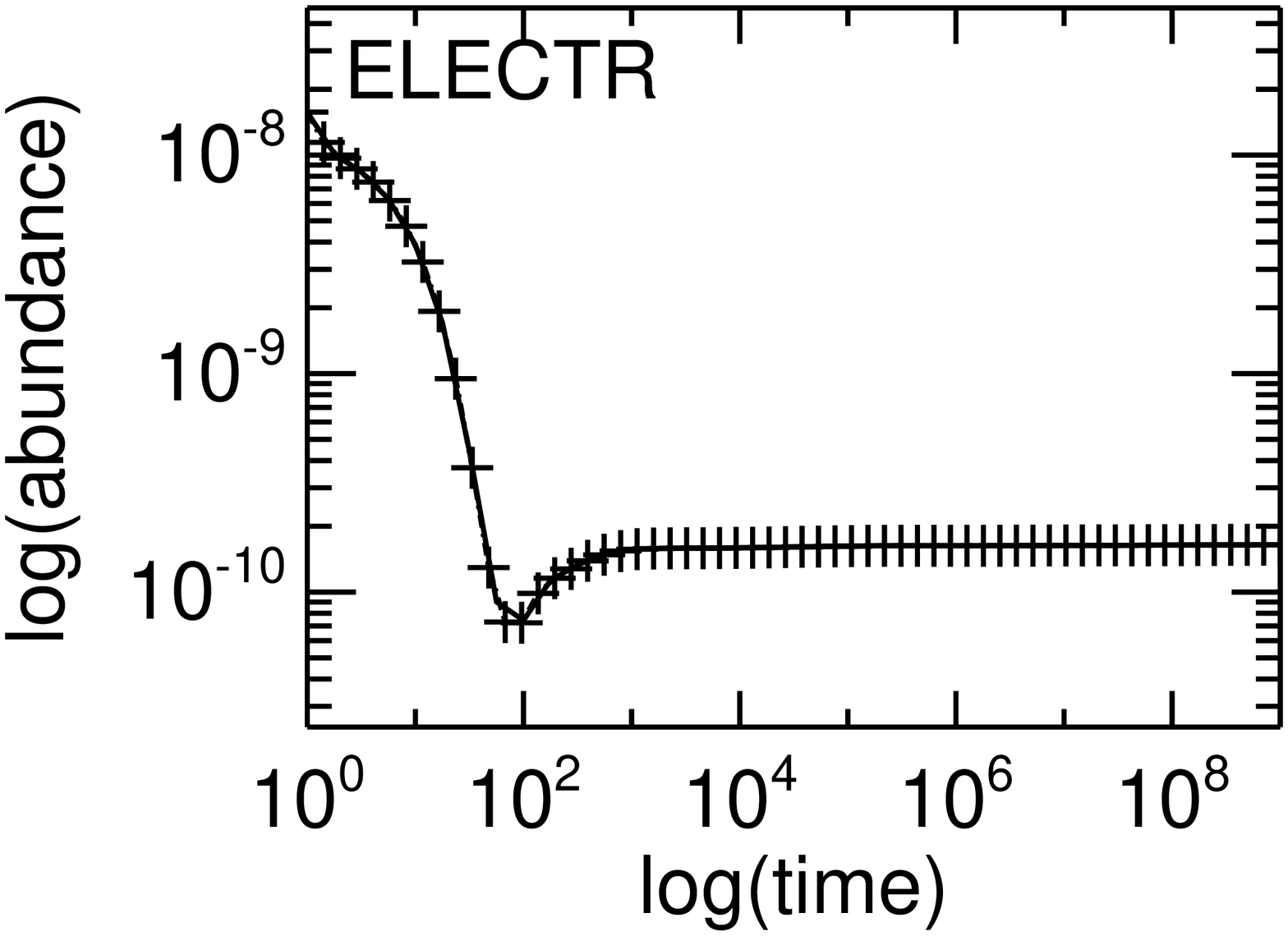}
\includegraphics[width=0.21\textwidth,clip=,angle=90]{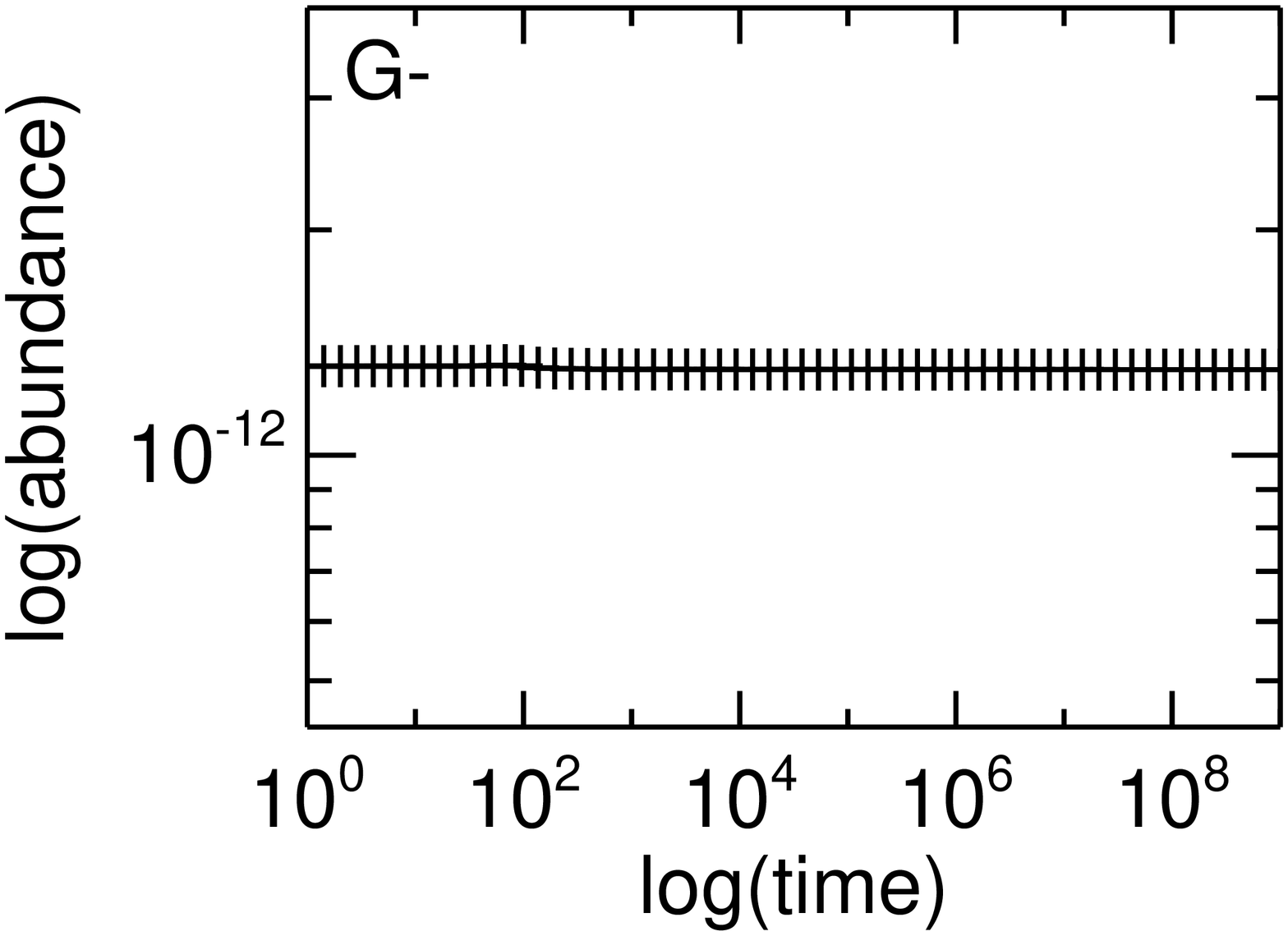}\\
\includegraphics[width=0.21\textwidth,clip=,angle=90]{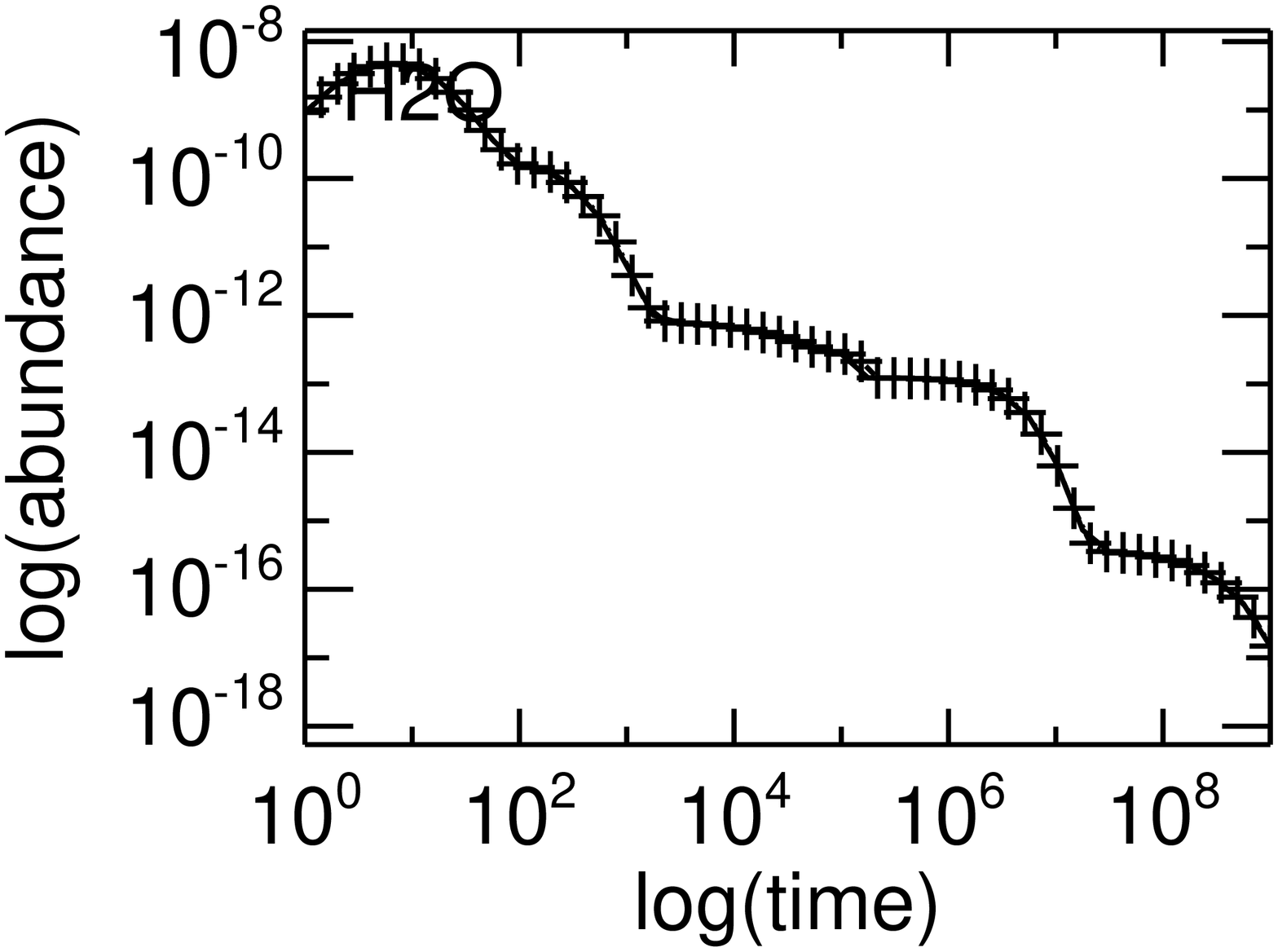}
\includegraphics[width=0.21\textwidth,clip=,angle=90]{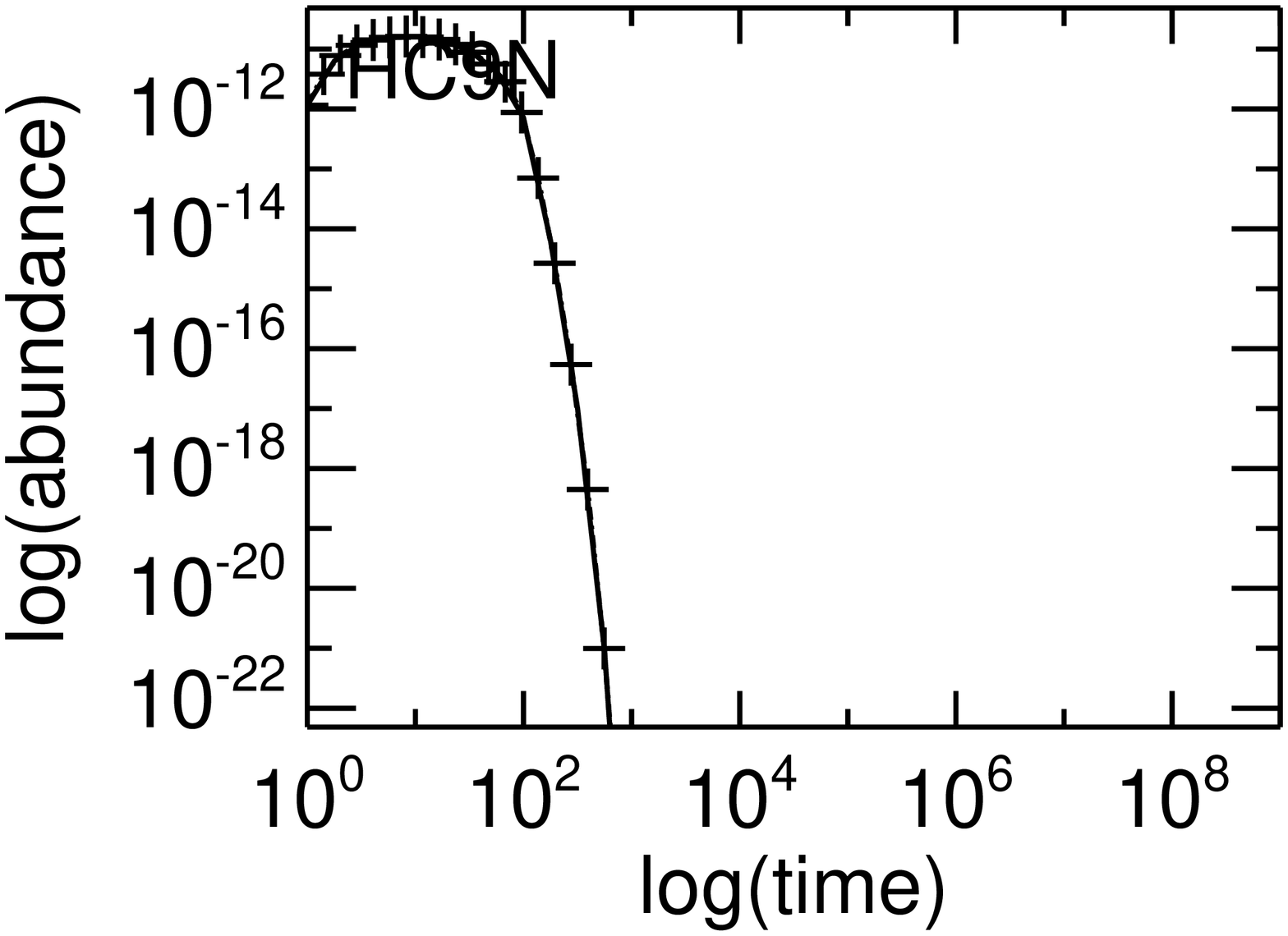}
\includegraphics[width=0.21\textwidth,clip=,angle=90]{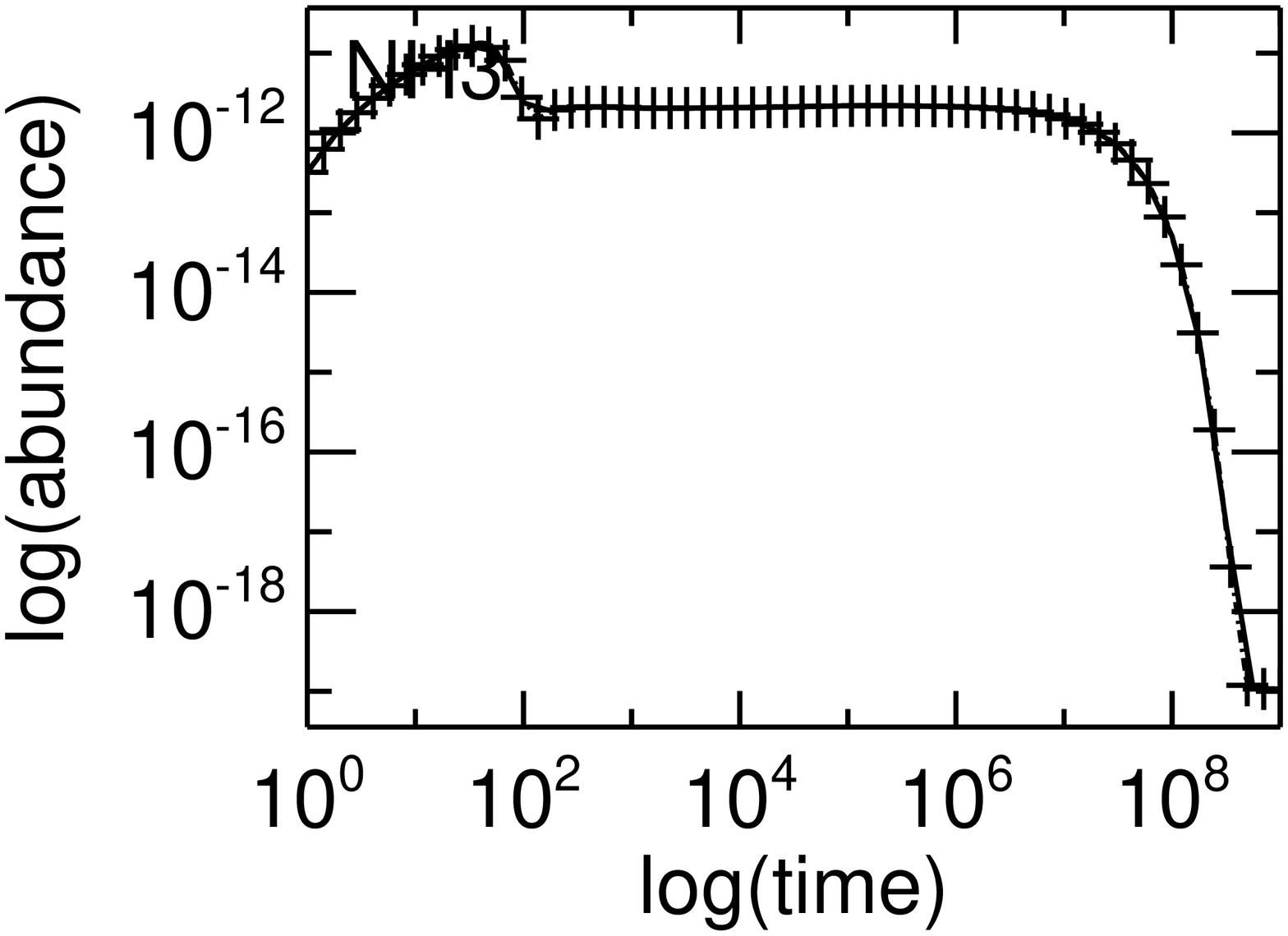}\\
\includegraphics[width=0.21\textwidth,clip=,angle=90]{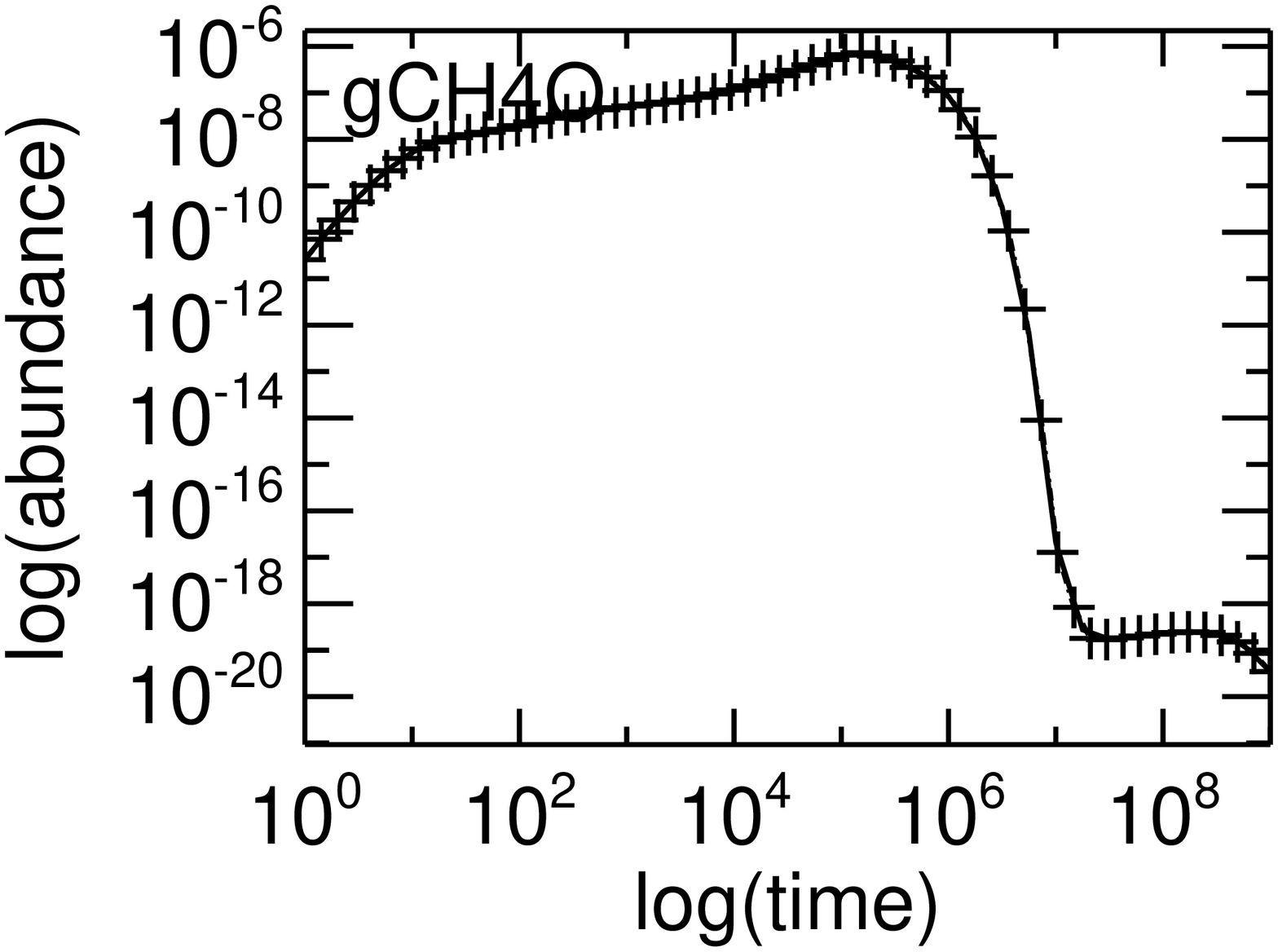}
\includegraphics[width=0.21\textwidth,clip=,angle=90]{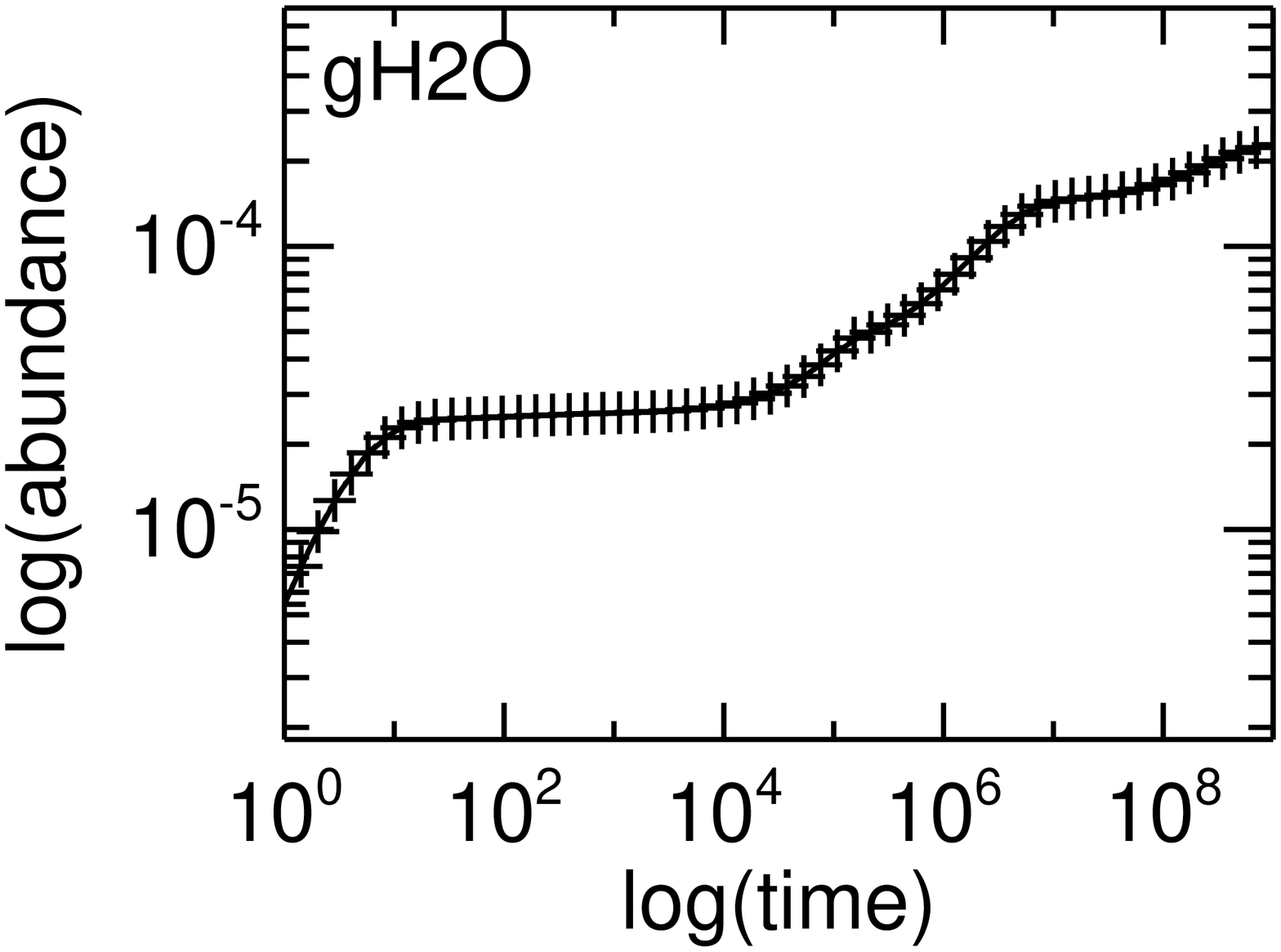}
\includegraphics[width=0.21\textwidth,clip=,angle=90]{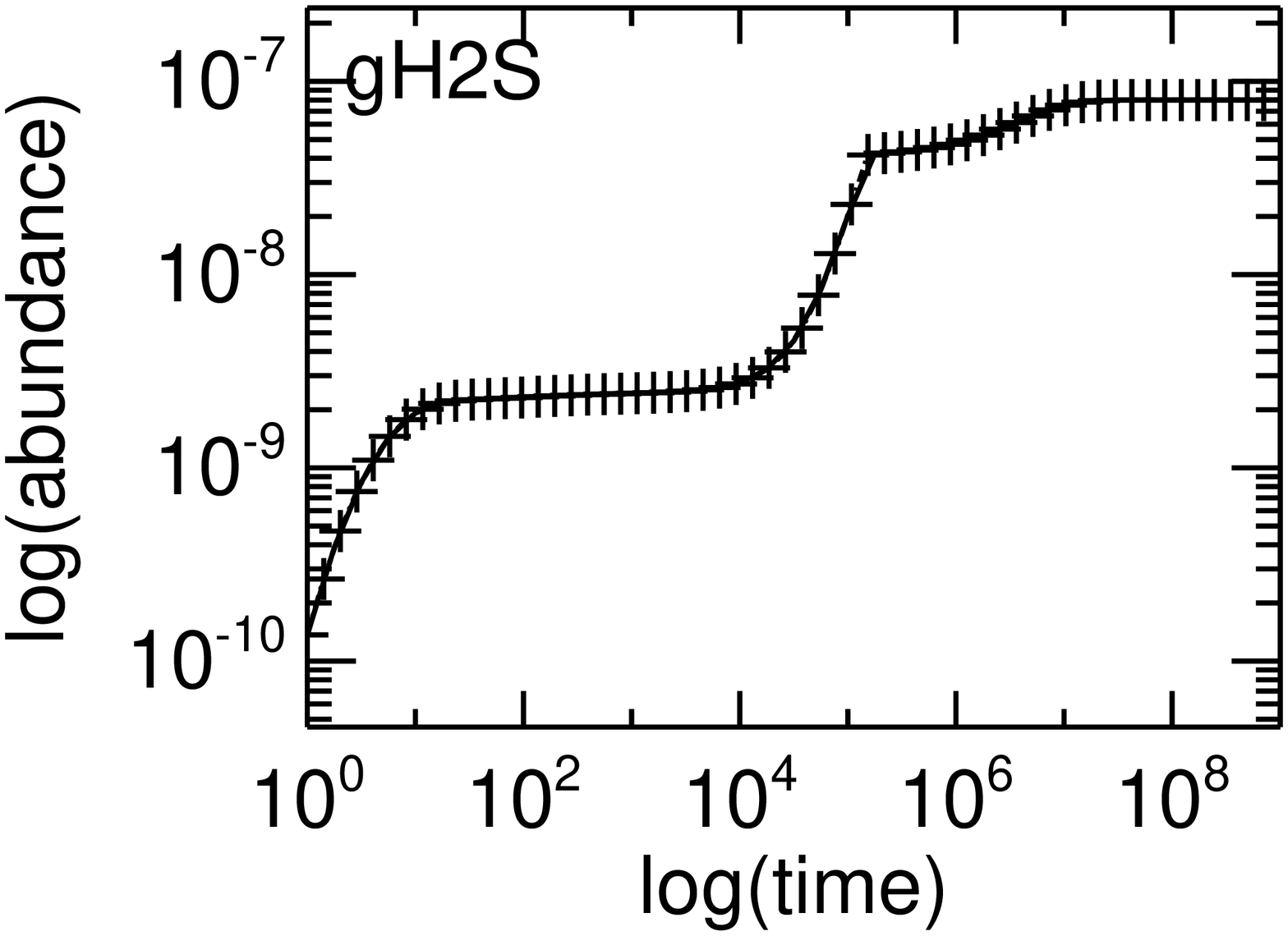}\\
\includegraphics[width=0.21\textwidth,clip=,angle=90]{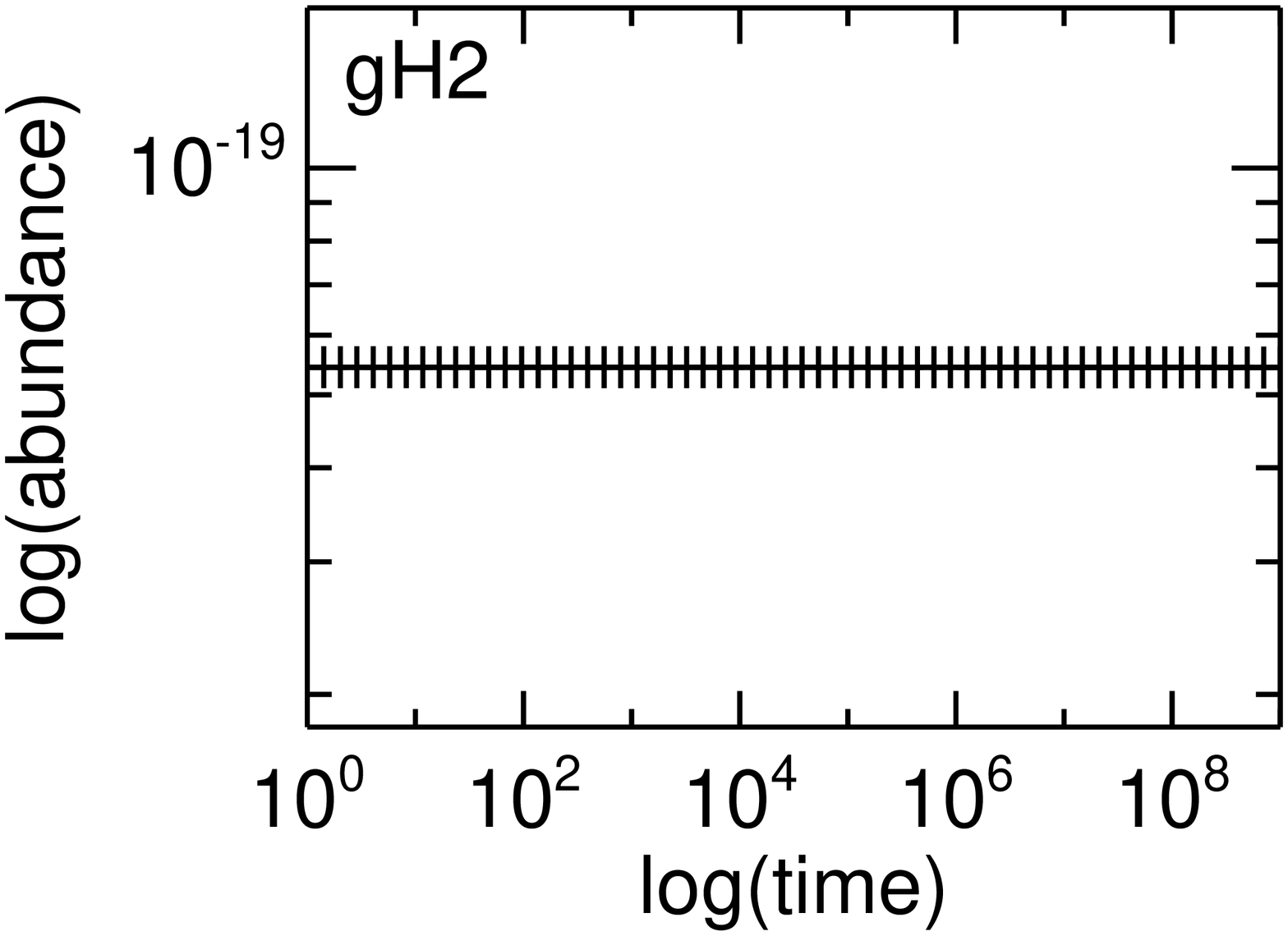}
\includegraphics[width=0.21\textwidth,clip=,angle=90]{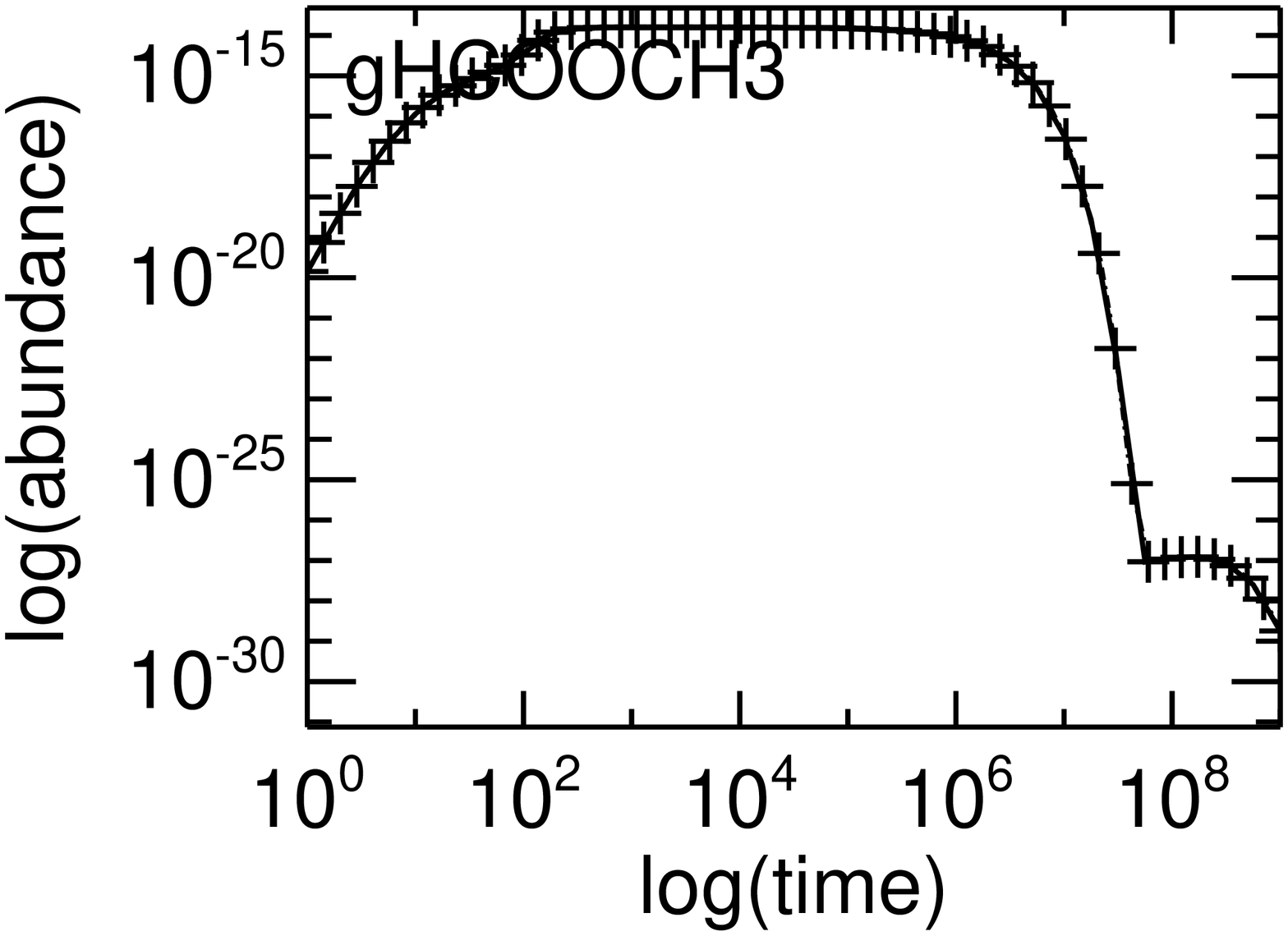}
\includegraphics[width=0.21\textwidth,clip=,angle=90]{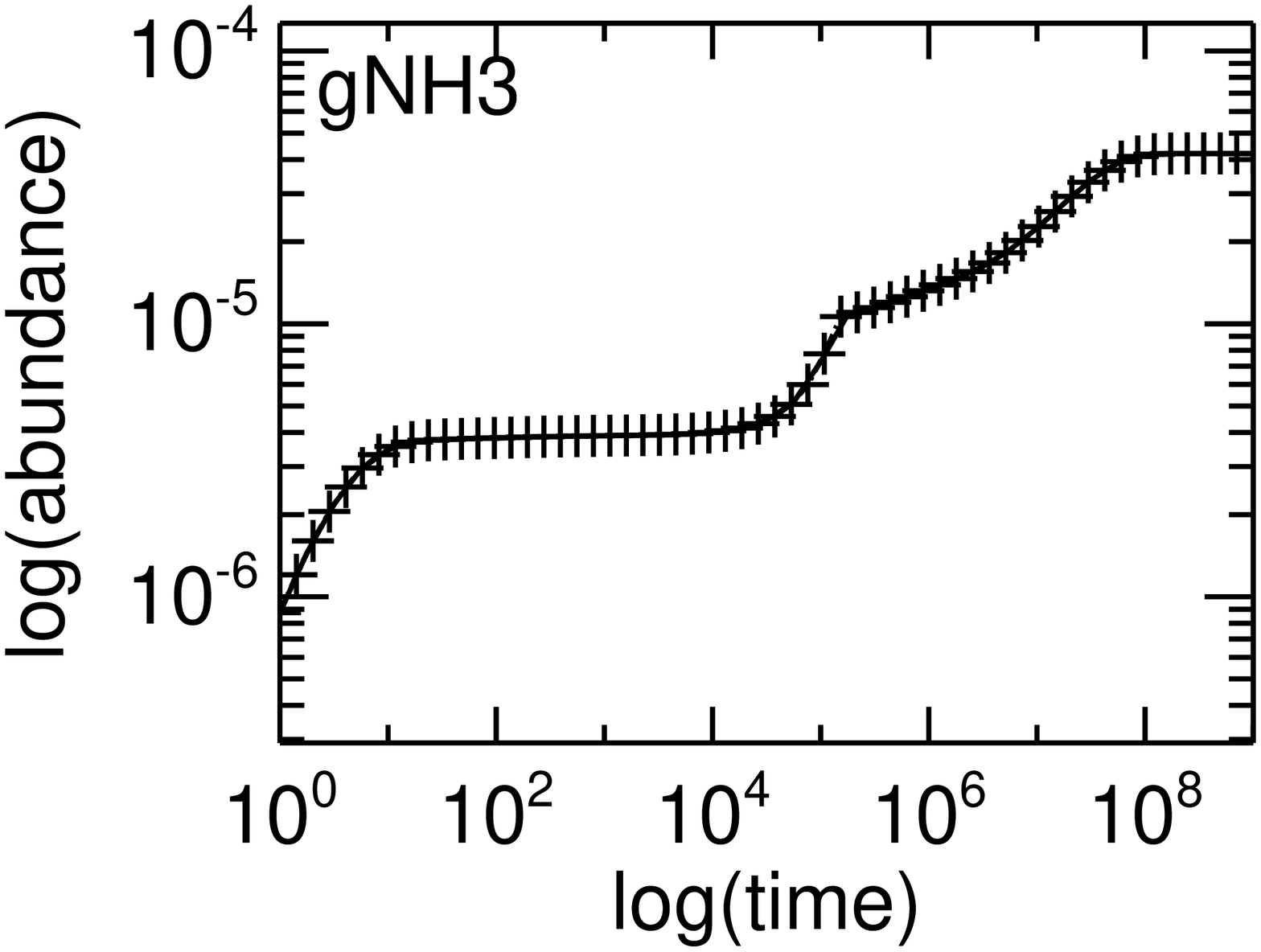}
\caption{The same as in Fig.~\ref{fig:TMC1} but
for the ``DISK1'' case (DM Tau, $r=100$~AU, midplane).}
\label{fig:DISK1}
\end{figure*}

\begin{figure*}
\includegraphics[width=0.21\textwidth,clip=,angle=90]{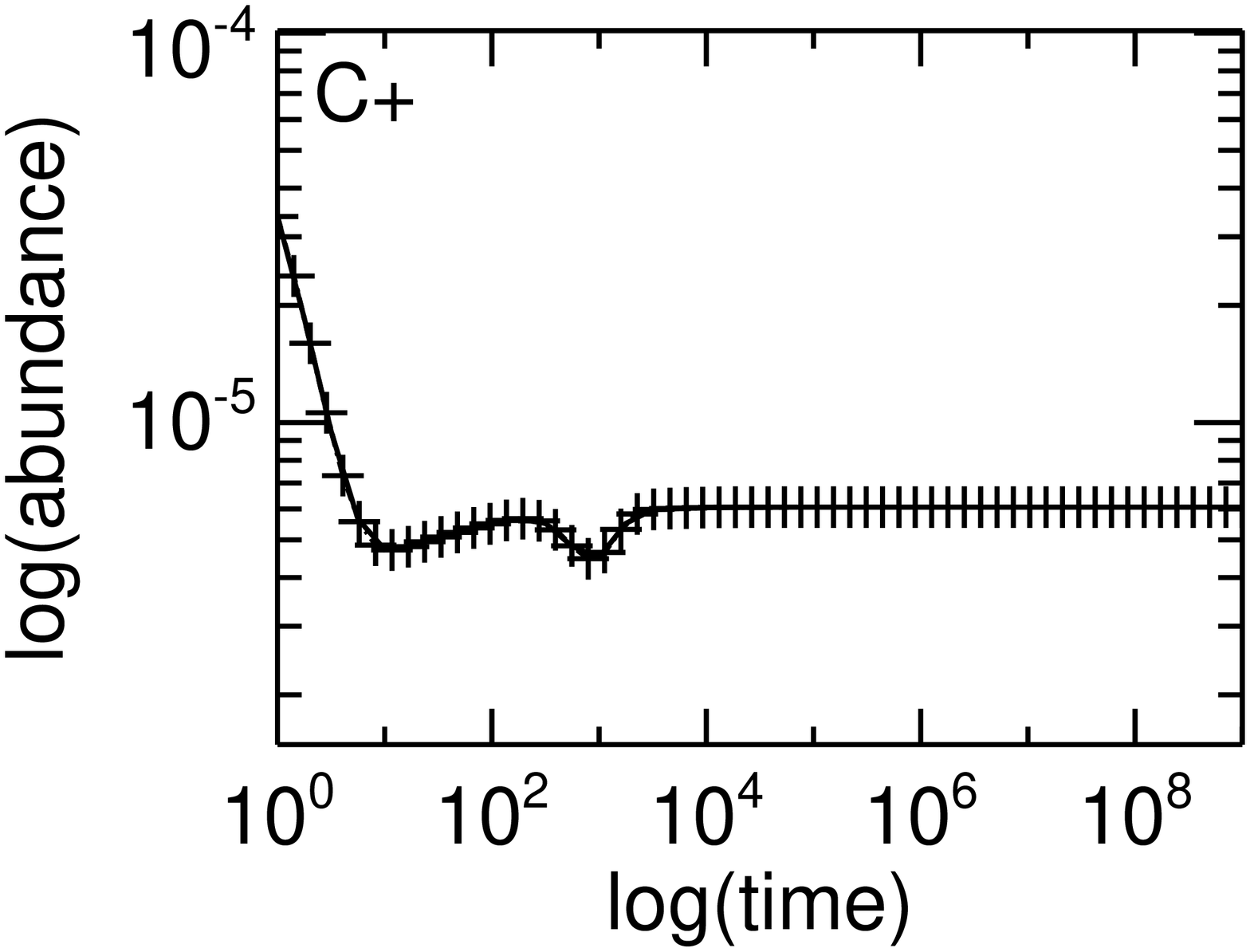}
\includegraphics[width=0.21\textwidth,clip=,angle=90]{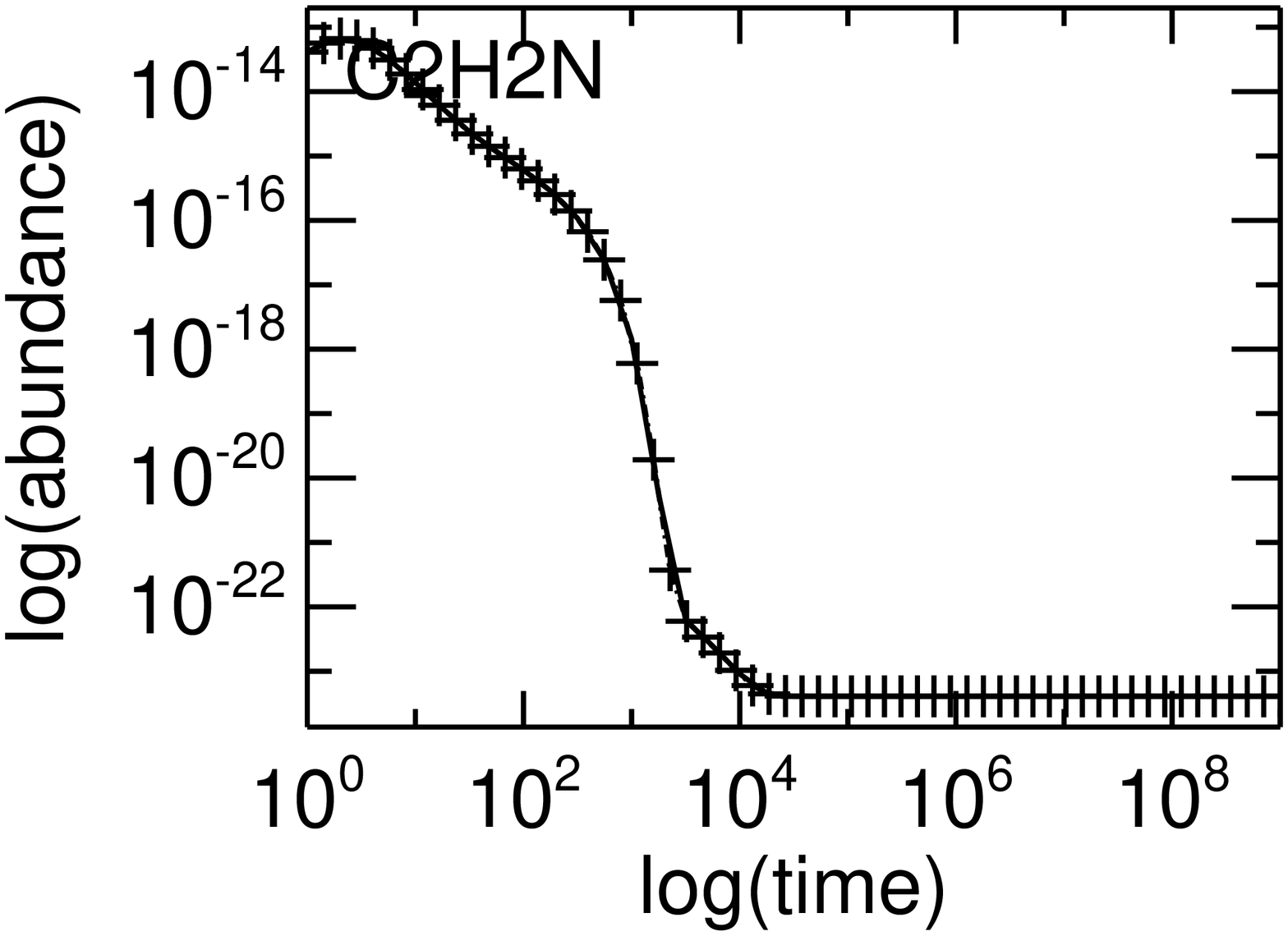}
\includegraphics[width=0.21\textwidth,clip=,angle=90]{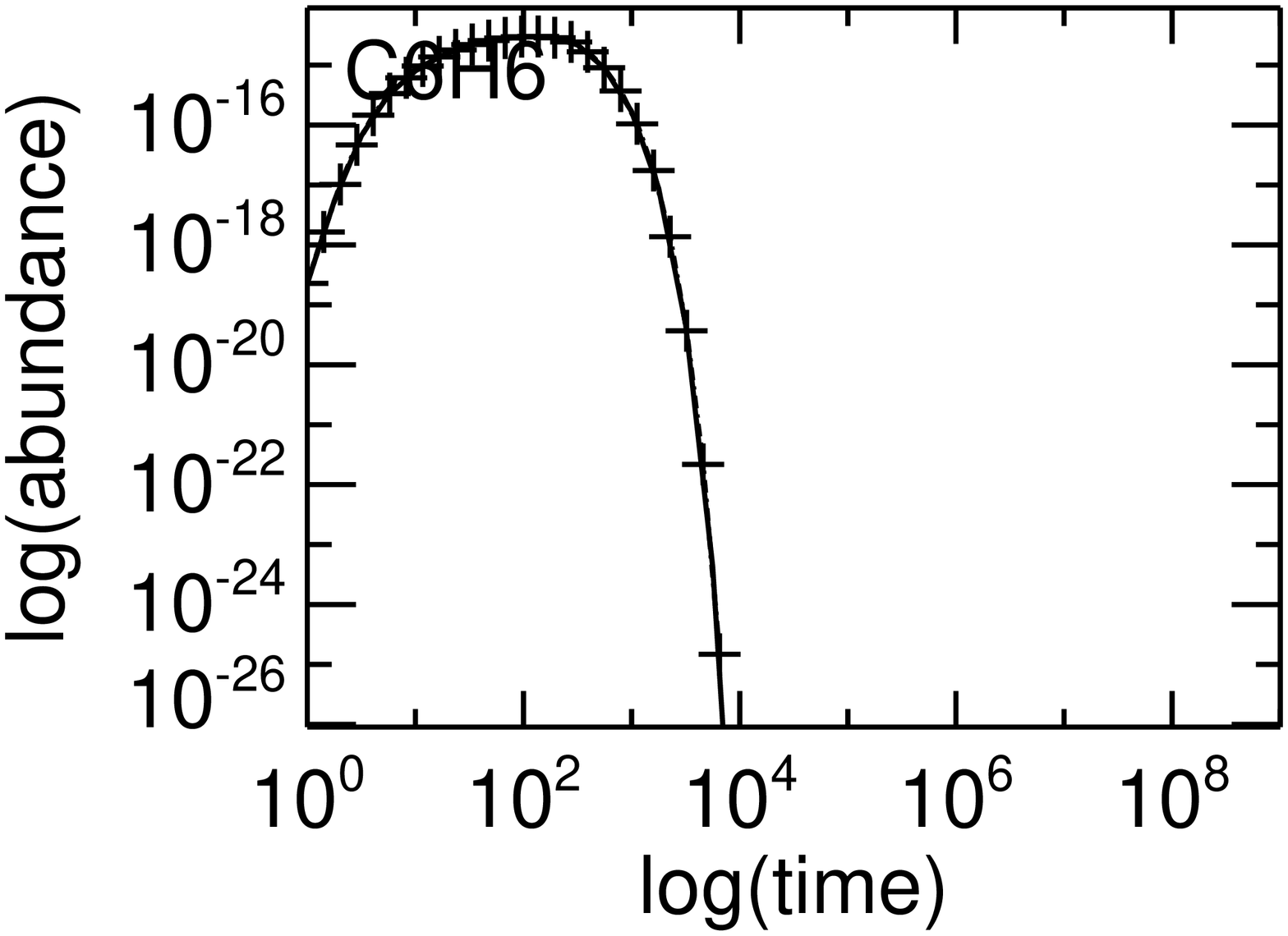}\\
\includegraphics[width=0.21\textwidth,clip=,angle=90]{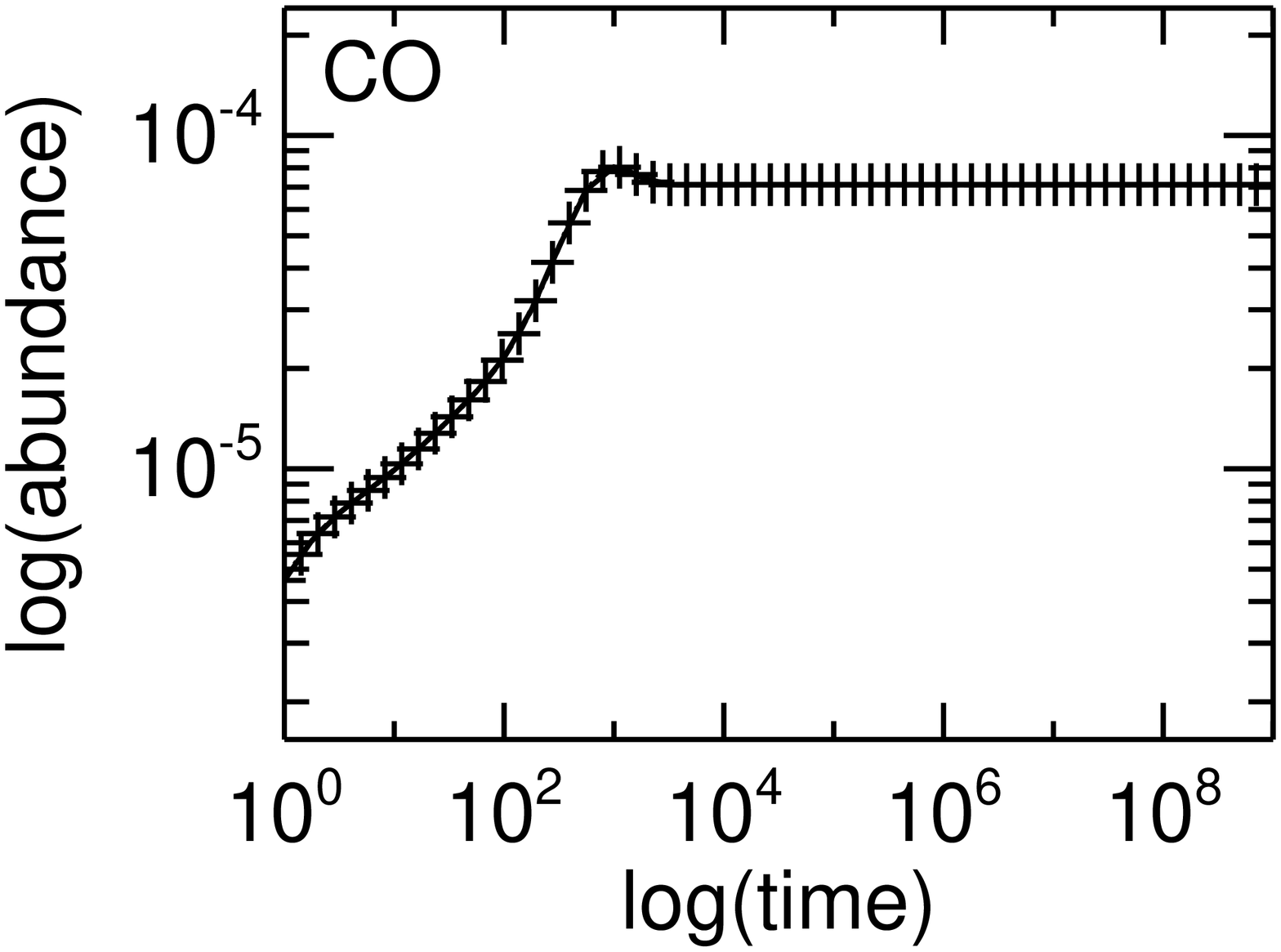}
\includegraphics[width=0.21\textwidth,clip=,angle=90]{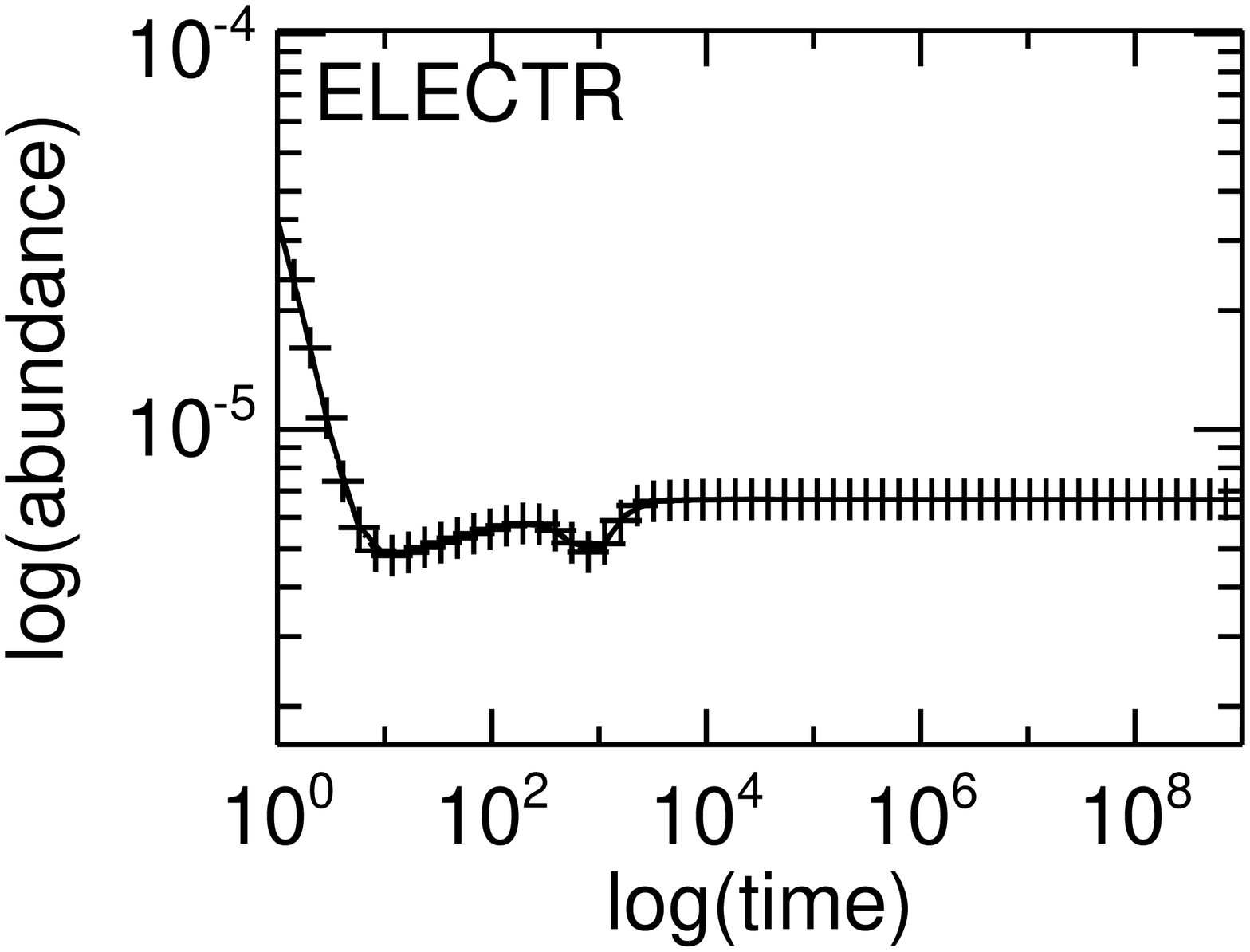}
\includegraphics[width=0.21\textwidth,clip=,angle=90]{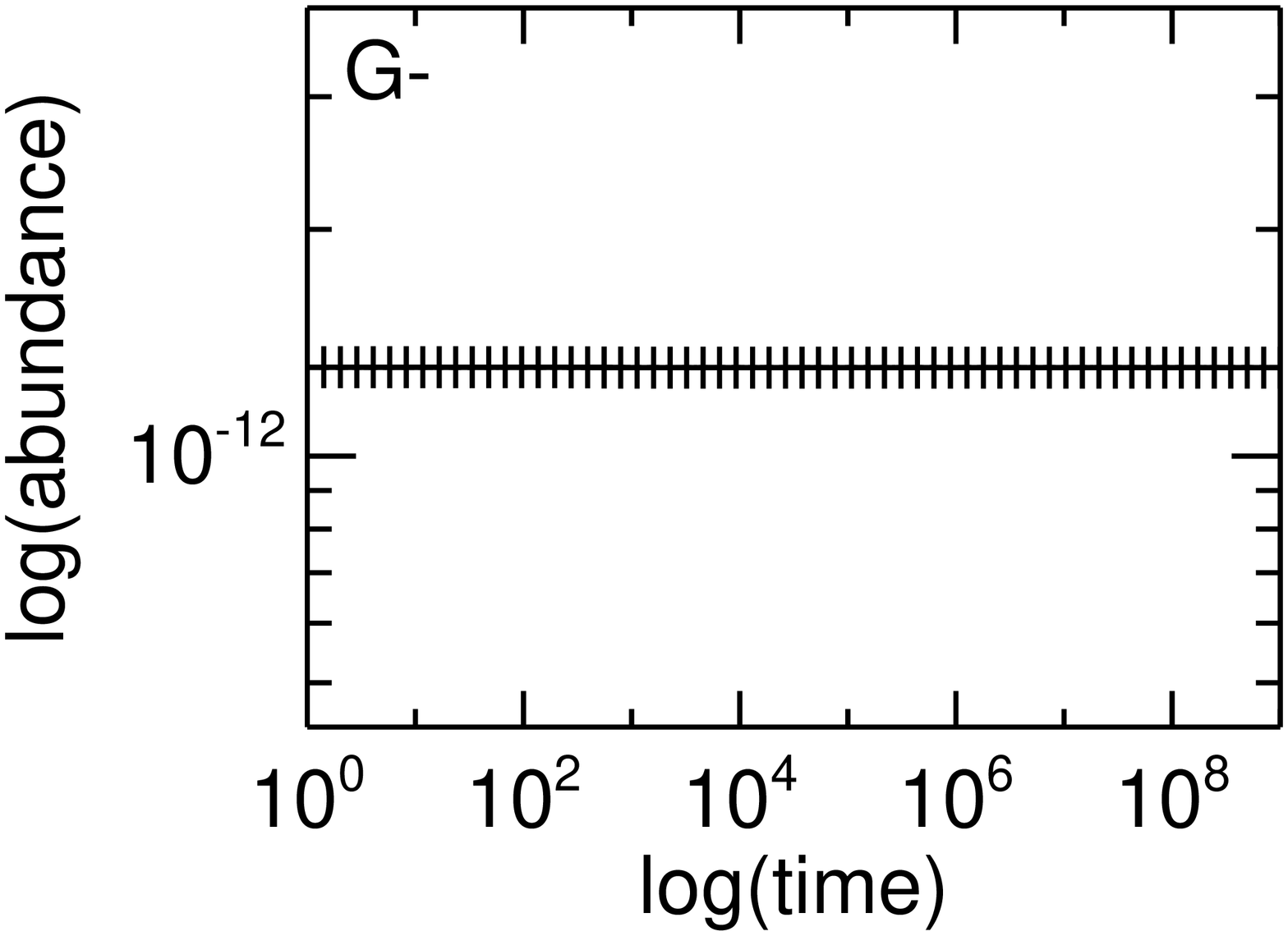}\\
\includegraphics[width=0.21\textwidth,clip=,angle=90]{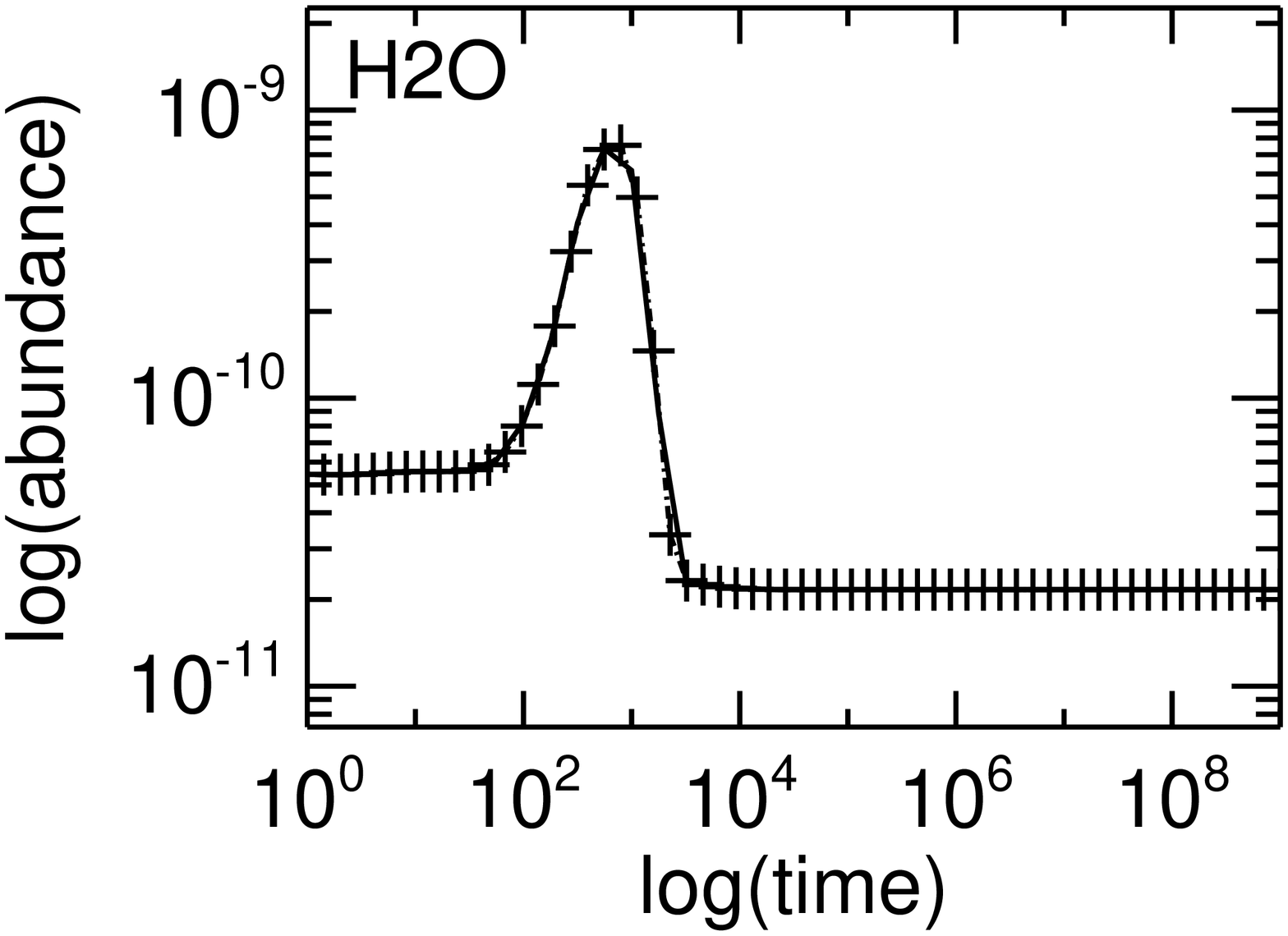}
\includegraphics[width=0.21\textwidth,clip=,angle=90]{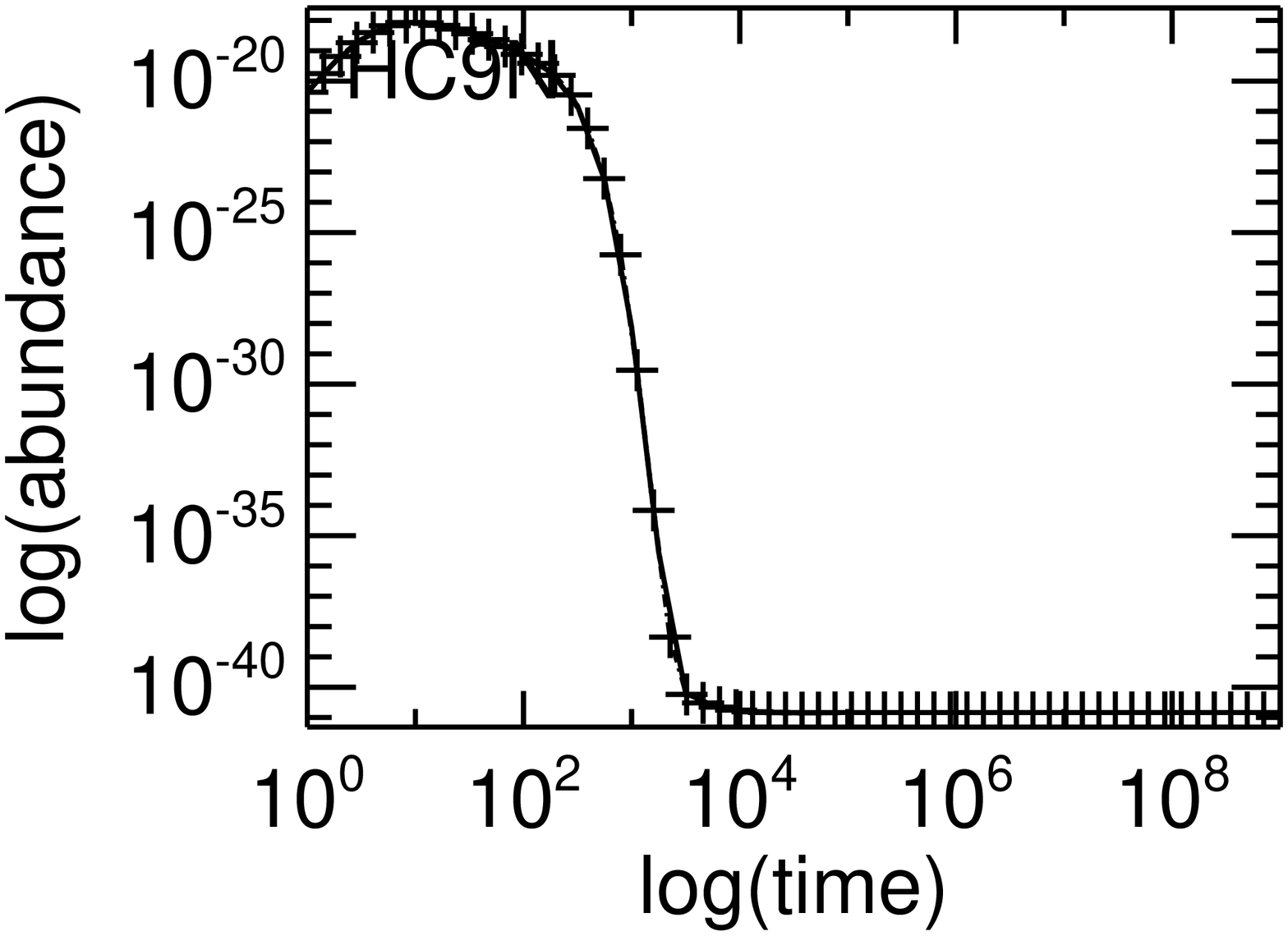}
\includegraphics[width=0.21\textwidth,clip=,angle=90]{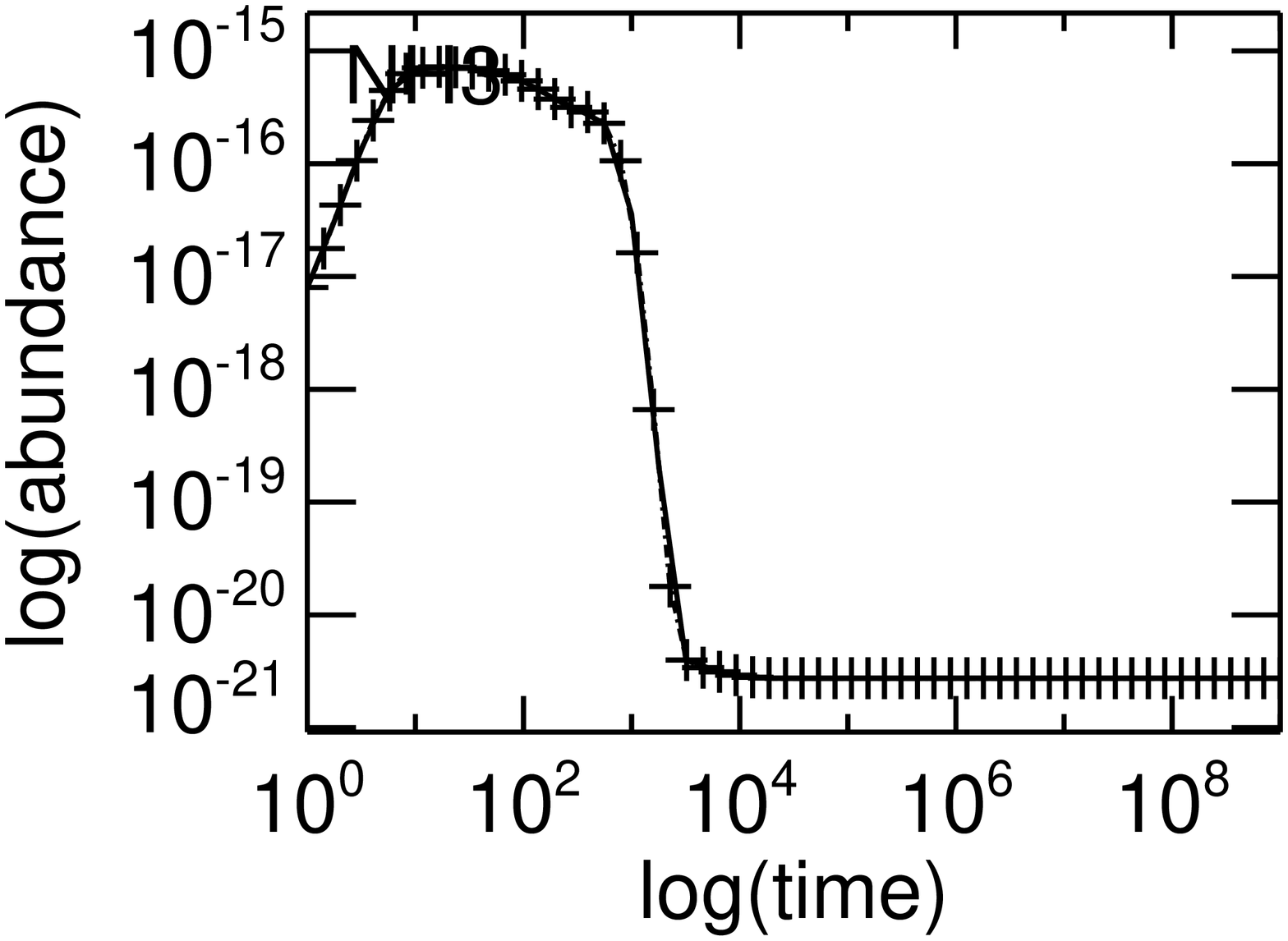}\\
\includegraphics[width=0.21\textwidth,clip=,angle=90]{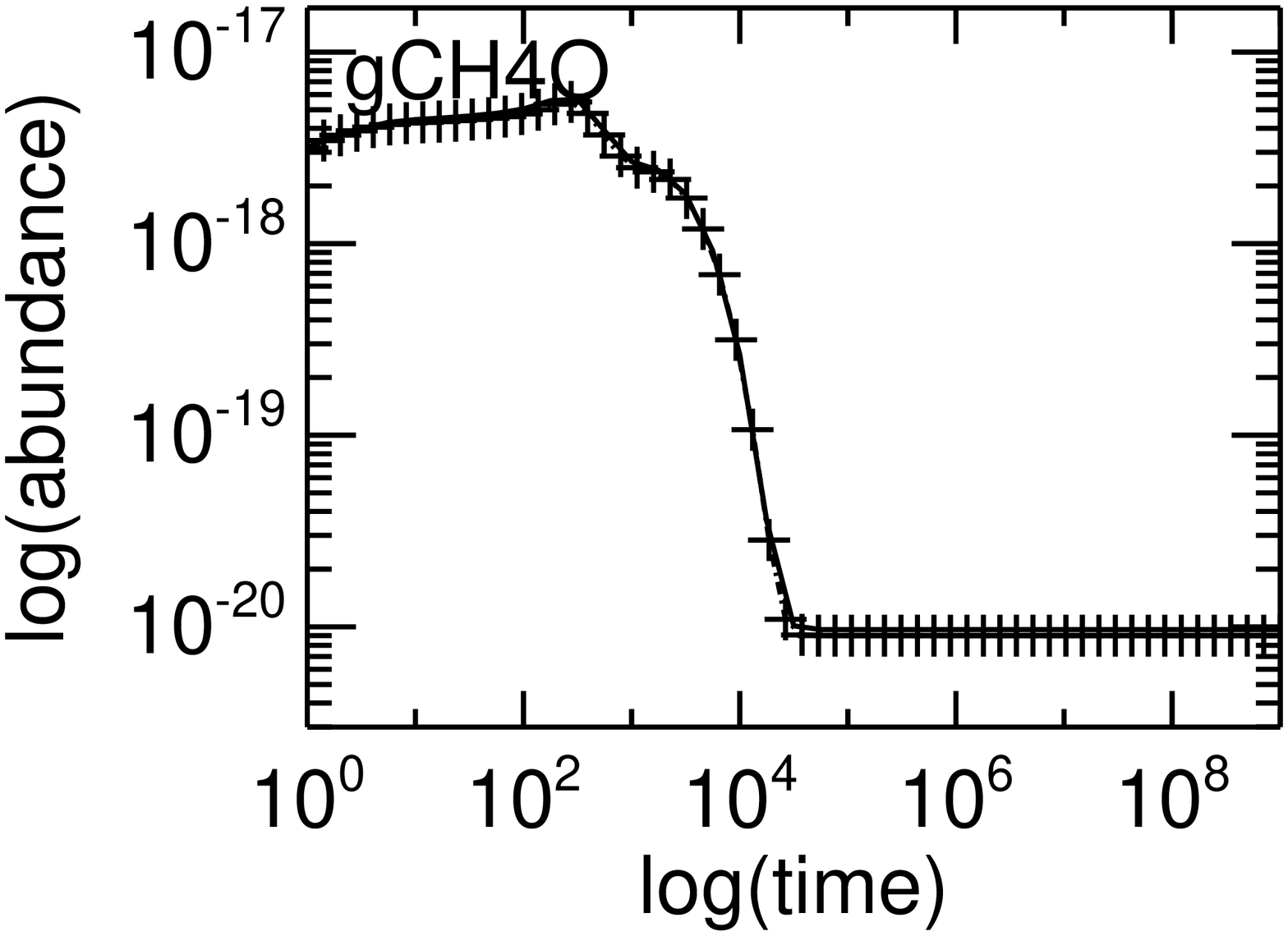}
\includegraphics[width=0.21\textwidth,clip=,angle=90]{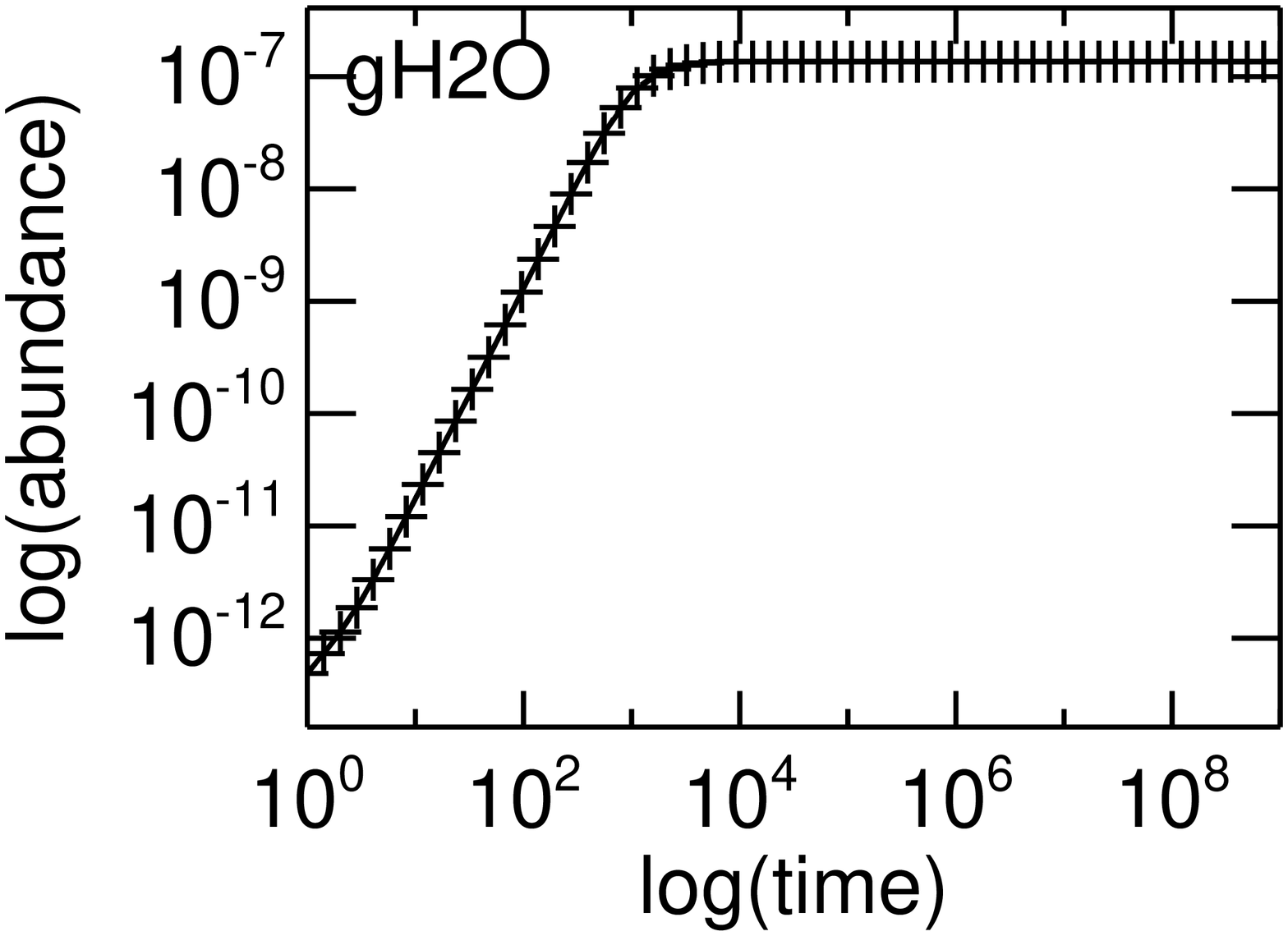}
\includegraphics[width=0.21\textwidth,clip=,angle=90]{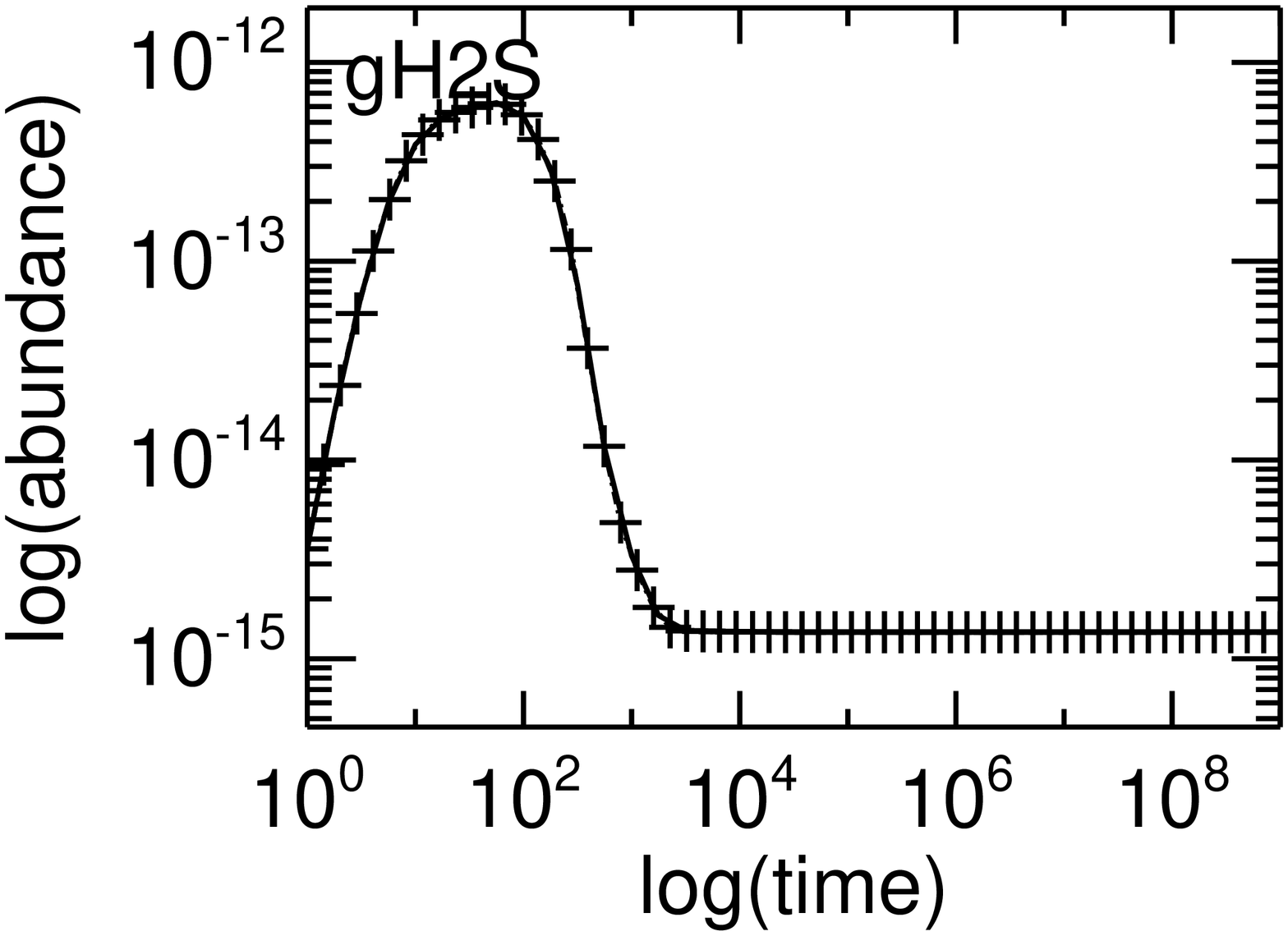}\\
\includegraphics[width=0.21\textwidth,clip=,angle=90]{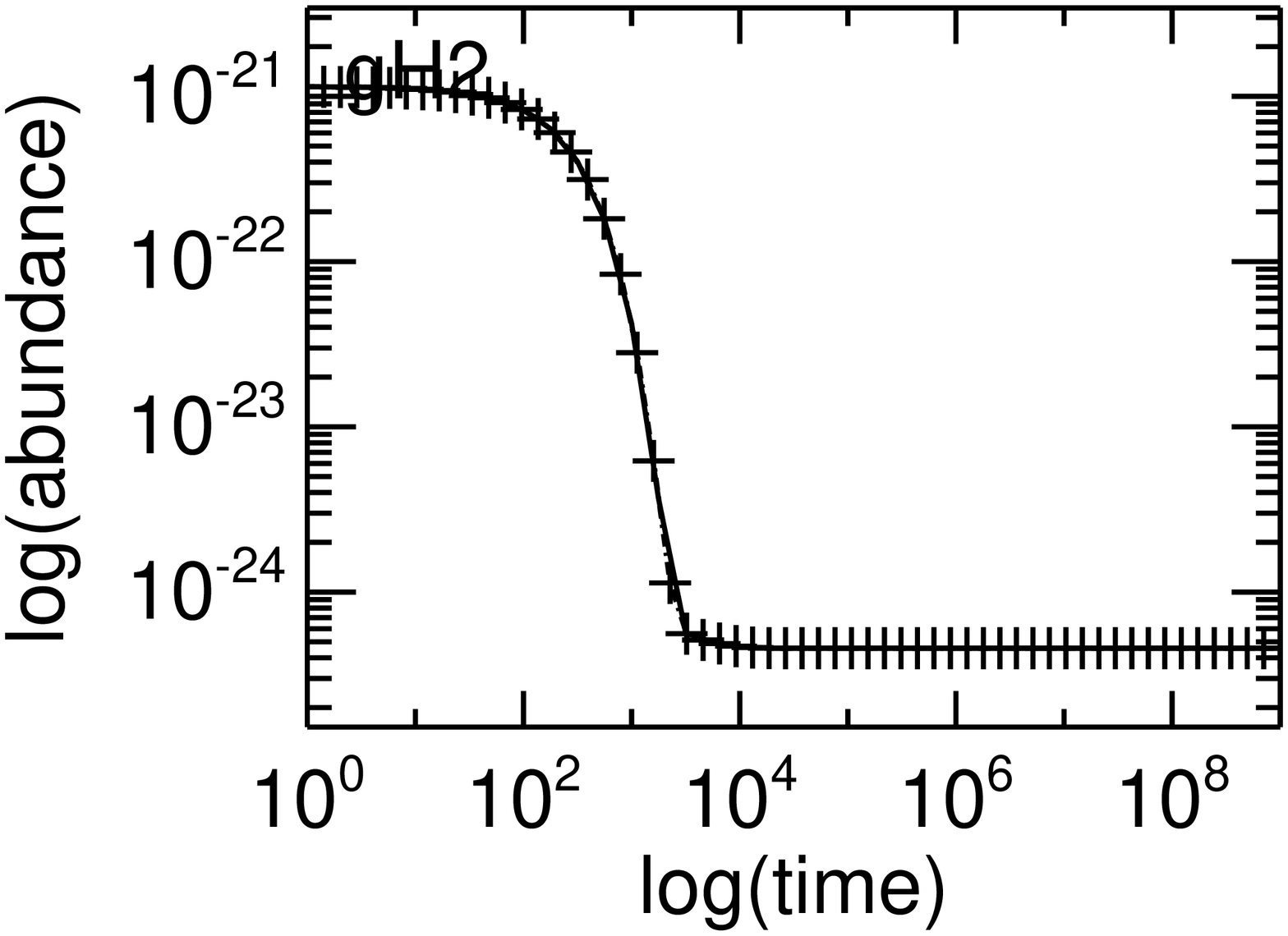}
\includegraphics[width=0.21\textwidth,clip=,angle=90]{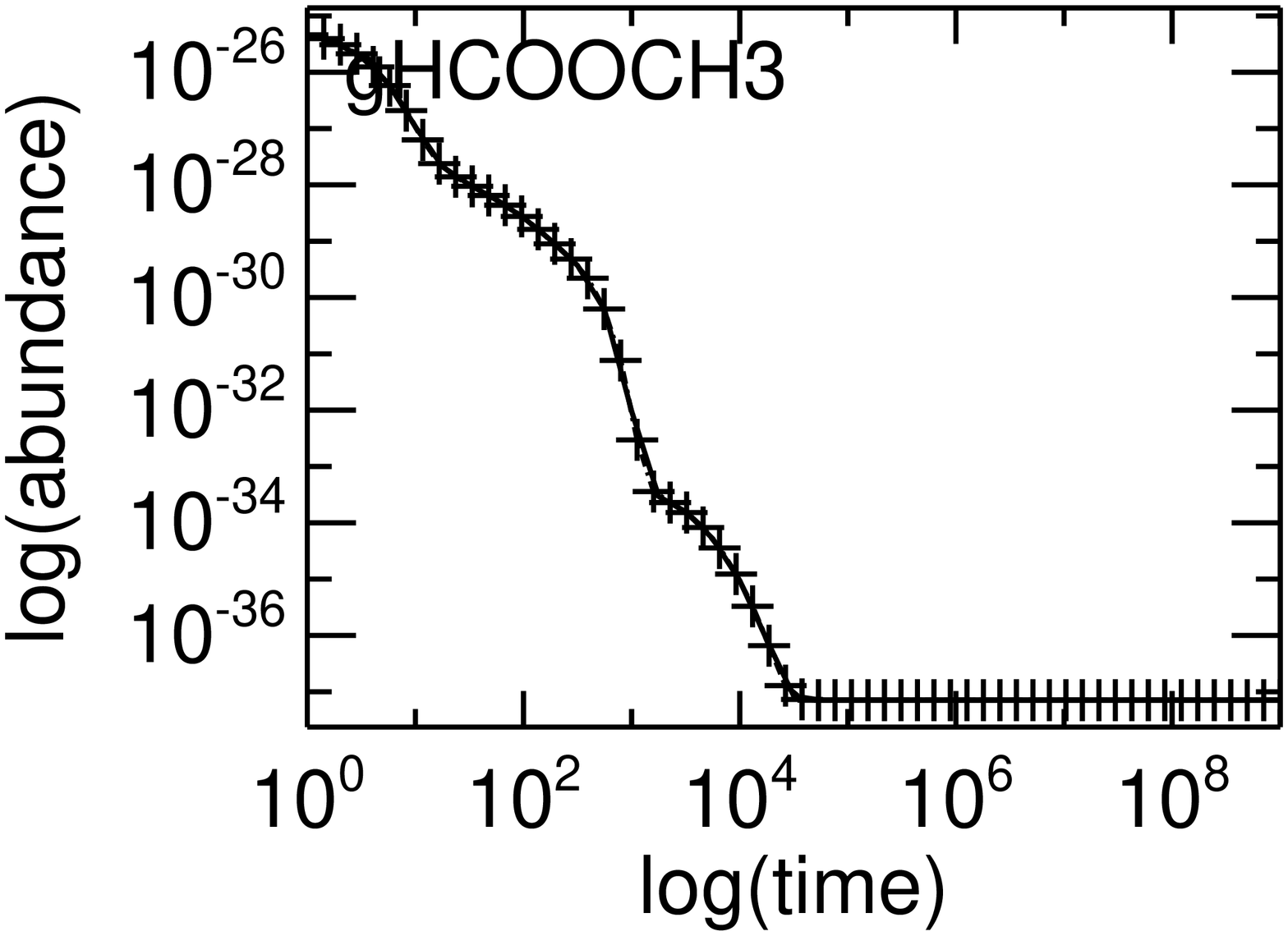}
\includegraphics[width=0.21\textwidth,clip=,angle=90]{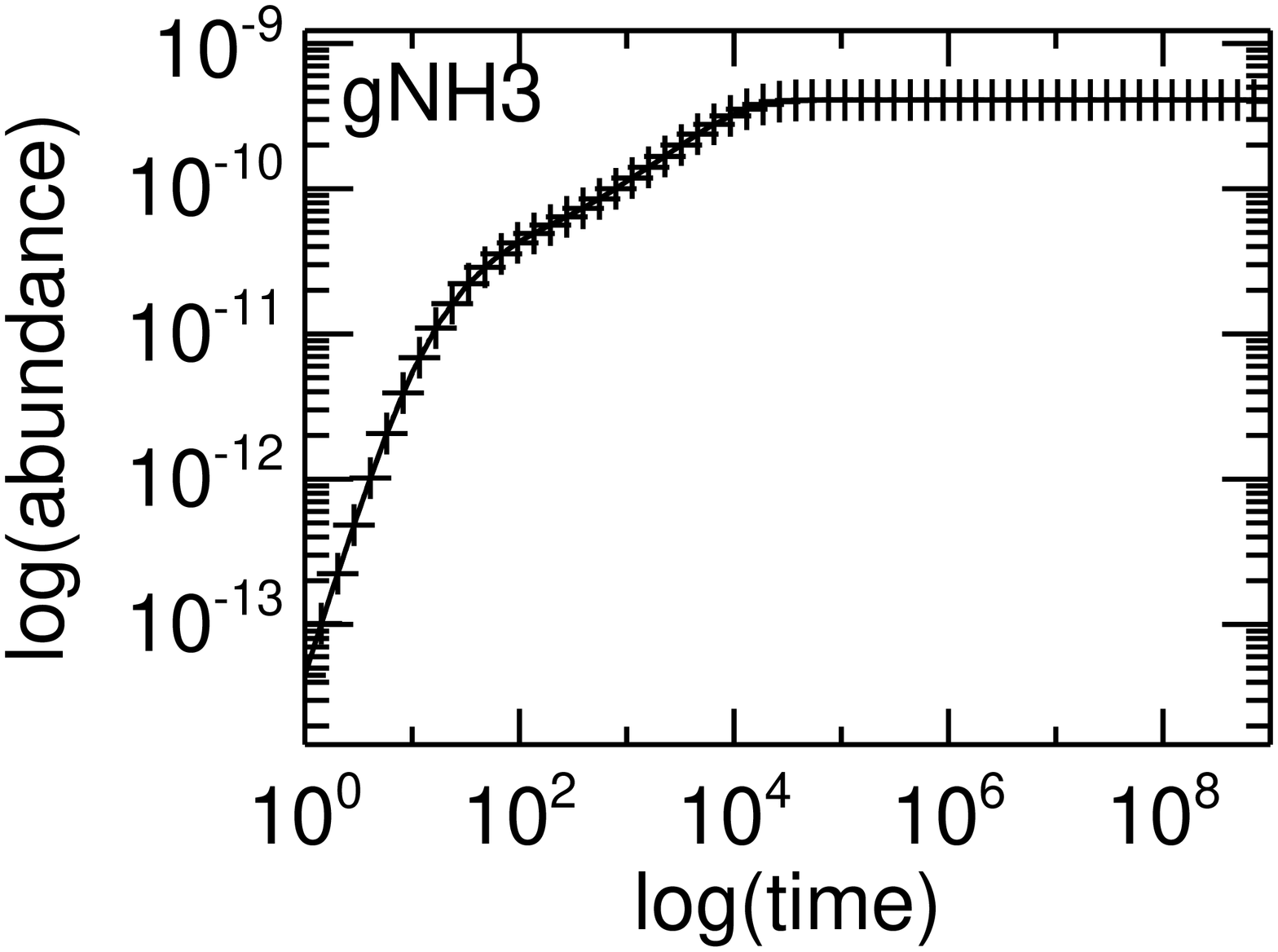}
\caption{The same as in Fig.~\ref{fig:DISK1} but
for the ``DISK2'' case (DM Tau, $r=100$~AU, warm molecular layer).}
\label{fig:DISK2}
\end{figure*}

\begin{figure*}
\includegraphics[width=0.21\textwidth,clip=,angle=90]{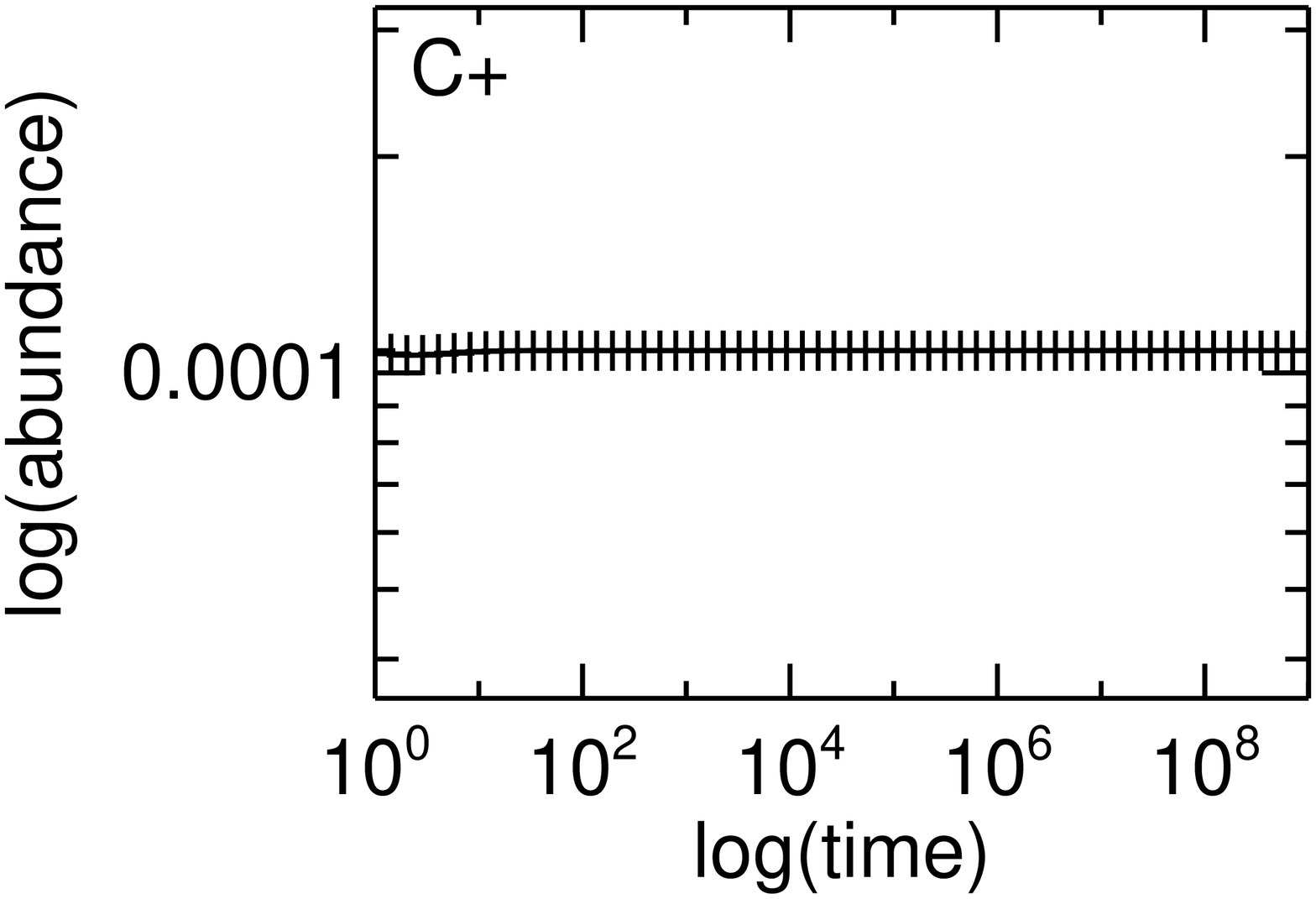}
\includegraphics[width=0.21\textwidth,clip=,angle=90]{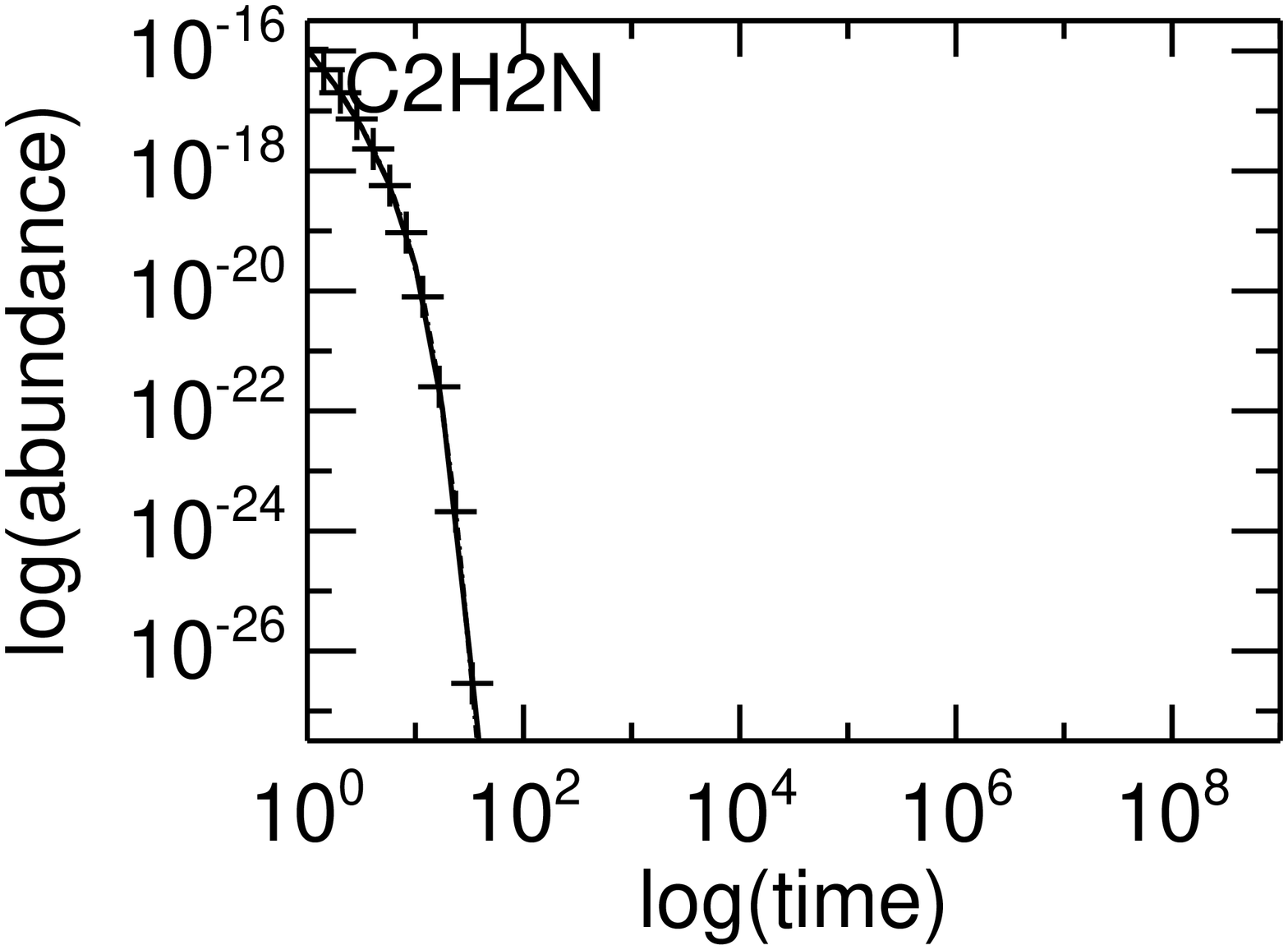}
\includegraphics[width=0.21\textwidth,clip=,angle=90]{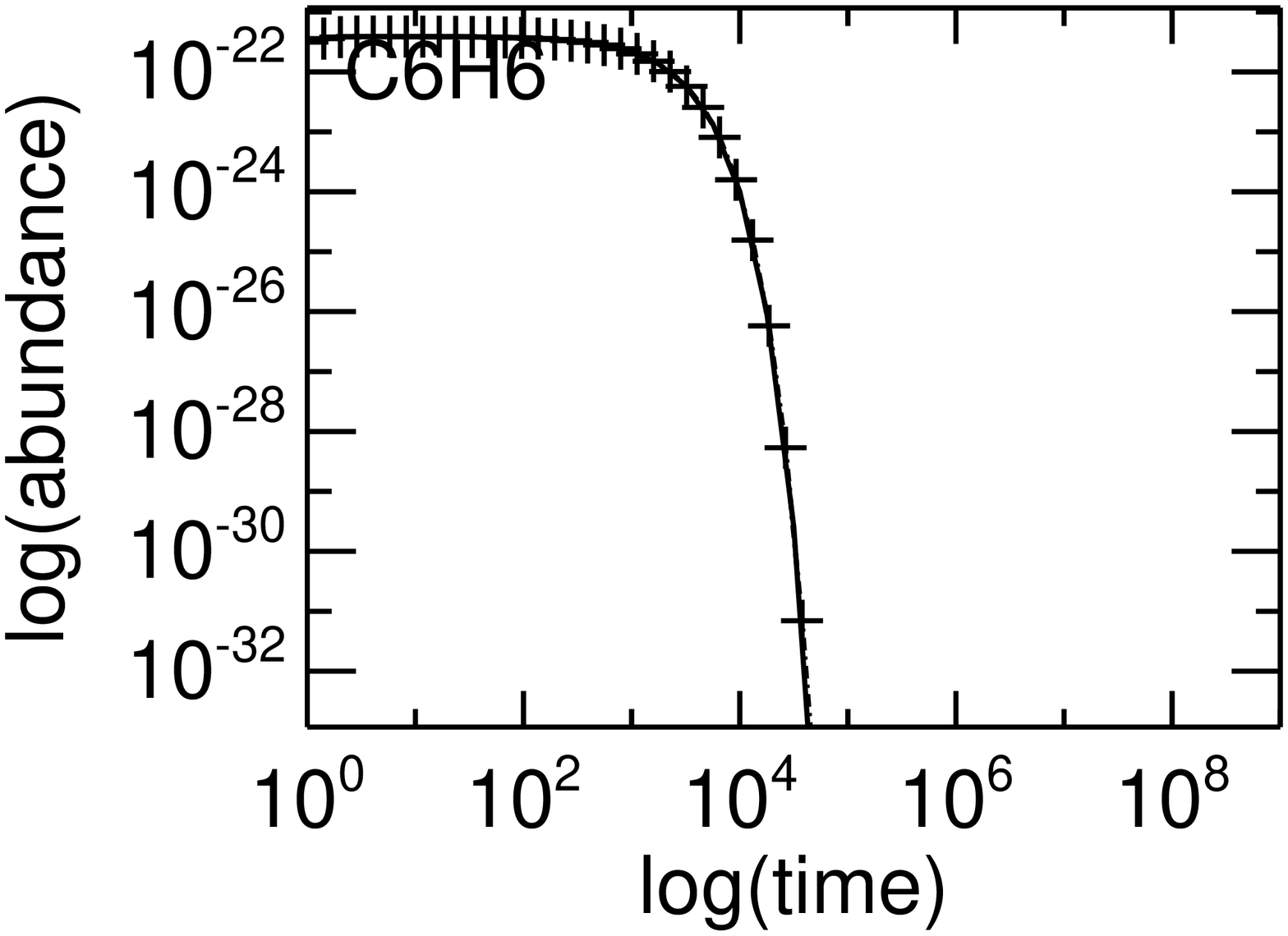}\\
\includegraphics[width=0.21\textwidth,clip=,angle=90]{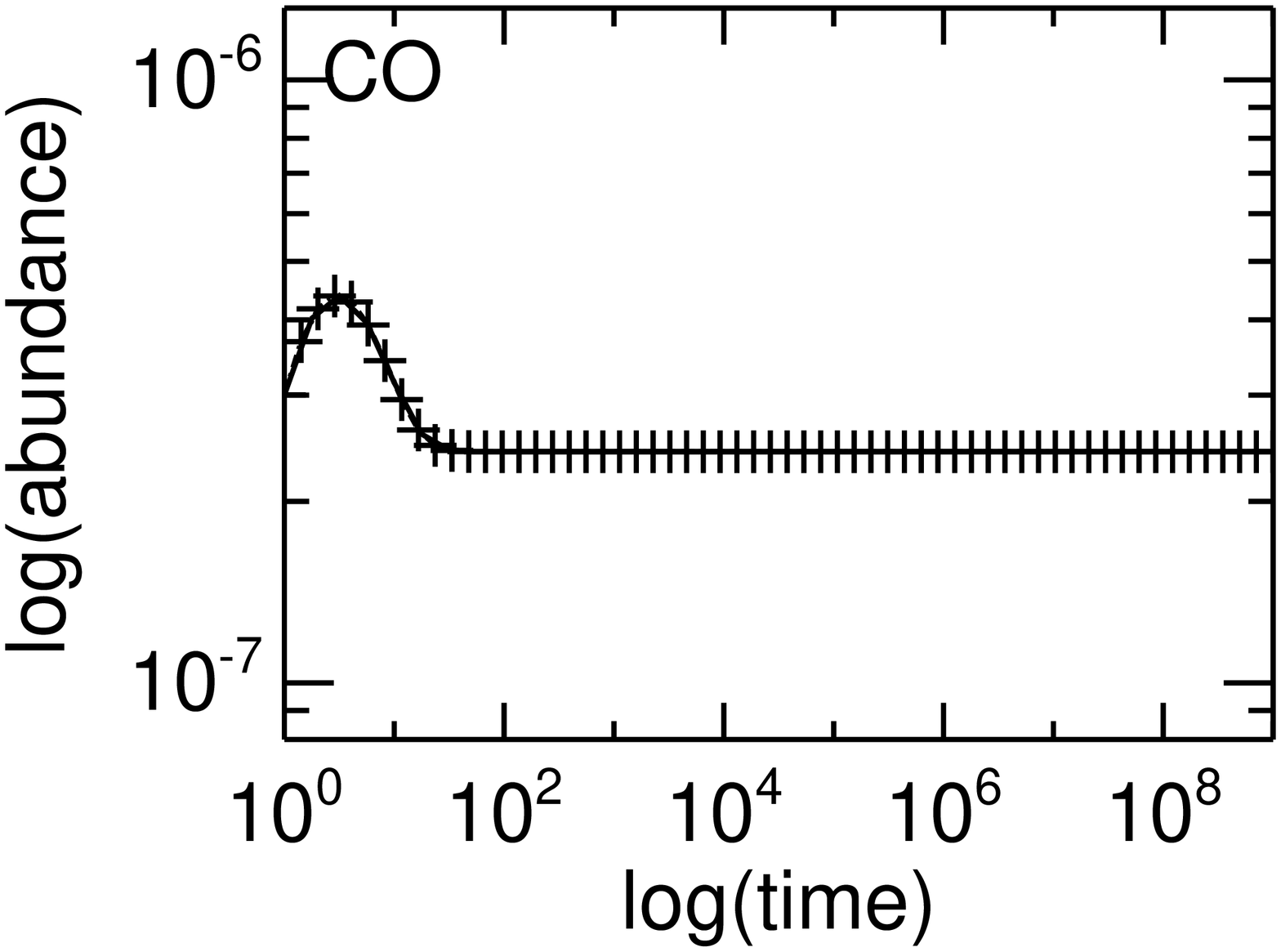}
\includegraphics[width=0.21\textwidth,clip=,angle=90]{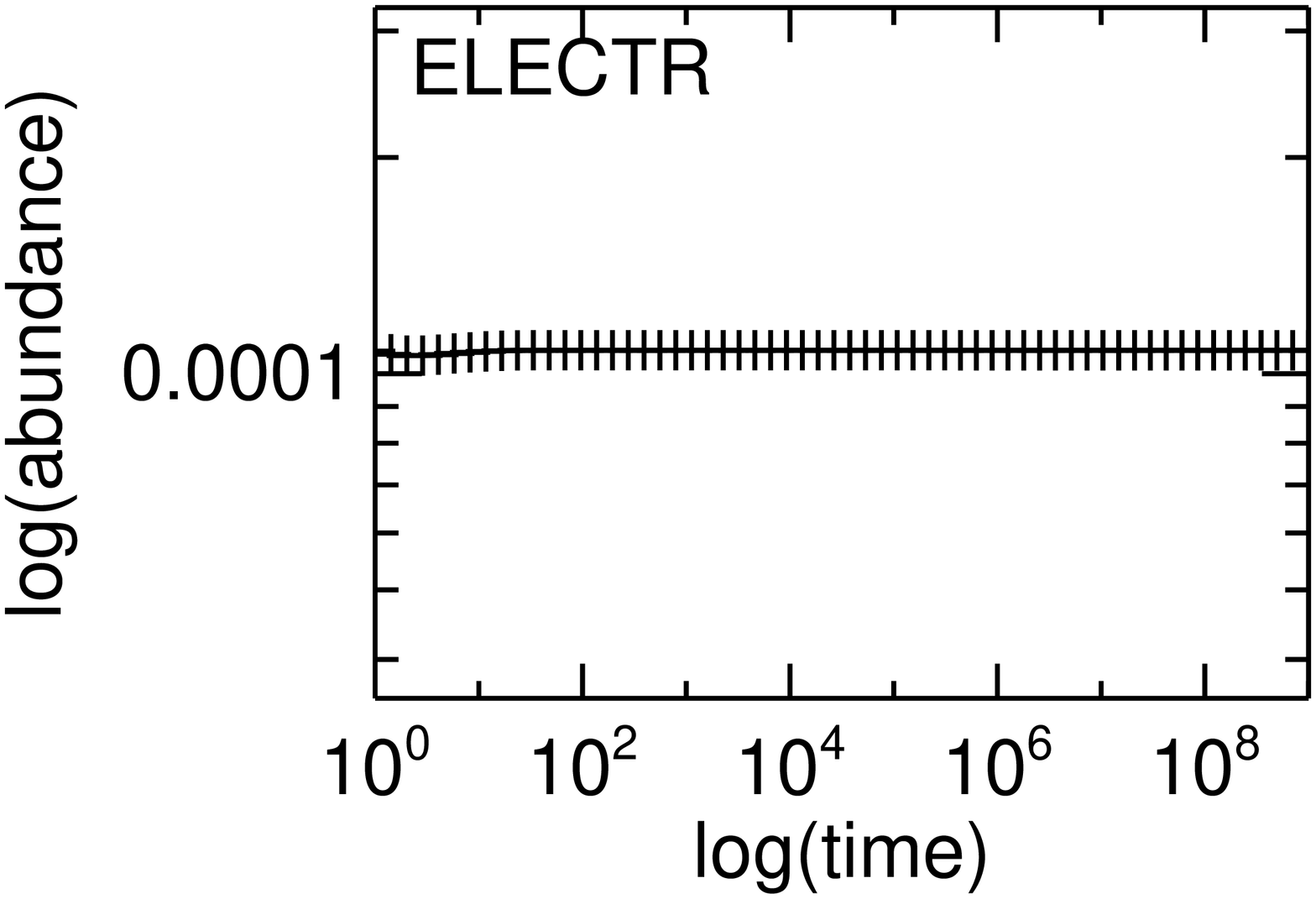}
\includegraphics[width=0.21\textwidth,clip=,angle=90]{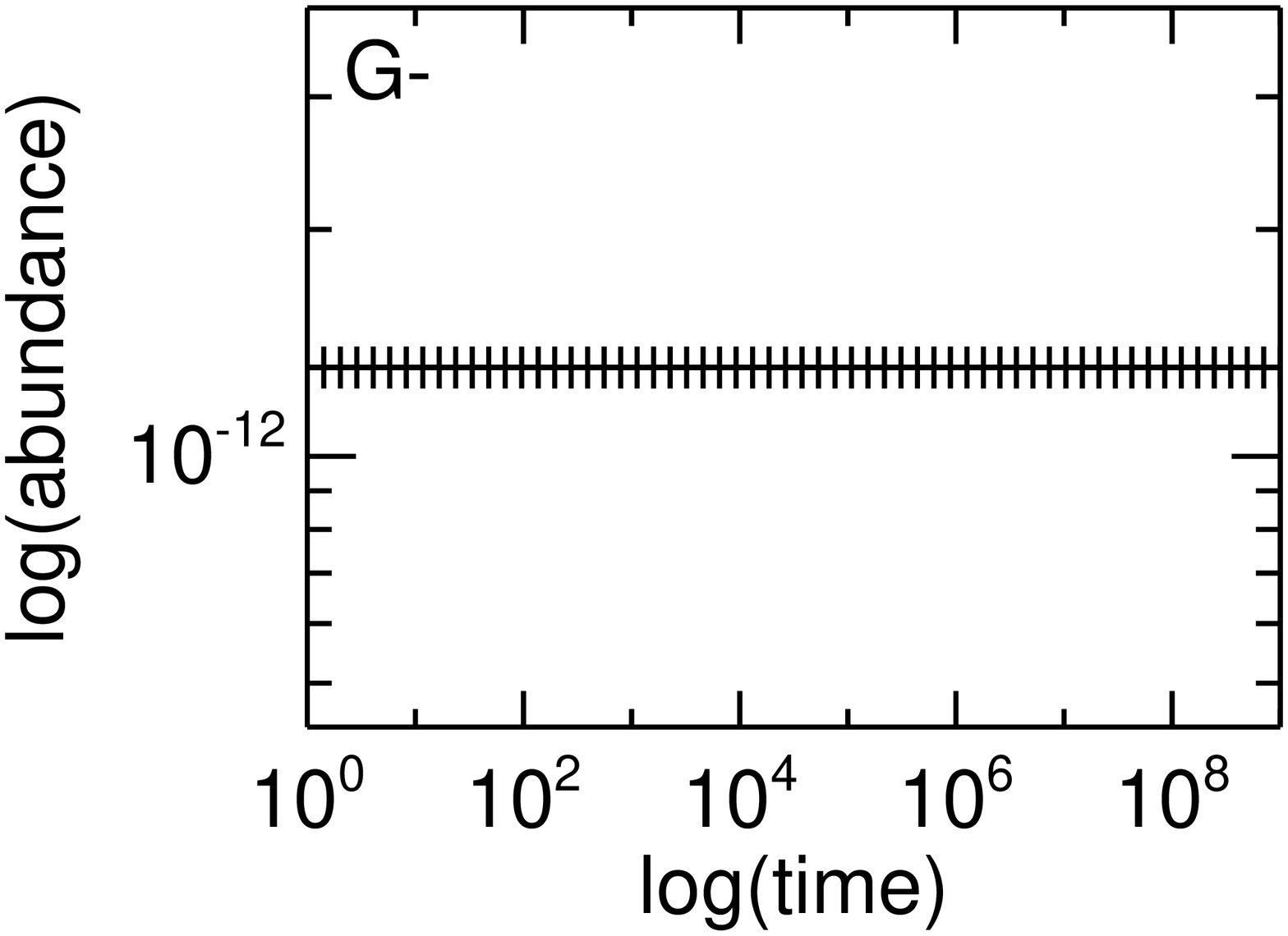}\\
\includegraphics[width=0.21\textwidth,clip=,angle=90]{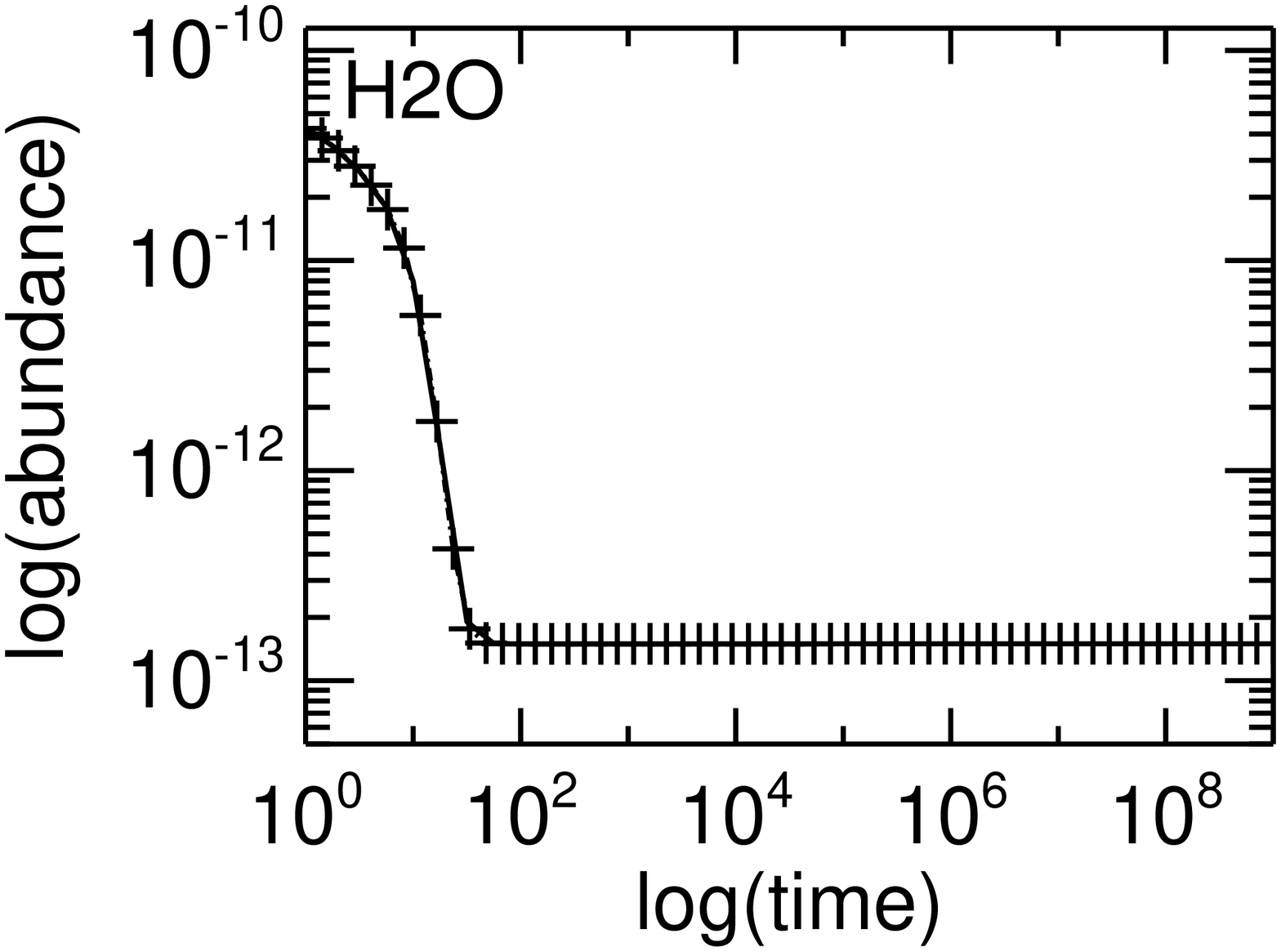}
\includegraphics[width=0.21\textwidth,clip=,angle=90]{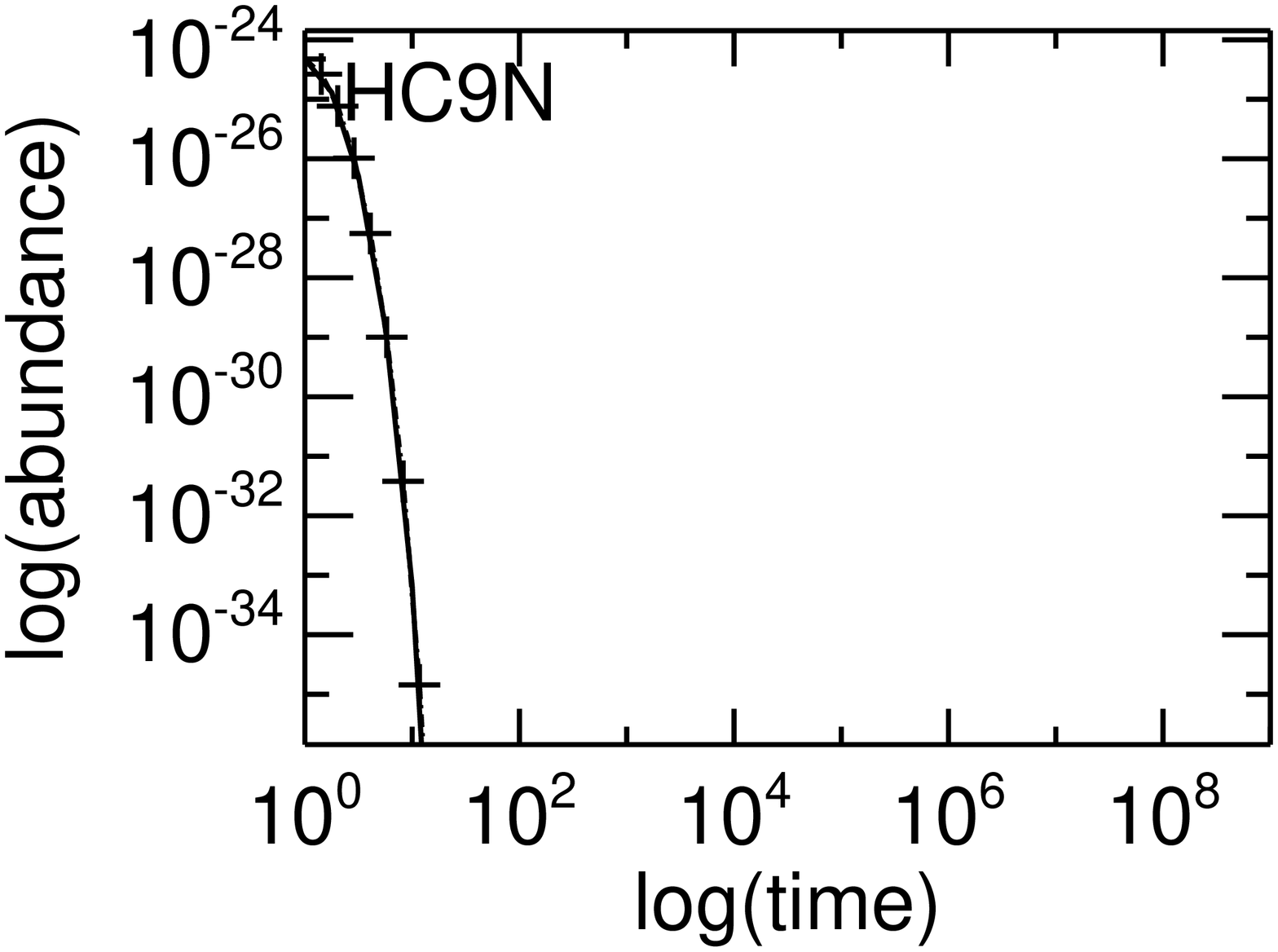}
\includegraphics[width=0.21\textwidth,clip=,angle=90]{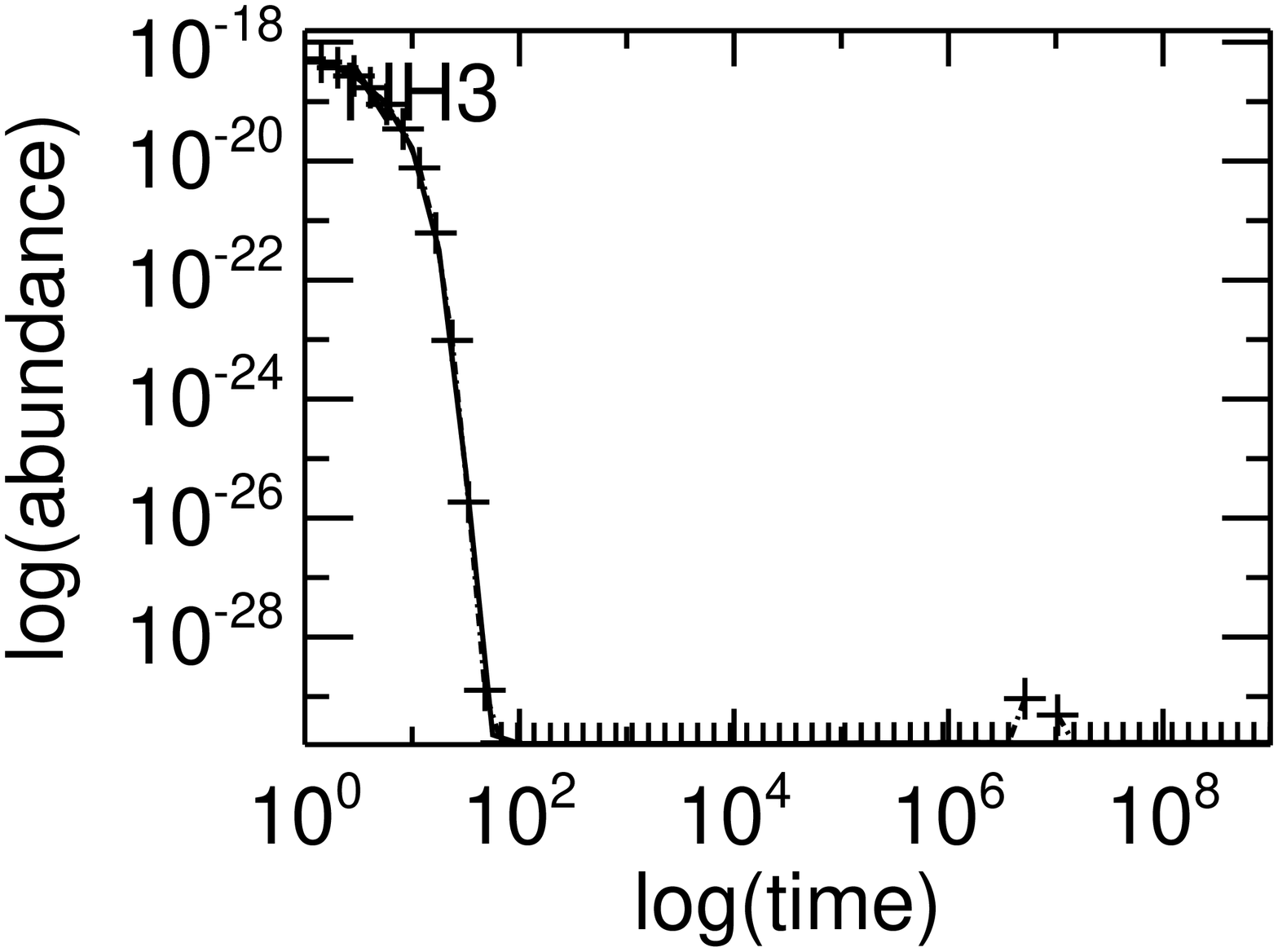}\\
\includegraphics[width=0.21\textwidth,clip=,angle=90]{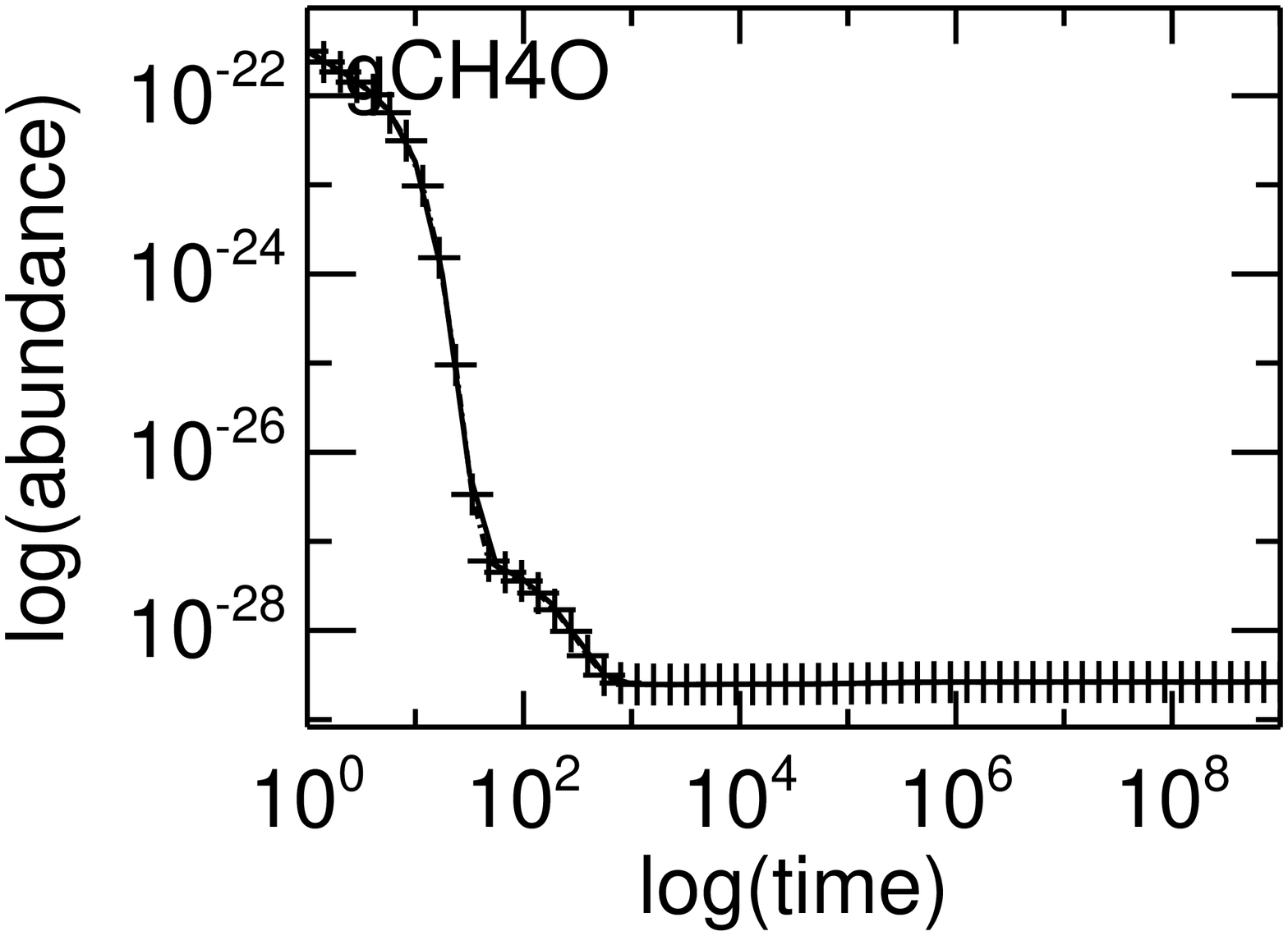}
\includegraphics[width=0.21\textwidth,clip=,angle=90]{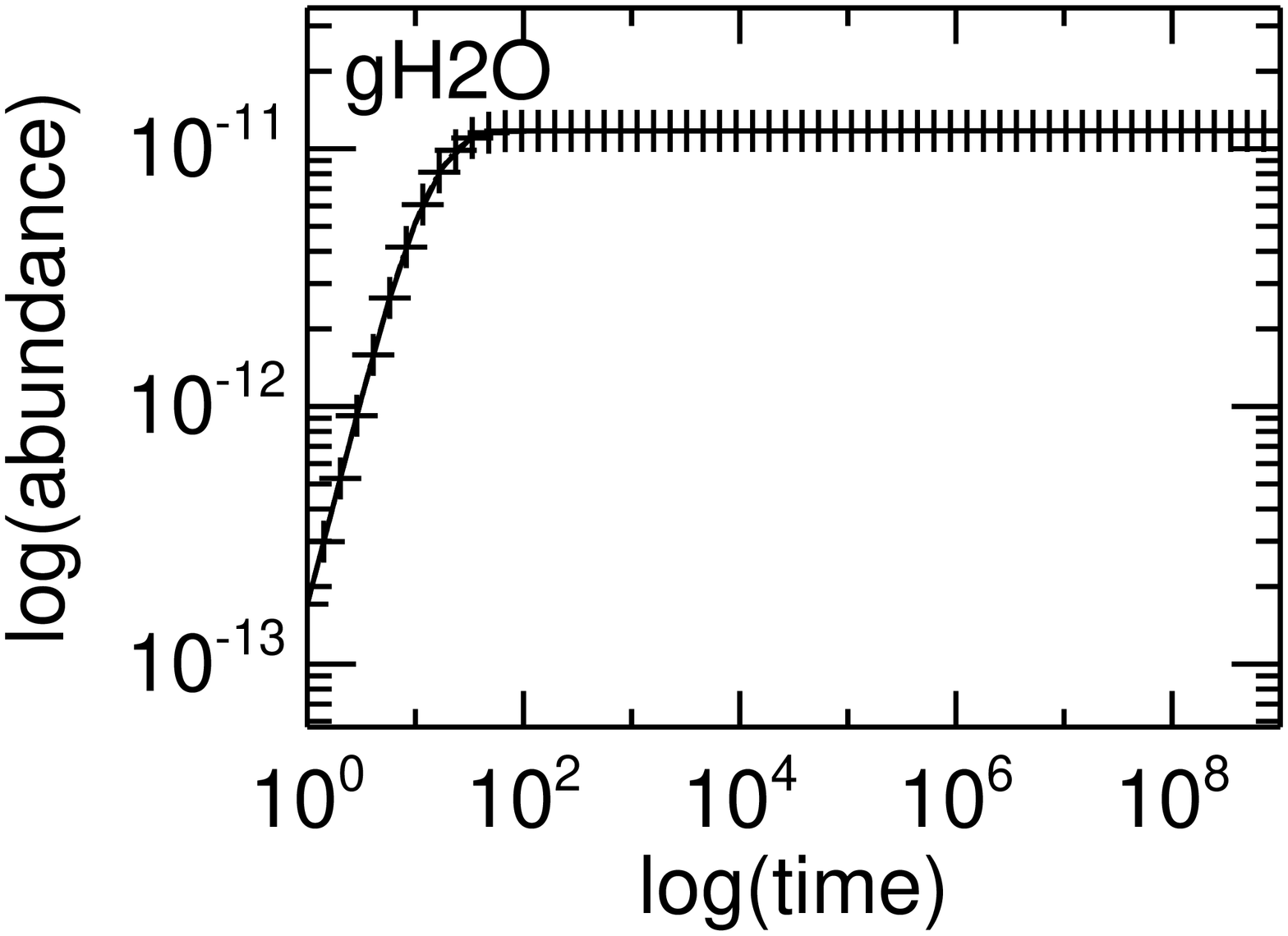}
\includegraphics[width=0.21\textwidth,clip=,angle=90]{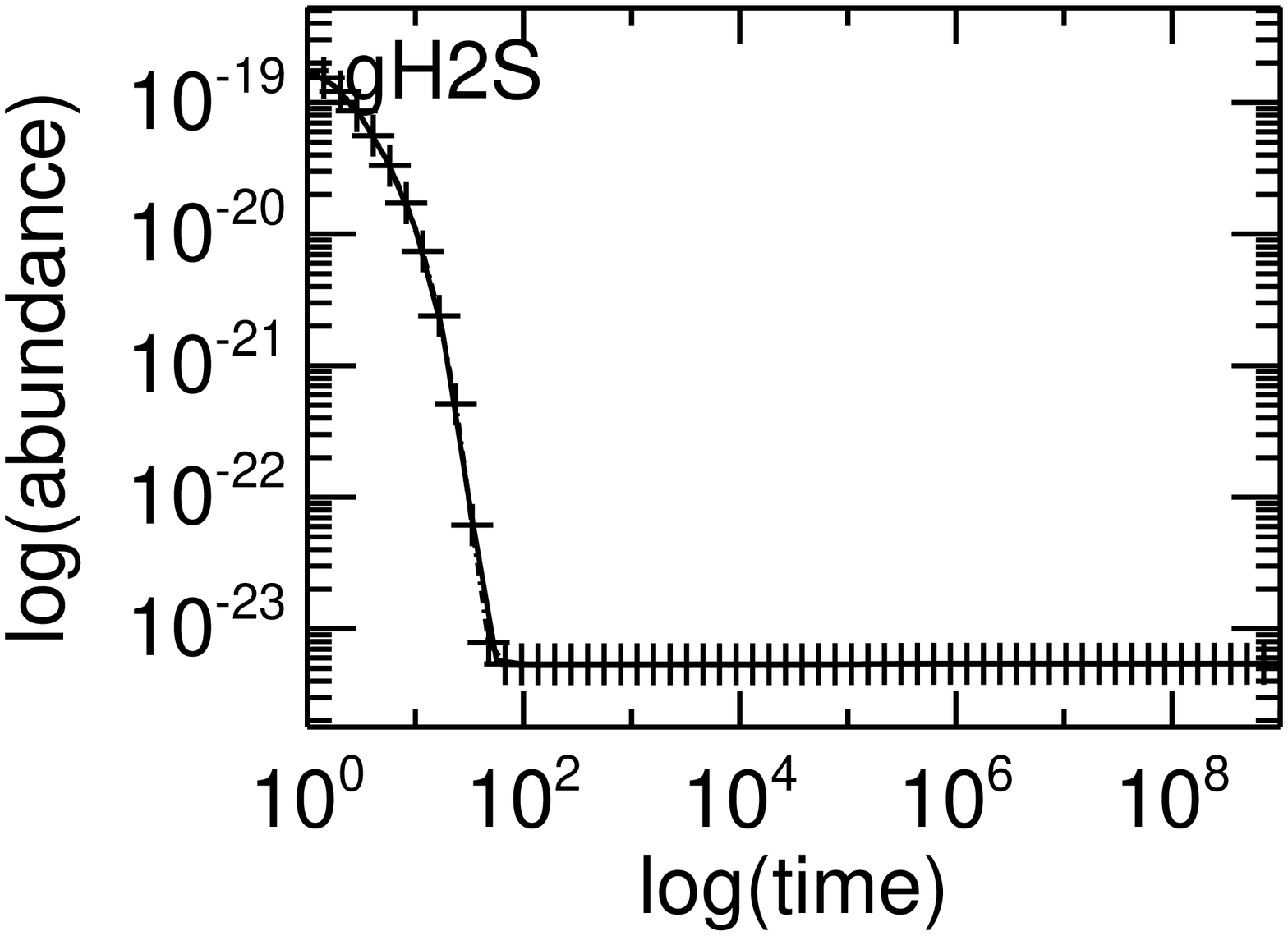}\\
\includegraphics[width=0.21\textwidth,clip=,angle=90]{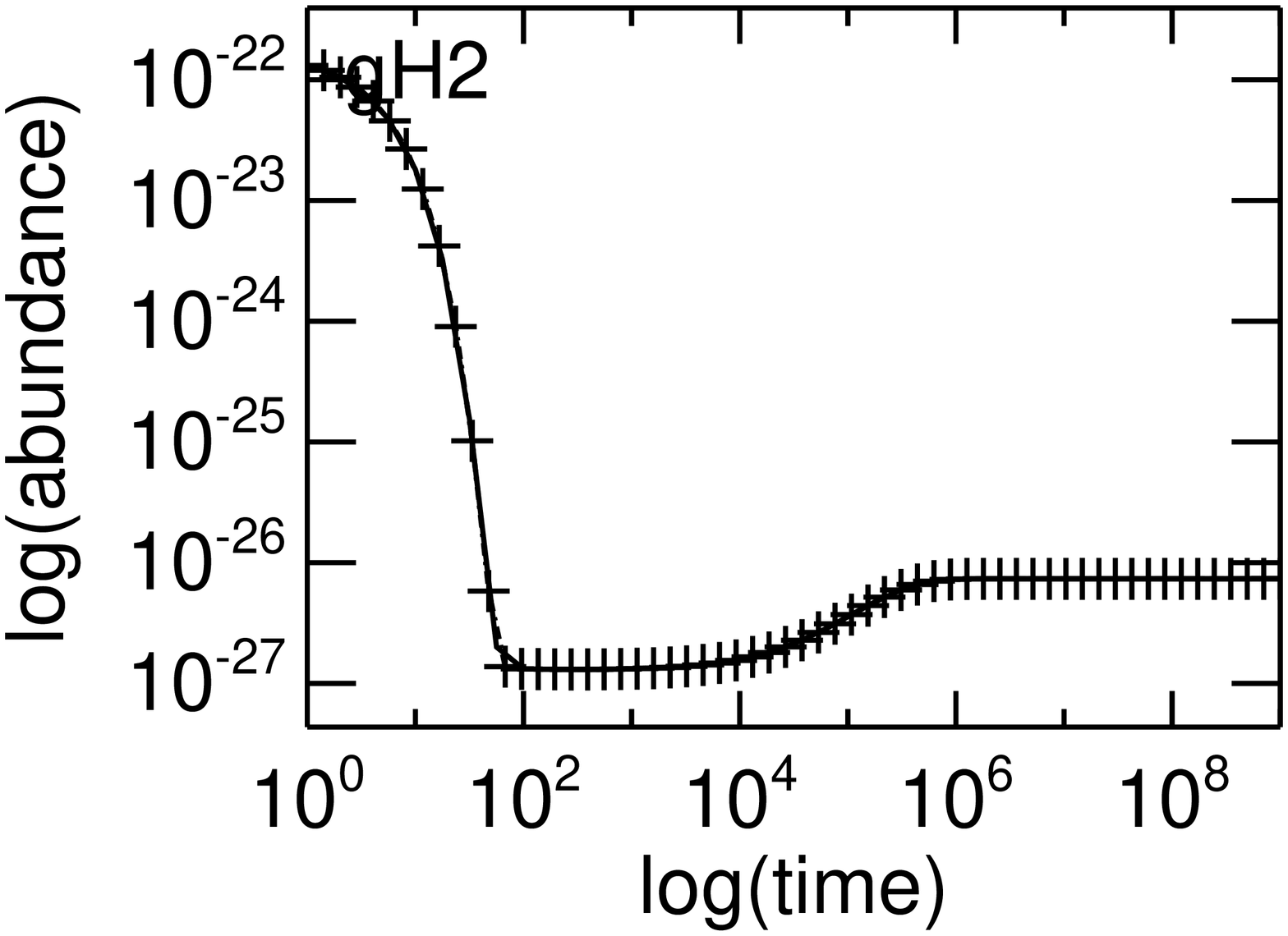}
\includegraphics[width=0.21\textwidth,clip=,angle=90]{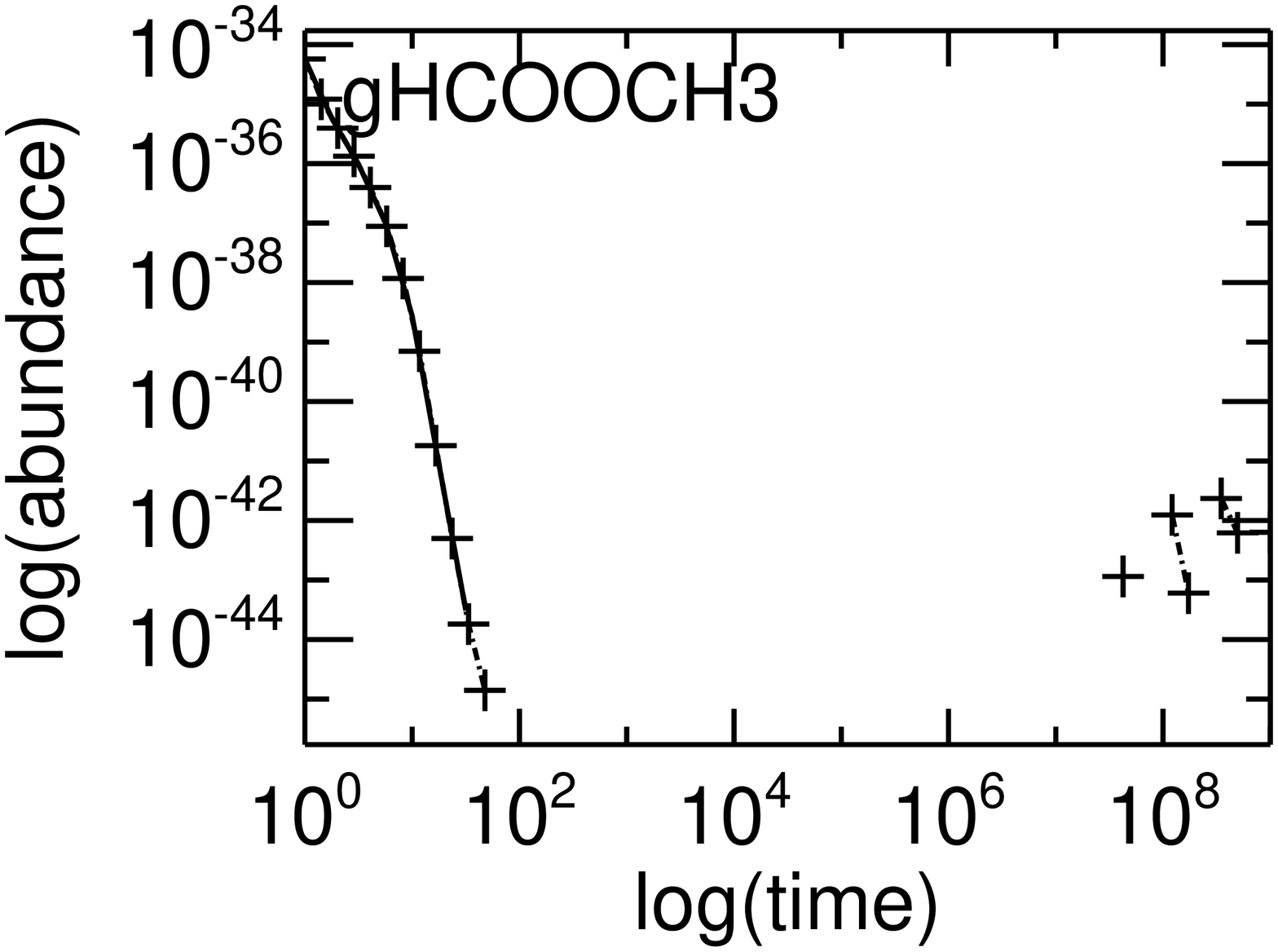}
\includegraphics[width=0.21\textwidth,clip=,angle=90]{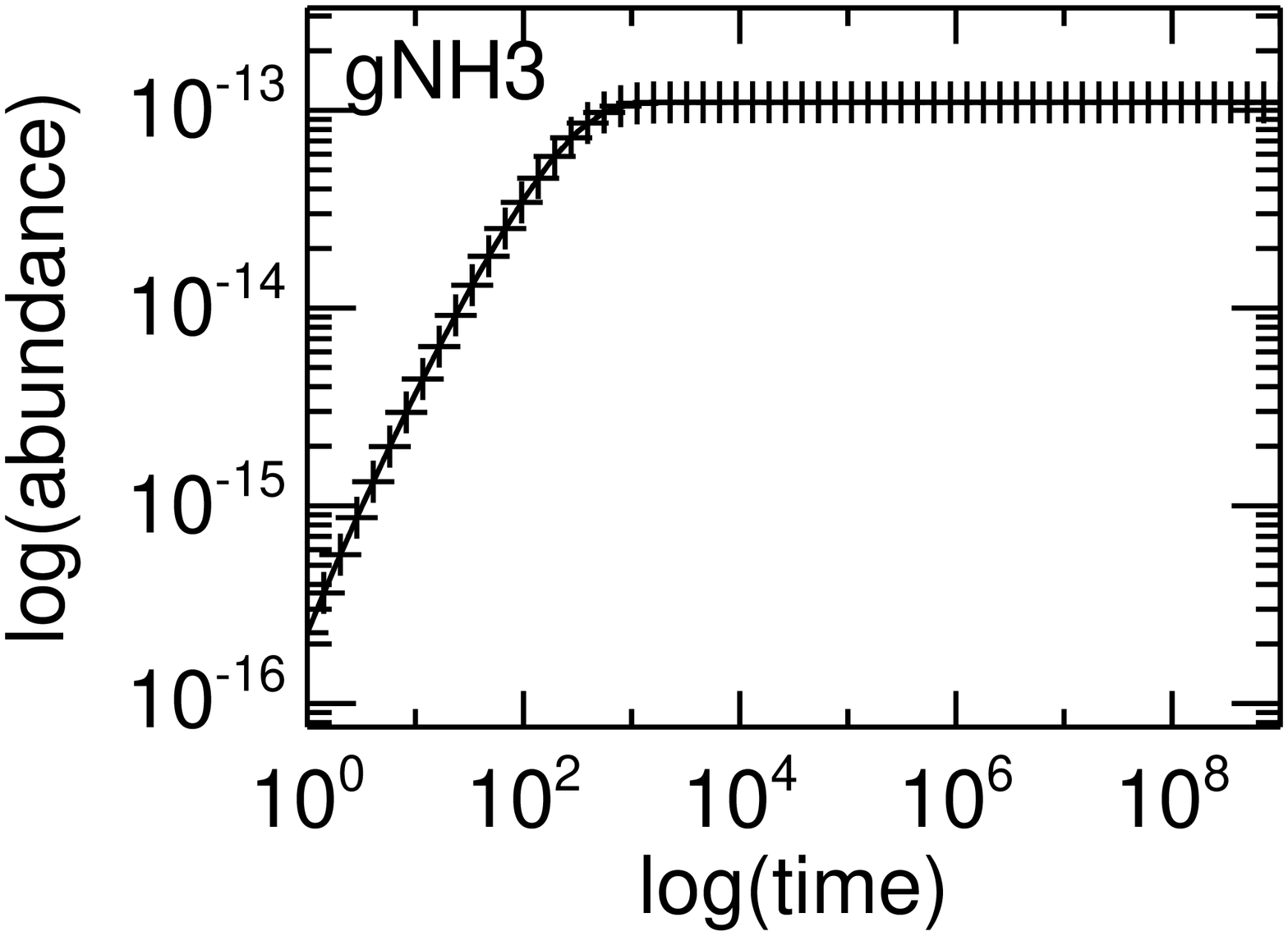}
\caption{The same as in Fig.~\ref{fig:DISK1} but
for the ``DISK3'' case (DM Tau, $r=100$~AU, atmosphere).}
\label{fig:DISK3}
\end{figure*}

Using these 5 representative benchmarking models and our extended chemical
network, time-dependent chemistry is calculated for the entire $10^9$~years of
evolution. A perfect agreement is achieved
between the Bordeaux and Heidelberg chemical models for all considered
physical conditions, all species, and all time moments. The results for assorted
chemical species representing various
chemical families and various degree of complexity are compared in
Figs.~\ref{fig:TMC1}--\ref{fig:DISK3}. The good agreement is in fact not surprising when only chemistry is concerned.
Both codes are based on the rate equations approach to model chemical
processes and use well documented and robust procedures to handle a multitude of
complex physico-chemical processes (e.g., photodissociation, cosmic ray
desorption, photoprocessing of ices, surface reactions).

This agreement required all the
constants, reaction rates, and parameters of the physical models between the two
astrochemical codes to match perfectly. There are many important parameters not always
mentioned in description of a chemical model, and which may
hamper a comparison of results. In what follows, we discuss major problems
that arose during course of our benchmarking study.

%Such a wonderful scientific achievement could only be possible thanks to
%exhausting and time-consuming efforts to bring into an agreement all the
%constants, reaction rates, and parameters of the physical models between the two
%astrochemical codes. There are many important parameters that often not
%mentioned in description of a chemical model or simply overlooked, and which may
%hamper a comparison of results. In what follows, we discuss major problems
%that arose during course of our benchmarking study.

The first priority is to check that both codes have exactly the same values of
fundamental constants expressed in the same physical units, and additional
useful constants (e.g., year in seconds, UV albedo of the dust, etc.).
Second, the ``standard'', well-defined physical parameters that can be
defined as constants in the codes (like the cosmic ray ionization rate, the FUV
IS, etc.) has to be checked. Our next major obstacle was different
definition of gas particle density: one group uses pure hydrogen nucleus density while
another uses mass density of the gas (with all heavy elements included).
After we began to use the same units for the FUV radiation field and the same
conversion factor between $N_{\rm H}$ and $A_{\rm V}$ our models showed perfect
agreement for a pure gas-phase chemical network.

As a next point, we added gas-grain interactions to our models. After grain
properties (shape, radius, density, surface density of sites, porosity,
material, etc.) are fixed, apparently the best approach is to add reactions
between atomic and molecular ions with charged grains, grain re-charging, and
electron sticking to grains. Next, one has to adopt the value of the sticking
coefficient of other gas-phase species.
An unexpected issue at that stage came from the fact that ``ALCHEMIC''
used atomic masses including isotopes while ``NAUTILUS'' used atomic
masses of major isotope only. Next, it is extremely important for disk chemical
models under comparison to have similar desorption and diffusion energies
of surface molecules since gas-grain interactions and build-up of thick complex
icy mantles are powerful in protoplanetary disks. A substantial portion of
time to reach the perfect agreement between the two models was spent to
properly implement desorption mechanisms. We did not consider UV
photodesorption along with thermal and CRP-driven desorption because it would
require a proper description of the UV radiation transport through the disk
model. As we relied on the same rate equations approach for surface chemistry,
the good agreement between the full models was reached soon. Here one has to be
careful with treatment of homogeneous reactions (i.e., involving the same
species, e.g. H $+$ H $\rightarrow$ H$_2$), as their rates by statistical
arguments are only half of those for heterogeneous reactions. Last, but not
least, for easy comparison of results it is important to specify in detail the
output format of the simulated data.

\section{Conclusion \& Summary}
\label{sum}
We present in this paper the results of several detailed benchmarks of the two
dedicated chemical codes for the Heidelberg and Bordeaux astrochemistry groups.
Both codes are used to model time-dependent chemical evolution of
molecular clouds, hot cores and corinos, and protoplanetary disks. The codes
use the recent osu\_03\_2008 gas-phase ratefile, supplemented with an extended
list of gas-grain and surface processes. A detailed
description of the codes, along with considered chemical processes and means to
compute their rates are presented. Five representative physical models are
outlined, and their chemical evolution is simulated.  The first case (``TMC1'')
is relevant to the chemistry of a dense cloud core, the second case (``Hot
Corino'') is similar to a warm star-forming region, e.g. an inner part of a hot core
or corino. The last three models correspond to various disk layers at
the radius of $\approx 100$~AU, representing distinct chemistry.
In all these benchmarking test runs, despite a large range of considered
physical parameters like temperature of gas and dust, density, and UV intensity,
we found perfect agreement between the codes. We however had to compare and bring 
into agreement step-by-step a
large number of parameters and description of chemical processes in ``ALCHEMIC''
and ``NAUTILUS''.
Moreover, this benchmark study allowed us to correct some minor
errors and helped us to become fully confident in our chemical simulations and
predictions. We hope that this study and detailed report on all components of
the codes and models will be helpful for other astrochemical models to improve
their quality. This is an essential step toward development of more
sophisticated chemical models of the ISM and disks, since ALMA, the
extremely powerful observational facility, will soon become operational.
With this instrument, the quality of molecular
interferometric maps will drastically improve, allowing observers and
astrochemists to work together in order to investigate the chemistry
of the planetary-forming zones in disks ($r < 30$~AU) in great detail.

\begin{acknowledgements}
%The authors are thankful to the anonymous referee for valuable comments and suggestions.
This research has made use of NASA's Astrophysics Data System.
DS acknowledges support by {\it Deutsche Forschungsgemeinschaft} through
SPP~1385: ``The first ten million years of the solar system - a
planetary materials approach'' (SE 1962/1-1).
SG, AD, VW, FH, and VP are
financially supported by the French Program ``Physique Chimie du
Milieu Interstellaire'' (PCMI).
\end{acknowledgements}

%\bibliographystyle{aa}
%\bibliography{apj-jour,refs}

% \appendix

%\section{Benchmark Figures}
%%%%%%%%%%%%%%%%%%%%%%%%%%%%%%%%%%%%%%%%%%%%%%%%%%%%%%%%%%%%%%%%%%%
% Figures:
%%%%%%%%%%%%%%%%%%%%%%%%%%%%%%%%%%%%%%%%%%%%%%%%%%%%%%%%%%%%%%%%%%%
\clearpage

\clearpage

\clearpage

\clearpage

\clearpage

%%%%%%%%%%%%%%%%%%%%%%%%%%%%%%%%%%%%%%%%%%%%%%%%%%%%%%%%%%%%%%%%%%%
% Tables:
%\section{Tables of physical and chemical Conditions}
%%%%%%%%%%%%%%%%%%%%%%%%%%%%%%%%%%%%%%%%%%%%%%%%%%%%%%%%%%%%%%%%%%%
% Physical conditions:
\clearpage
\begin{table}
\begin{minipage}[t]{\columnwidth}
\caption{Physical models}
\label{tab:phys_mods}
\centering
\renewcommand{\footnoterule}{}  % to avoid a line before footnotes
\begin{tabular}{lllll}
\hline\hline
Model & T & n(H+2H$_2$) & Av & $\chi$ \footnote{FUV, $6~<~h\nu~<~ 13.6$~eV, (Draine~1978)} \\
      & [K] & [cm$^{-3}$] & [mag] & \\
\hline
TMC1     & 10    & 2(4)\footnote{A(-B) means $\rm A\times $10$^{-B}$}     & 10    & $\chi_0$ \\
HOT CORE & 100   & 2(7)     & 10   & $\chi_0$ \\
DISK1    & 11.4  & 5.41(8)  & 37.1  & 428.3$\chi_0$ \\
DISK2    & 45.9  & 2.59(7)  & 1.94  & 393.2$\chi_0$ \\
DISK3    & 55.2  & 3.67(6)  & 0.22  & 353.5$\chi_0$ \\
\hline
\end{tabular}
\end{minipage}
\end{table}

\begin{table}
\begin{minipage}[t]{\columnwidth}
\caption{Constants and fixed parameters}
\label{tab:const}
\centering
\renewcommand{\footnoterule}{}  % to avoid a line before footnotes
\begin{tabular}{ll}
\hline\hline
Name & Value \\
\hline
$k_B$ & $1.38054\times 10^{-16}$ erg K$^{-1}$ \\
$\hbar$ & $1.05459\times 10^{-27}$~erg s$^{-1}$ \\
$\zeta_{\rm CR}$ & $1.3\times 10^{-17}$~s${-1}$ \\
$a_g$ & 0.1~$\mu$m \\
$\rho_d$ & 3 g cm$^{-3}$ \\
$m_{d/g}$ & 0.01 \\
$N_S$ & $1.5\times 10^{15}$ sites cm$^{-2}$ \\
$S$ & $1.885\times 10^6$ \\
$m_p$ & $1.66054\times 10^{-24}$ g \\
$\mu$ & 1.43 amu \\
$T_{crp}$ & 70 K \\
$f$ & $3\times 10^{-19}$ \\
$b$ & 1\AA \\
$T_{\rm diff}/T_{\rm des}$ & 0.77 \\
\hline
\end{tabular}
\end{minipage}
\end{table}

% Initial abundances:
%\clearpage
\begin{table}
\begin{minipage}[t]{\columnwidth}
\caption{Initial abundances}
\label{tab:init_abund}
\centering
\renewcommand{\footnoterule}{}  % to avoid a line before footnotes
\begin{tabular}{ll}
\hline\hline
Species & $n(X)$/$n_{\rm H}$\\
\hline
He    & 9.00(-2)\footnote{A(-B) means $\rm A\times $10$^{-B}$} \\
H$_2$ & 5.00(-1) \\
C$^+$ & 1.2(-4) \\
N     & 7.6(-5) \\
O     & 2.56(-4) \\
S$^+$ & 8.00(-8) \\
Si$^+$& 8.00(-9) \\
Na$^+$& 2.00(-9) \\
Mg$^+$& 7.00(-9) \\
Fe$^+$& 3.00(-9) \\
P$^+$ & 2.00(-10)\\
CL$^+$& 1.00(-9) \\
\hline
\end{tabular}
\end{minipage}
\end{table}

\end{document}